\documentclass[12pt]{article}
\usepackage{thumbpdf}
\usepackage[pdftex,
        colorlinks=true,
        urlcolor=rltblue,               
        anchorcolor=rltbrightblue,
        filecolor=rltgreen,             
        linkcolor=rltred,               
        menucolor=webblue,
        citecolor=webgreen,
        pdftitle={A Document With An Image},
        pdfauthor={Udo K. Schuermann, Ringlord Technologies},
        pdfsubject={A Demonstration with Extra Cheese},
        pdfkeywords={latex pdf gimp batch thumbnails hyperlinks},
        pagebackref,
        pdfpagemode=None,
        bookmarksopen=true]{hyperref}
\usepackage[pdftex]{graphicx}
\usepackage{color}
\usepackage{amsmath}
\usepackage{amssymb}
\usepackage{euscript}
\usepackage{multirow}
\usepackage{color}
\usepackage{colortbl}
\usepackage{calc}

\renewcommand\arraystretch{1.45}

\newcommand{\figtxt}[1]{\footnotesize{#1}}

\makeatletter
\newlength\twolinebox@linelength
\newlength\twolinebox@columnheight
\newcommand{\twolinebox}[2]{%
   \setlength{\twolinebox@linelength}%
             {\maxof{\widthof{#1}}{\widthof{#2}}}%
   \setlength{\twolinebox@columnheight}{\heightof{#1}+\depthof{#1}+0.2em+0.4em/2+\heightof{0}/2}%
   \raisebox{0pt}[\twolinebox@columnheight][\heightof{\vbox{\vskip0.2em\hbox to 
   \twolinebox@linelength {#1\hfil}\vskip0.4em\hbox to 
   \twolinebox@linelength {#2\hfil}}}+\depthof{\vbox{\vskip0.2em\hbox to 
   \twolinebox@linelength {#1\hfil}\vskip0.4em\hbox to 
   \twolinebox@linelength {#2\hfil}}}-\twolinebox@columnheight+0.2em]{\vbox to 
   \twolinebox@columnheight{\vskip0.2em\hbox to 
   \twolinebox@linelength {#1\hfil}\vskip0.4em\hbox to 
   \twolinebox@linelength {#2\hfil}}}%
}

\newcommand{\tw}      {\textwidth}

\newcommand{\ie}[0]   {\textit{i.e.}}
\newcommand{\eg}[0]   {\textit{e.g.}}

\newcommand{\mm}[1]   {\mathrm{#1}}
\newcommand{\mbf}[1]  {\mathbf{#1}}

\newcommand\WINHAC[0] {\textsf{WINHAC}}

\newcommand\LHAPDF[0] {LHAPDF}

\newcommand\Pythia[0] {\textsc{Pythia}}

\newcommand{\Cen}     {\mathrm{cen}}
\newcommand{\Min}     {\mathrm{min}}
\newcommand{\Max}     {\mathrm{max}}

\newcommand{\DfDx}[2]     {\frac{d\,{#1}}{d\,{#2}}}
\newcommand{\flatDfDx}[2] {d\,{#1}/d\,{#2}}
\newcommand{\cotan}       {\mathrm{cotan}\,}

\newcommand{\percent}{\,\%}

\newcommand{\MeV}    {\,\mathrm{MeV}}
\newcommand{\GeV}    {\,\mathrm{GeV}}
\newcommand{\TeV}    {\,\mathrm{TeV}}

\newcommand{\plus}   {{+}}
\newcommand{\minus}  {{-}}

\newcommand{\GF}     {{G_\mathrm{F}}}
\newcommand{\MW}     {{M_W}}

\newcommand{\Wp}     {{W^+}}
\newcommand{\BFWp}   {\mathbf{\Wp}}

\newcommand{\MWp}    {{M_\Wp}}

\newcommand{\Wm}     {{W^-}}   
\newcommand{\BFWm}   {\mathbf{\Wm}}

\newcommand{\MWm}    {{M_\Wm}}

\newcommand{\lm}     {{l^-}}
\newcommand{\lp}     {{l^+}}

\newcommand{\yW}     {{y_W}}
\newcommand{\pTW}    {{p_{T,W}}}

\newcommand{\pTl}    {{p_{T,l}}}
\newcommand{\rhol}   {{\rho_{l}}}
\newcommand{\pTlp}   {{p_{T,\lp}}}
\newcommand{\pTlm}   {{p_{T,\lm}}}
\newcommand{\pTnu}   {{p_{T,\nu}}}
\newcommand{\mTlnu}  {{m_{T,\,l\,\nu_l}}}

\newcommand{\ETmiss} {{\slashii E_T}}
\newcommand{\etal}   {{\eta_l}}

\newcommand{\thetal}       {{\theta_l}}

\newcommand{\costhetaWlwrf}{{\cos\theta_{W,l}^{\ast}}}
\newcommand{\thetaWlwrf}   {{\theta_{W,l}^{\ast}}}

\newcommand{\val}     {{\mathrm{(v)}}}

\newcommand{\ppbar}   {{p\,\pbar}}
\newcommand{\pbar}    {{\bar p}}
\newcommand{\pp}      {{p\,p}}
\newcommand{\dd}      {{d\,d}}

\newcommand{\kT}      {{k_T}}
\newcommand{\pT}      {{p_T}}

\newcommand{\FlatDsigmaDobs}[1]
                             {{d\,\sigma}/{d\,{#1}}}

\newcommand{\Vckmsqr}[2]{{\left|V_{#1#2}\right|}^2}

\newcommand{\dof}       {{\mathrm{dof}}}
\newcommand{\chiD}      {{\chi^2}}
\newcommand{\chiDmin}   {{\chi_\mathrm{min}^2}}
\newcommand{\Asym}[1]   {\mathrm{Asym}^{(+,-)}\left(#1\right)}
\newcommand{\FlatAsym}[1]
                        {\mathrm{Asym}^{(+,-)}(#1)}
\newcommand{\DAsym}[1]  {\mathrm{DAsym}^{(+,-)}\left({#1}\right)}
\newcommand{\rec}       {{\mathrm{(rec.)}}}
\newcommand{\smear}     {{\mathrm{(smr.)}}}

\newcommand{\DeltaPM}   {\Delta_{(+,-)}}

\newcommand{\es}        {\varepsilon}

\def\slashii#1{\setbox0=\hbox{$#1$}            
  \dimen0=\wd0                                 
  \setbox1=\hbox{\sl/} \dimen1=\wd1            
  \ifdim\dimen0>\dimen1                        
     \rlap{\hbox to \dimen0{\hfil\sl/\hfil}}   
     #1                                        
  \else                                        
     \rlap{\hbox to \dimen1{\hfil$#1$\hfil}}   
     \hbox{\sl/}                               
  \fi}

\definecolor{rltbrightred}{rgb}{1,0,0}
\definecolor{rltred}{rgb}{0.75,0,0}
\definecolor{rltdarkred}{rgb}{0.5,0,0}
\definecolor{rltbrightgreen}{rgb}{0,0.75,0}
\definecolor{rltgreen}{rgb}{0,0.5,0}
\definecolor{rltdarkgreen}{rgb}{0,0,0.25}
\definecolor{rltbrightblue}{rgb}{0,0,1}
\definecolor{rltblue}{rgb}{0,0,0.75}
\definecolor{rltdarkblue}{rgb}{0,0,0.5}
\definecolor{webred}{rgb}{0.5,.25,0}
\definecolor{webblue}{rgb}{0,0,0.75}
\definecolor{webgreen}{rgb}{0,0.5,0}
\definecolor{Black}{rgb}{0,0,0}
\definecolor{Greymax}{rgb}{0.65,0.65,0.65}
\definecolor{Greycen}{rgb}{0.75,0.75,0.75}
\definecolor{Greymin}{rgb}{0.85,0.85,0.85}
\definecolor{hl}{rgb}{
                0.909803922,       
                0.82745098,               
                0.909803922}

\begin{document}

\begin{titlepage}
 
\begin{flushright}
\bf Final version June/2009
\end{flushright}
\vspace{2mm}

\begin{center}
{\LARGE\bf Measurement of $\;\mbf{\MWp - \MWm}$ \vspace{2mm}\\
             at~LHC} \\
\end{center}

\vspace{2mm}

\begin{center}
{\large\bf   F.~Fayette$^{a}$, 
                   M.~W.~Krasny$^{a}$, 
                   W.~P\l{}aczek$^{b,a}$ 
                  {\rm and}  A.~Si\'odmok$^{b,a}$  }

{\em $^a$LPNHE, Pierre et Marie Curie Universit\'es Paris VI et Paris VII,\\ 
         Tour 33, RdC, 4, pl. Jussieu, 75005 Paris, France}\\  \vspace{2mm}
{\em $^b$Marian Smoluchowski Institute of Physics, Jagiellonian University,\\
         ul.\ Reymonta 4, 30-059 Cracow, Poland}\\ \vspace{2mm}

\end{center}

\vspace{2mm}
\begin{abstract}

This paper is the second of the series of papers proposing 
{\it dedicated strategies} for precision measurements of the Standard
Model parameters at the LHC. 
The common feature of these strategies 
is their robustness with respect to the systematic measurement 
and modeling error sources.
Their impact on the precision of the measured parameters is 
reduced using dedicated observables and dedicated 
measurement procedures  which exploit flexibilities 
of the collider and detector running modes. 
In the present paper we focus our attention on the 
measurement of the charge asymmetry of the $W$-boson mass.
This measurement is of primordial importance for the LHC experimental 
program, both as a direct test of the charge-sign-independent coupling of the 
$W$-bosons to the matter particles and as a necessary first step towards the precision 
measurement of the charge-averaged $W$-boson mass. 
We propose and evaluate the LHC-specific strategy to measure
the mass difference between the positively and negatively charged $W$-bosons,
$\MWp-\MWm$. We show that its  present precision can 
be improved at the LHC by a factor of $20$. 
We argue that such a precision is beyond the reach of 
the standard measurement and calibration 
methods imported to the LHC from the Tevatron program.   
\end{abstract}

\vspace{0mm}
\begin{center}
{\it Final accepted version for the European Physical Journal C}
\end{center}
 
\vspace{1mm}
\begin{flushleft}
{\bf Final version,\\ 
     June~2009}
\end{flushleft}  

\vspace{2mm}
\footnoterule
\noindent
{\footnotesize
$^{\star}$The work is partly supported by the program of the French--Polish
cooperation between IN2P3 and COPIN No.\ 05-116 
and 
by the EU Marie Curie Research Training Network grant 
under the contract No.\ MRTN-CT-2006-035505.
}

\end{titlepage}

\section{Introduction}\label{s_introduction}

As demonstrated by Gerhard L\"uders and Wolfgang Pauli \cite{Pauli}, any Lorentz-invariant quantum field theory 
obeying the principle of locality must be CPT-invariant.
For theories with spontaneous symmetry breaking, 
the requirement of the Lorentz-invariance concerns both the interactions of the 
fields and the vacuum properties. 
In the CPT-invariant quantum field theories masses of particles and their antiparticles are equal.

The Standard Model is CPT-invariant. In this model the $\Wp$ and  $\Wm$ bosons are constructed  
as each-own antiparticles, which couple to leptons with precisely the same $SU(2)$ strength,  $g_{W}$.
The hypothesis of the exact equality of their masses is pivotal for the present understanding of 
the electroweak sector of the Standard Model.
It is rarely put in doubt -- even by those who consider the CPT invariant Standard Model as 
only a transient model of particle interactions.
However, from a purely  experimental perspective, even such a basic assumption must be checked 
experimentally to  the highest achievable precision.

The most precise, indirect experimental constraint on equality of the  
 $\Wp$ and  $\Wm$ masses can be derived from the measurements of the life-time asymmetries 
of positively and negatively charged muons \cite{Yao:2006px}. These measurements, if interpreted 
within the Standard Model framework, verify  the equality  of the masses of the  $\Wp$ and  $\Wm$ bosons 
to the precision of $1.6\,$MeV. Such a precision cannot be reached in direct measurements of 
their mass difference.
The experimental uncertainty of the directly-measured  mass difference, 
$\MWp-\MWm=-200 \pm 600\,$MeV~\cite{Yao:2006px}, 
is about $400$ times higher. Recently 
the CDF collaboration~\cite{Aaltonen:2007ps}  
measured $\MWp - \MWm$ to be $257 \pm 117\,$MeV
in the electron decay channel,  and  $286 \pm 136\,$MeV  in the muon decay channel.
These measurements provide to this date  the best model-independent verification of the equality of the masses of the two charge 
states of the $W$-bosons. They are compatible with the charge symmetry hypothesis. 
It is worth stressing,   that the present precision of the direct measurement of
the {\it charge-averaged} mass of the $W$-boson, $M_{W}=80.398 \pm 0.025\,$GeV, 
derived under the assumption of  the equality of the masses, 
is by a factor of $10$ better than the precision of the direct individual measurements 
of the masses of its charged states.

Apart from the obligatory precision test of the CPT-invariance of
the spontaneously broken gauge-theory with a priori unknown  
vacuum properties, we would like to measure $\MWp - \MWm$ at the LHC for the three following reasons.   
Firstly, we would like to constrain possible future  
extensions of the Standard Model in which the effective coupling of the Higgs particle(s) to the $W$-boson depends 
upon its charge. Secondly, contrary to the Tevatron case,  the measurement of the charge-averaged mass at the LHC  
cannot be dissociated from, and must be preceded by, the  measurement of the masses of the $W$-boson charge states.
Therefore, any effort to improve the precision of the direct measurement of the 
charge-averaged mass of the $W$-boson and, as a consequence, the  indirect constraint on  the 
mass of the Standard Model
Higgs boson,  must be, in our view,  preceded  by a  precise understanding of the $W$-boson charge asymmetries.  
Thirdly, we would like to measure the $W$-boson 
polarisation asymmetries at the LHC. Within  the Standard Model framework the charge asymmetries provide 
an important indirect access to the polarisation asymmetries. 
This is a direct consequence of both the CP conservation in the gauge-boson 
sector and the purely ($V-A$)-type of the C- and P-violating coupling 
of the $W$-bosons to fermions. Any new phenomena contributing  to the $W$-boson 
polarisation asymmetries at the LHC must thus be reflected in the observed charge asymmetries.

The optimal strategies for measuring the charge-averaged mass of the $W$-boson
and for measuring directly the masses  of its charge-eigenstates are bound to be different. 
Moreover, the optimal strategies are bound to  be different at the LHC and  at
the Tevatron. 

At the Tevatron, producing equal numbers of the
$\Wp$ and $\Wm$ bosons, the measurement strategy was optimised
to achieve the best precision for the charge-averaged mass of the $W$-boson. For example, 
the  CDF collaboration \cite{Aaltonen:2007ps}   
traded off  the requirement of the precise relative control of the detector response 
to positive and negative particles over the full detector fiducial volume for a  weaker 
requirement of a precise relative control of charge-averaged
biases of the detector response in the left and right sides of the detector.
Such a strategy provided the most precise  measurement of the charge-averaged $W$-boson mass, 
but large measurement errors of the 
charge-dependent $W$-boson masses.  

If  not constrained by the beam transfer systems, 
the best dedicated, bias-free strategy for measuring  $\MWp - \MWm$ 
in proton--antiproton colliders would  be rather straightforward. 
It  would boil down to collide, for a fraction of time, 
the direction-interchanged beams of protons and anti-protons, associated with a simultaneous change   
of the sign of the solenoidal $B$-field in the detector fiducial volume.  Such a measurement 
strategy cannot be realised at the Tevatron, leaving to the LHC collider the task of improving 
the measurement precision. 

The statistical precision of the future measurements of the $W$-boson properties at the LHC 
will be largely superior to that achieved at the Tevatron. On the other hand,
it will be difficult to reach comparable or smaller systematic errors. 
The measurements of the $W$-boson mass and its charge 
asymmetry can no longer be factorised and optimised independently. 
The flavour structure of the beam particles will have to be controlled with a significantly 
better precision at the LHC than  at the  Tevatron. While being of limited importance for the  $\MW$ measurement  
at Tevatron, the  present knowledge of the momentum-distribution asymmetries of: (1) down and up valence 
quarks, and (2) of charm  and strange  quarks will limit significantly 
the achievable measurement precision.   
The `standard candles', indispensable for  precise experimental control of the reconstructed  lepton momentum scale -- 
the $Z$-bosons and `onia' -- will be less powerful in the case of  proton--proton collisions 
with respect to  the net-zero-charge proton--antiproton collisions. 
Last but not least, extrapolation of the strong interaction effects 
measured in the  $Z$-boson production  processes, to the processes of $W$-boson production will be 
more ambiguous due to an increased contribution of the bottom and charmed quarks.

Earlier studies of the prospects of the charge-averaged $W$-boson mass measurement by the 
CMS~\cite{CMS_W} and ATLAS~\cite{Atlas_W} collaborations ignored   
the above LHC-collider-specific effects and arrived at rather optimistic estimates
of the achievable measurement precision at the LHC. 
In our view, in order to improve the 
precision of the Tevatron experiments,  
both for the average and for the charge-dependent masses of the $W$-boson,    
some novel, dedicated strategies, adapted to the LHC environment must be developed. 
Such strategies will  have to employ    
full capacities of the collider and of the detectors in  reducing the impact of 
the theoretical, phenomenological and measurement uncertainties on the 
precision of the $W$-boson mass measurement.

This series of papers propose and evaluate the coherent  
strategies of measuring the mass of the $W$-boson, its width $\Gamma _W$ and their charge asymmetries at the LHC. 
In the introductory paper \cite{Krasny:2007cy} we have proposed the strategies which optimize 
the use of the $Z$-bosons  as a `standard candle' for the $W$-boson production processes. 
In the present one we propose 
and evaluate the LHC dedicated strategy for measuring the charge asymmetry of the $W$-boson mass. 
The  strategies for measuring the $W$-boson width, $\Gamma _W$, and its charge asymmetry, and 
the $W$-boson mass under the assumption of $\MWp=\MWm$ will be presented in separate papers. 
Such a sequence of papers reflects the order in which these measurements will, in our view,  have to be done.

This paper is divided into the following sections.
In Section~\ref{s_tools} the software tools which we have developed for 
our analysis are presented. 
In Section~\ref{s_charge_asym}
the sources of the charge asymmetries in the $W$ boson production 
and decay processes are discussed,  with particular emphasis on their distinctive features 
at $\pp$ and  $\ppbar$ colliders.   
Our measurement strategy for the LHC is presented  in 
Section~\ref{s_measurement_method}.
In Section~\ref{s_analysis_strategy}
we present an  analysis method used in quantitative evaluation  of 
the achievable measurement precision of the $W$-mass charge asymmetry.     
Modeling of the dominant systematic errors affecting the measurement
is discussed in Section~\ref{s_model_impact_sys}.  
Their  impact on the achievable precision of  $\MWp - \MWm$ is analysed 
in Section~\ref{sec_results}.
Finally, in Section~\ref{s_summary}, a summary of the results
and the outlook are presented.

\section{Tools}\label{s_tools}

\subsection{Monte Carlo event generator \WINHAC{}}\label{ss_tools_WINHAC}

The main tool that has been used for the study presented in this paper is
\WINHAC{}~\cite{Placzek2003zg,CarloniCalame:2004qw,Gerber:2007xk,Bardin:2008fn}. 
It is a dedicated Monte Carlo (MC) event generator for precision description
of the single $W$-boson production
and  decay at hadron colliders. It has been thoroughly tested and
cross-checked with independent calculations%
~\cite{Placzek2003zg,CarloniCalame:2004qw,Golonka:2005pn,Bardin:2008fn}.
This MC program has already been used in our previous studies of the experimental prospects 
for exploring the electroweak symmetry breaking mechanism and for 
the precision measurements of the Standard
Model parameters at the LHC \cite{Krasny:2005cb,Krasny:2007cy}.

The recent version of \WINHAC{}, 1.30~\cite{WINHAC:MC}, features the 
exclusive Yennie--Frautschi--Suura exponentiation~\cite{yfs:1961}
of the QED effects including the ${\cal O}(\alpha)$ electroweak corrections for
the charged-current Drell--Yan process at the parton level,
see Ref.~\cite{Bardin:2008fn} for more details. 
This parton-level process is convoluted with the parton distribution
functions (PDFs) provided by the \LHAPDF{} package~\cite{Whalley:2005nh}.
This package includes a large set of recent PDF parametrisations by
several groups. \WINHAC{} has been  interfaced with 
the \Pythia{}~6.4~\cite{Sjostrand:2006za} MC event generator, 
which provides the modeling of the QCD/QED 
initial-state parton shower as well as the hadronisation.

The current version of \WINHAC{} does not include explicit NLO QCD corrections
to the hard process matrix elements. All the QCD effects are generated by 
\Pythia{}. Therefore, the QCD precision of \WINHAC{}
is of the LO-improved type, the same as of \Pythia{}. On the other hand,
the QED/EW corrections to the hard process are included in \WINHAC{} to the 
accuracy of  the YFS ${\cal O}(\alpha)$ exponentiation. 
They are in agreement at the per-mille level with independent 
calculations, in particular the ones provided by the Monte Carlo programs 
{\sf HORACE} and {\sf SANC}~\cite{CarloniCalame:2004qw,Bardin:2008fn}.
In all studies presented in this paper the QED/EW corrections have 
been switched off. In general, these missing corrections  are as important as the 
missing NLO QCD  corrections. Once carefully matched, all the above radiative corrections 
will have to be included while interpreting  the future LHC $W$-boson data in the framework 
of the Standard Model. However, for the evaluation of the dedicated measurement strategies proposed 
in this paper they are of secondary importance. Moreover, they can be factorized out,  
and applied externally to the data analysis procedure. 
The goal  to minimize them and to control them, as much as possible, experimentally
will be the driving  principle while  defining  the observables 
used in this paper and while optimizing their dedicated measurement procedures.  

The following set of collider modes  have been implemented:
proton--proton,
proton--antiproton, proton--nucleus and nucleus--nucleus. 
For collisions involving nuclei 
the nuclear-shadowing effects~\cite{Eskola:1998df,Eskola:1998iy} 
can be switched on.

The parton-level matrix elements are calculated numerically from 
spin amplitudes~\cite{Placzek2003zg}. This allows for studies of the 
spin-dependent effects in the charged-current Drell--Yan process. For example,
\WINHAC{} provides options for separate  generations of processes with pure-transversely 
or pure-longitudinally polarised $W$-bosons at the Born level.

In addition to the charged-current Drell--Yan process, \WINHAC{} 
includes the neutral-current Drell--Yan process
(with $\gamma+Z$ bosons in the intermediate state), 
presently at the Born level only. 
For a more precise description of the  latter process,
including the full set of the QED/EW radiative corrections, 
a dedicated MC event generator, called 
{\sf ZINHAC}~\cite{ZINHAC:MC}, is being developed. We plan to use these
twin MC generators for precision studies/analyses of the
Drell--Yan processes at the LHC.

For the presented study the version 1.23 of \WINHAC{} has been used.
It is equivalent to the version 1.30
for all the aspects addressed in this paper.

\subsection{Event generation, simulation and data storage}\label{ss_simulation}

The studies reported in this paper have been performed  
for an  integrated luminosity of  $10\,\mathrm{fb}^{-1}$.
By the time of reaching such a luminosity, approximately 
$1.1\times 10^8~\Wp$ and $0.8\times10^8~\Wm$ bosons  will be produced at the LHC. 
In order to optimize the strategy of measuring their masses, the generation and simulation
of ${\cal O}(100)$ event-samples was required. Each of these samples  
corresponds either to a  specific bias  in the detector response, 
or to a  specific  theoretical (phenomenological)  modeling method 
of the $W$-boson production processes.
In addition,  a large set of unbiased event-samples, reflecting variable values of 
the masses of the $\Wp$ and $\Wm$ bosons, was simulated.    
For an assessment of the impact of the systematic biases on the 
overall measurement precision, each of the above event-samples must 
contain at least  $10^8$ events in order to match the systematic and 
 statistical measurement precisions. 

The analysis presented in this paper is thus based on the total sample of ${\cal O}(10^{10})$ 
$W$-boson production events.
Generating, simulating and handling of such 
a large event sample was a major challenge.  In essence, we stored the simulated 
data samples in histograms, with the bin sizes adjusted to the detector resolution, rather 
than in the \texttt{NTuple}-like event summaries. This allowed us, instead of storing the
Terabyte event summaries,  to confine our analysis within a ``laptop-size'' 
data-storage space.  

The events have been simulated using a parametrised average response of the detector.
In the studies presented in this paper we used the response functions of the
ATLAS detector \cite{AtlasTDRvol1:1999fq}.
The systematic biases in  the detector response have been modeled and 
simulated in the dedicated generation and simulation runs rather than by
the re-weighting procedures. 
These aspects are presented in a more detail in Ref.~\cite{Florent_PHD}.

\subsection{Conventions}\label{ss_conventions_florent}

Both Cartesian and cylindrical coordinates are used in this paper.
The beam-collision point defines the origin of the coordinate systems.
Colliding beams move along the $z$ axis, $+y$ points upward, and $+x$ to 
the center of the LHC ring.
The following convention is used for the cylindrical coordinates:
$r$ is the radius in the $x-y$ plane,
$\phi$ the azimuthal angle with respect to the $+x$~direction,
and $\theta$ the polar angle with respect to the $+z$~direction.
Unit vectors along these different directions are denoted as $\vec e_i$,  where $i$ 
can stands for $x$, $y$, $z$.

\section{Charge asymmetries}
\label{s_charge_asym}

In this section we discuss the basic features of the charged current Drell--Yan
process for the following three modes of the high-energy collisions: proton--proton ($\pp$), 
proton--antiproton ($\ppbar$) and deuteron--deuteron ($\dd$). 
We analyze separately the production and decay mechanisms  of the $W$-bosons.
We compare the $\pp$ collisions with the 
$\ppbar$ and $\dd$ ones in order to understand the effects specific to 
the proton--proton collisions. This analysis will allow us to optimise the LHC dedicated 
strategy for the precision 
measurement of the mass difference between the $W^+$ and $W^-$ bosons at the LHC.

\subsection{Observables}
\label{ss_conventions}
We consider the charged-current Drell--Yan processes, \ie{} the single $W$-boson production
with leptonic decays at hadron colliders:
\begin{equation}
  p+p\:\: (p + \bar{p},\; d + d) \;\longrightarrow\; W^{\pm} + X \;\longrightarrow\; l^{\pm} + 
  \overset{(-)}{\nu_l} + X\,, 
  \label{e_ccDY_proc}
\end{equation}
where $l=e,\,\mu$. 

The commonly chosen observables for the above processes are the charged lepton
transverse momentum $\pTl$ and pseudorapidity $\etal$, defined as
\begin{eqnarray}
  \pTl  &=& \sqrt{p_{x,l}^2 + p_{y,l}^2}\,,\\ 
  \etal &=& -\ln\left(\tan(\thetal/2)\right)\,,
  \label{eq_obs_lept}
\end{eqnarray}
where $\thetal$ is the polar angle of the outgoing charged lepton in the laboratory frame.

Although the $W$-boson four-momentum cannot be directly measured at hadron colliders
(due to escaping neutrino), in our generator-level studies we use also the $W$-boson
(pseudo)observables -- its transverse momentum $\pTW$ and rapidity $\yW$:
\begin{eqnarray}
\pTW &=& \sqrt{p_{x,W}^2 + p_{y,W}^2}\,,\\
\yW  &=& \frac{1}{2}\,\ln\left(\frac{E_W+p_{z,W}}{E_W-p_{z,W}}\right)\,.
\label{eq_obs_W}
\end{eqnarray}

The charge asymmetry is used to scrutinize the differences between the $\Wp$ and $\Wm$ mediated 
processes. For a given scalar observable $a$, the charge asymmetry $\mathrm{Asym}^{(+,-)}(a)$ 
is defined like
\begin{equation}
    \Asym{a} \; =\; 
    \frac
        { \flatDfDx{\sigma^\plus}{a} - \flatDfDx{\sigma^\minus}{a} }
        { \flatDfDx{\sigma^\plus}{a} + \flatDfDx{\sigma^\minus}{a} },
        \label{eq_def_charge_asym}
\end{equation}
where $\plus$ and $\minus$ refer to the electric charge of the $W$~boson, or the final state
charged lepton,  and $\flatDfDx{\sigma^\pm}{a}$ is the differential cross section of an observable $a$.

All the results discussed below have been obtained using  the Born-level hard process 
implemented in  \WINHAC{} and convoluted with the initial-state QCD/QED parton shower 
generated by  \Pythia{} routines.
The samples of $200$ million weighted events have been generated 
for each of the $W$-boson charge.
The nucleon-nucleon centre-of-mass energy, $\sqrt s$, is assumed to be $14\TeV$ 
for the $\pp$ and $\ppbar$ collisions and  $7\TeV$ for the $\dd$ collisions.

\subsection{Production of     $\;\mbf{\Wp}$ and   $\;\mbf{\Wm}$}
\label{ss_W_prod}
In this subsection we discuss the $W$-boson production mechanism.
The observables analysed here are: the $W$-boson rapidity $\yW$, and
its transverse momentum, $\pTW$.
\begin{figure}[!h] 
  \begin{center}
    \includegraphics[width=0.495\tw]{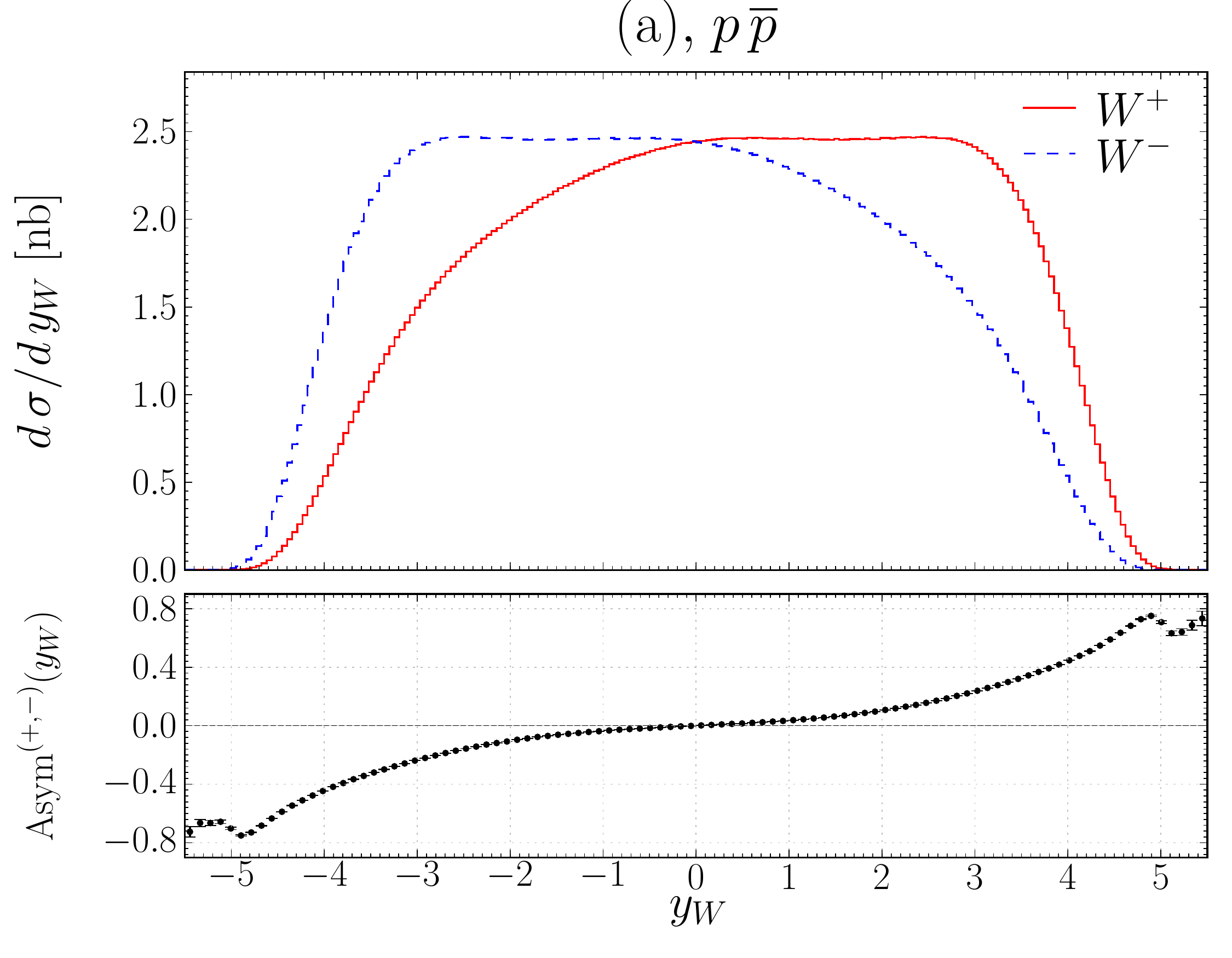}
    \hfill
    \includegraphics[width=0.495\tw]{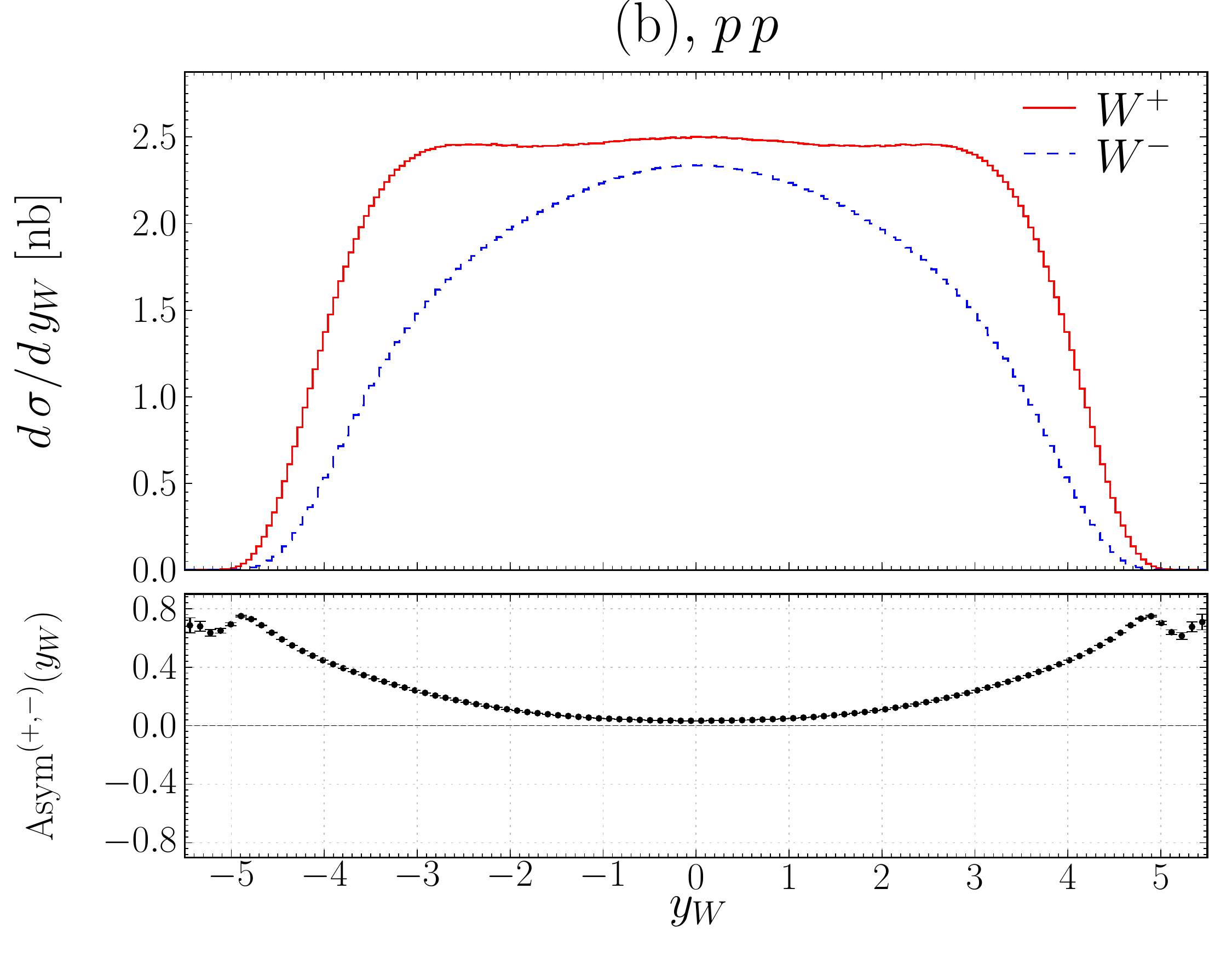}
    \vfill
    \includegraphics[width=0.495\tw]{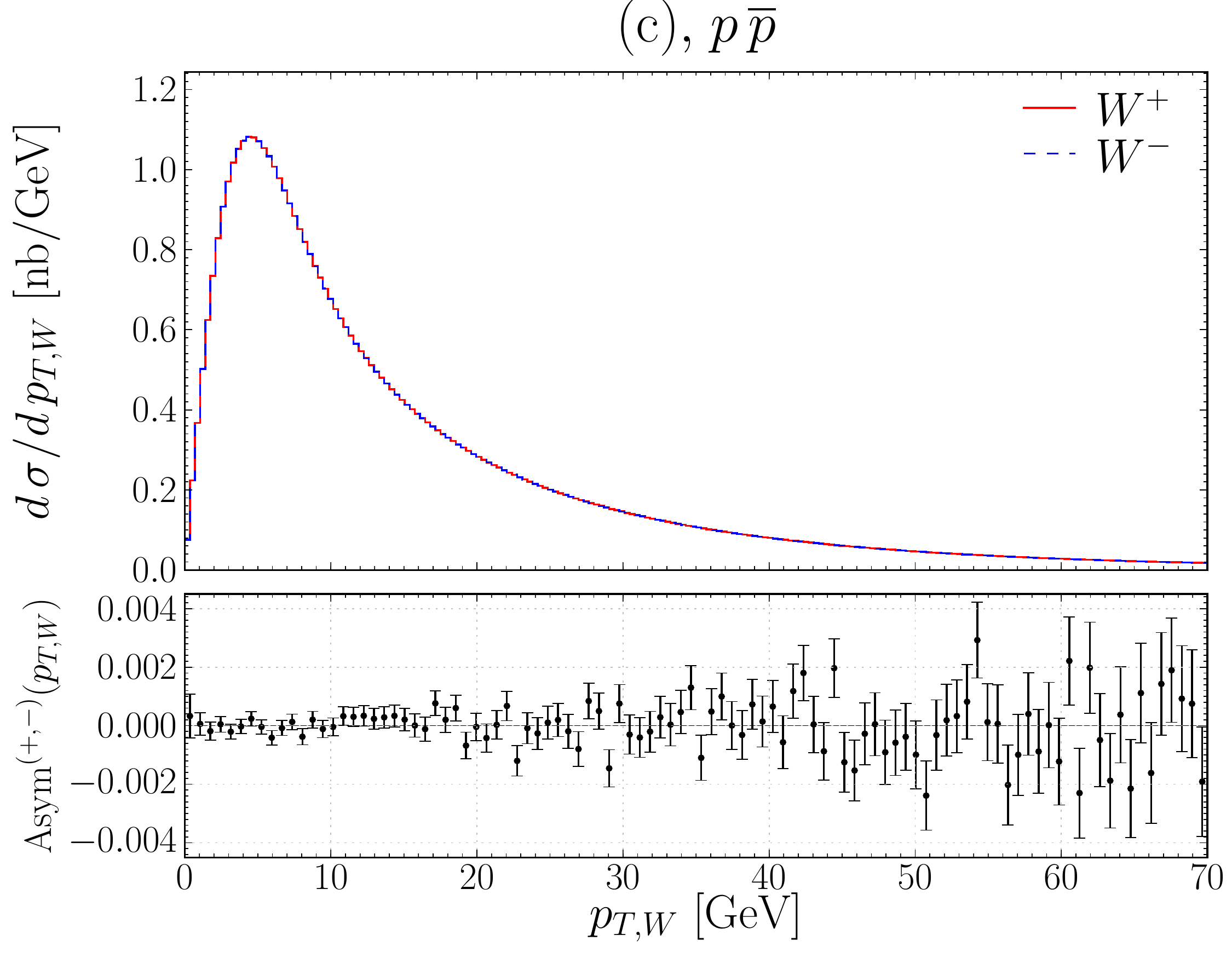}
    \hfill
    \includegraphics[width=0.495\tw]{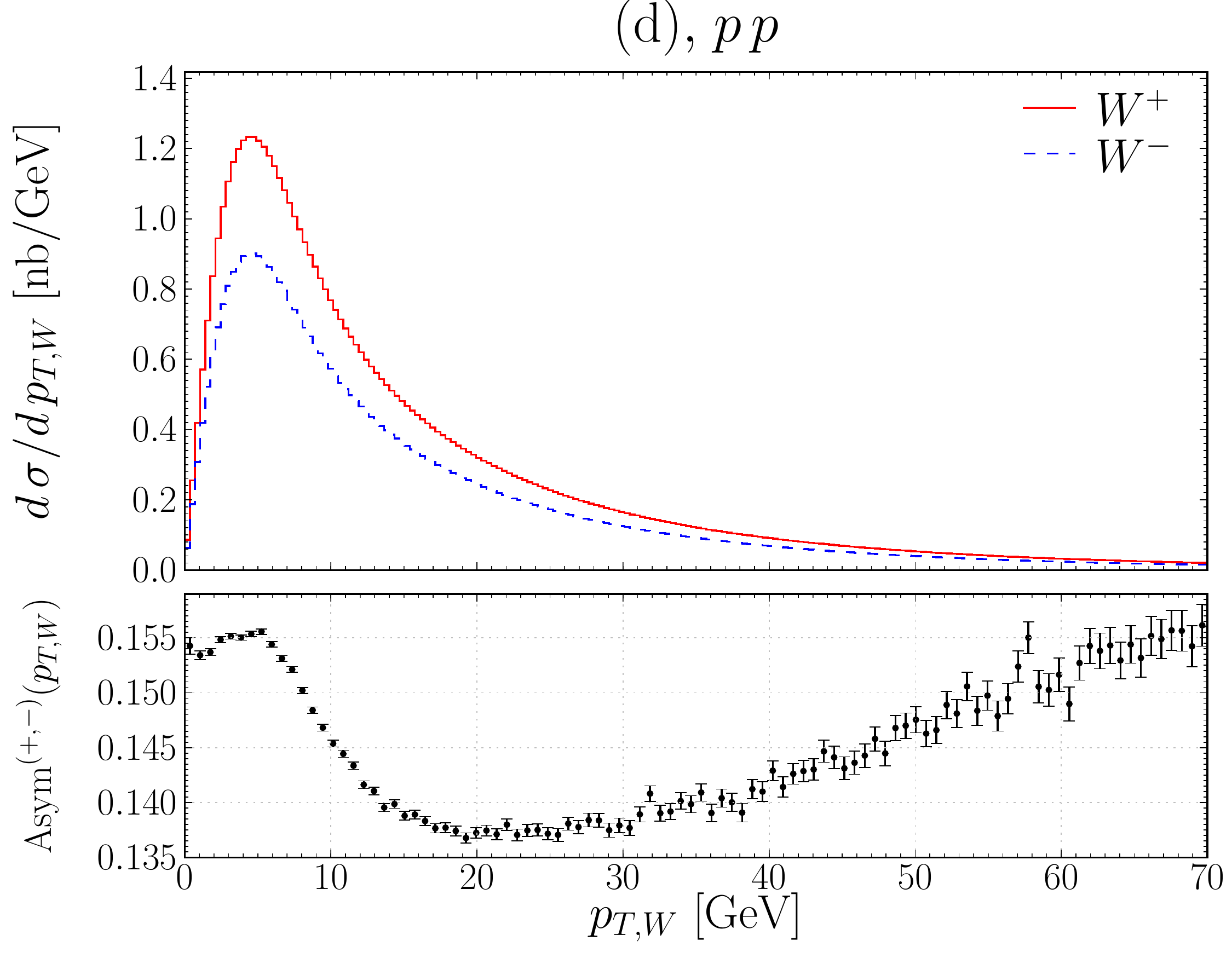}
    \caption[]
            {\figtxt{The rapidity $\yW$ and transverse momentum $\pTW$ distributions of the $W$-bosons and  their 
            charge asymmetries for the $\ppbar$ (a, c) and $\pp$ (b, d) collisions.}}
            \label{fig_yW_pTW_ppb_pp}
  \end{center} 
\end{figure}

In Fig.~\ref{fig_yW_pTW_ppb_pp} we present their distributions for the 
$W^+$- and $W^-$-bosons,  as well as their charge asymmetries for the two collision modes: 
$\ppbar$ and $\pp$. 
In the $\ppbar$ collisions the total production cross-sections and its $\pTW$
distribution are independent of the $W$-boson charge. 
The rapidity distributions for two charge states 
are mirror reflected for positive and negative rapidities.
This is illustrated in the left-hand-side (LHS) plots, (a) and (c).

The right-hand-side (RHS)  plots, (b) and (d), show the corresponding distributions for 
the $\pp$ collisions.
In this case the rapidity distributions are symmetric w.r.t.\ $\yW=0$.
Their shapes, however, are different for the $W^+$ and $W^-$ bosons. This reflects mainly the difference
in the valence $u$ and $d$ quark content of the proton. The $\yW$ distribution for the $W^+$-bosons is
higher, wider and flatter with respect to  the one for the $W^-$-bosons 
because there are twice as many the valence $u$
quarks as the valence $d$ quarks, and the former carry, on average, a higher 
fraction of the parent proton momentum. 
The charge asymmetry of the $\pTW$ distribution
reflects the differences in the relative cross-sections but also, what will be crucial 
for the studies presented in this paper, in the shape of their distributions.  
The nontrivial shape of the  asymmetry of the $\pTW$ distribution reflects
the flavour asymmetries in the distributions of quarks producing the $\Wp$ and $\Wm$ bosons.
The latter are predominantly driven by the $u$--$d$ quark asymmetries and are amplified by the non-equality of the $s$- and $c$-quark masses. 
We analyze the flavour structure of the $W$-bosons charge asymmetries 
by writing explicitly the simplified Born-level formulae for the total cross-section asymmetries 
for three types of collisions: $\ppbar$, $\pp$ and $\dd$:
\begin{eqnarray}
  \left(\sigma^{\plus} - \sigma^{\minus}\right)_{\ppbar}(s) 
  &=& 0, \label{eq_sWp_sWm_ppar}\\
  \left(\sigma^{\plus} - \sigma^{\minus}\right)_{\pp}(s) 
  &\propto& \int dx_q dx_{\bar{q}}\, \Big\{
  \Vckmsqr{u}{d}\left[ u^{(v)}(x_q)\,\bar d(x_{\bar{q}}) - d^{(v)}(x_q)\,\bar u(x_{\bar{q}}) \right] 
  \nonumber\\
  & & \hspace{20mm}
  +\, u^{(v)}(x_q) \left[ \Vckmsqr{u}{s}\,\bar s(x_{\bar{q}}) + \Vckmsqr{u}{b}\,\bar b(x_{\bar{q}}) 
  \right]
  \nonumber\\
  & & \hspace{20mm}
  - \,\Vckmsqr{c}{d}\,d^{(v)}(x_q)\,\bar c(x_{\bar{q}})\Big\}\,
   \sigma_{q \bar{q}}(\hat{s}), 
  \label{eq_sWp_sWm_pp}\\
  \left(\sigma^{\plus} - \sigma^{\minus}\right)_{\dd}(s)
  &\propto& \int dx_q dx_{\bar{q}}\;
  u^{(v)}(x_q) \left[ \Vckmsqr{u}{s}\,\bar s(x_{\bar{q}}) - \Vckmsqr{c}{d}\,\bar c(x_{\bar{q}}) 
  \right. 
  \nonumber\\
  & & \hspace{29mm}
  \left.
        +\; \Vckmsqr{u}{b}\,\bar b(x_{\bar{q}}) \right]\,
  \sigma_{q \bar{q}}(\hat{s}).
  \label{eq_sWp_sWm_dd}
\end{eqnarray}
In these formulae $d,u,s,c,b$ on the RHS denote the PDFs of the 
corresponding quark flavours and the superscript $(v)$ 
stands for valence, $V_{ij}$ is the CKM matrix element for the $i$ and $j$ flavours, 
while $\sigma_{q \bar{q}}(\hat{s})$ is the parton-level cross section for the $W$-boson production
with $\hat{s} =  x_qx_{\bar{q}}s$.
In the formulae for the $\dd$ collisions we assumme the isospin symmetry in the 
valence quark sector:   $u^{(v)} =  d^{(v)}$.
For pedagogical reasons, the formulae for  the $\pp$ and $\dd$ collisions are simplified.
We have omitted the 
explicit dependence of the PDFs on the factorisation scheme, on the  the transverse momenta $\kT$
of annihilating partons present both in the ``$\kT$-unintegrated'' PDFs, and in  
the partonic cross-sections (via $\kT$ dependence of  $\hat{s}$). 
All the above effects  have been taken into account in our Monte-Carlo studies.  
In our analysis partons have both the primordial transverse momenta 
and the perturbatively generated ones as  modeled by the initial-state
parton shower of the \Pythia{} generator. Their transverse momenta depend on the Bjorken $x$ of
the  annihilating  quark (antiquark)  and,  for heavy quarks (here $c$ and $b$),  
also on their  masses  (see Ref.~\cite{Sjostrand:2006za} for more details).%

As one can see from the above equations, the charge asymmetry
disappears for the $\ppbar$ collision mode (if, as assumed in the presented  studies, the masses 
of $\Wp$ and $\Wm$ bosons are equal). This collision 
mode is thus the optimal one for measuring  $\MWp - \MWm$. 
Any deviation from the equality of masses  
would result in non-zero asymmetries, regardless  of the level of 
understanding of the flavour and momentum structure of the beam particles.

For the  $\pp$ collision mode  several effects,  reflecting the present understanding 
of the partonic content of the beam particles (in particular, 
the understanding of the momentum distribution of valence quarks),  
contribute to the charge asymmetry of the $\pTW$ distribution and may  
mimic the $\MWp \neq \MWm$ effects. 

For the $\dd$ collisions the asymmetry is particularly simple. 
It is driven by the Cabibbo-suppressed 
difference of the distribution of the strange and charm quarks, weighted
by the distributions of the valence quarks. We have introduced to our discussion 
the $\dd$ collision mode at this point in order to analyze  the relative importance of 
the valence quark and `$s-c$' effects.  

In the following,  we analyze the numerical importance of various terms 
appearing in the above equations.

\begin{figure}[!ht] 
  \begin{center}
    \includegraphics[width=0.495\tw]{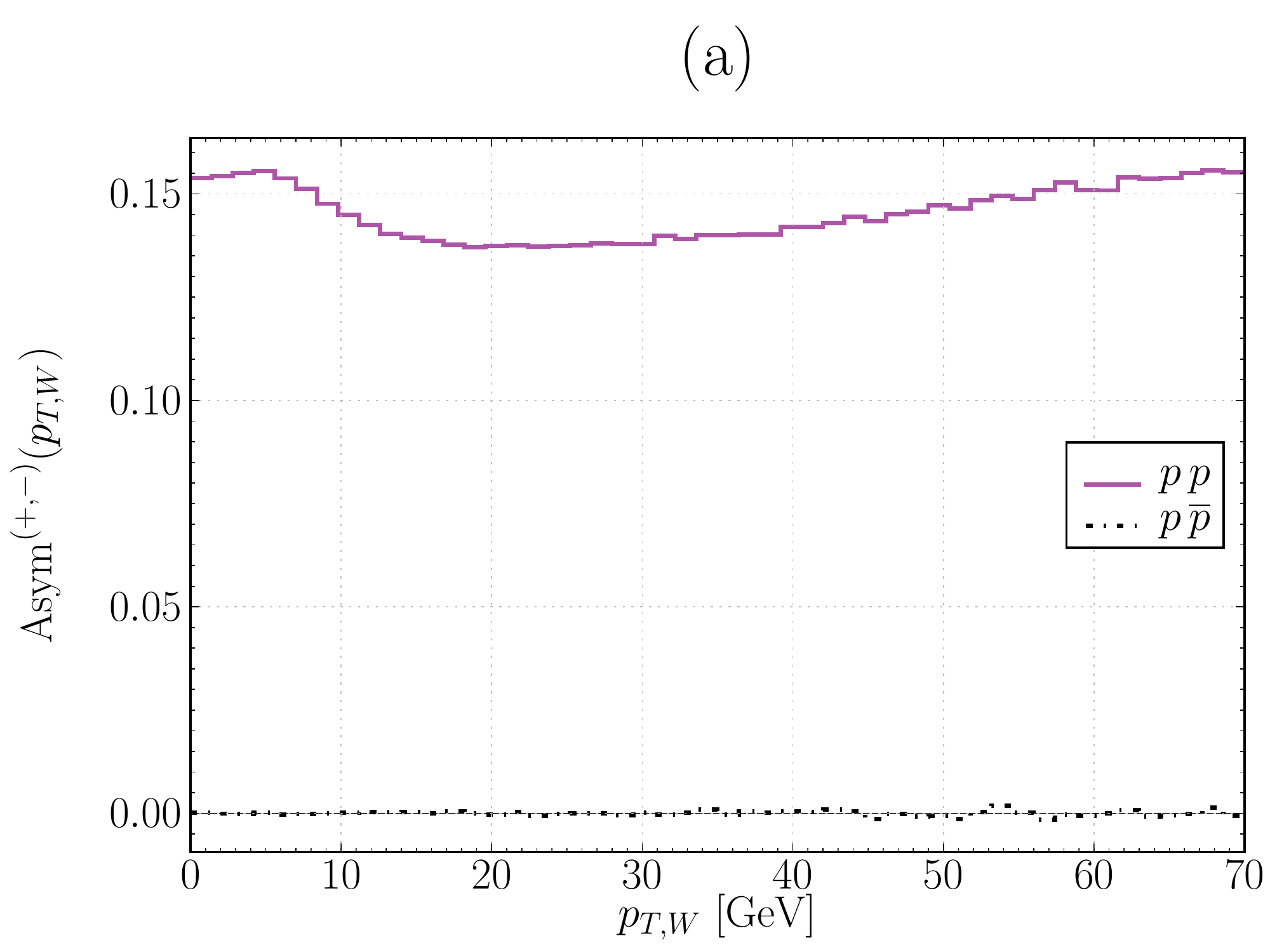}
    \hfill
    \includegraphics[width=0.495\tw]{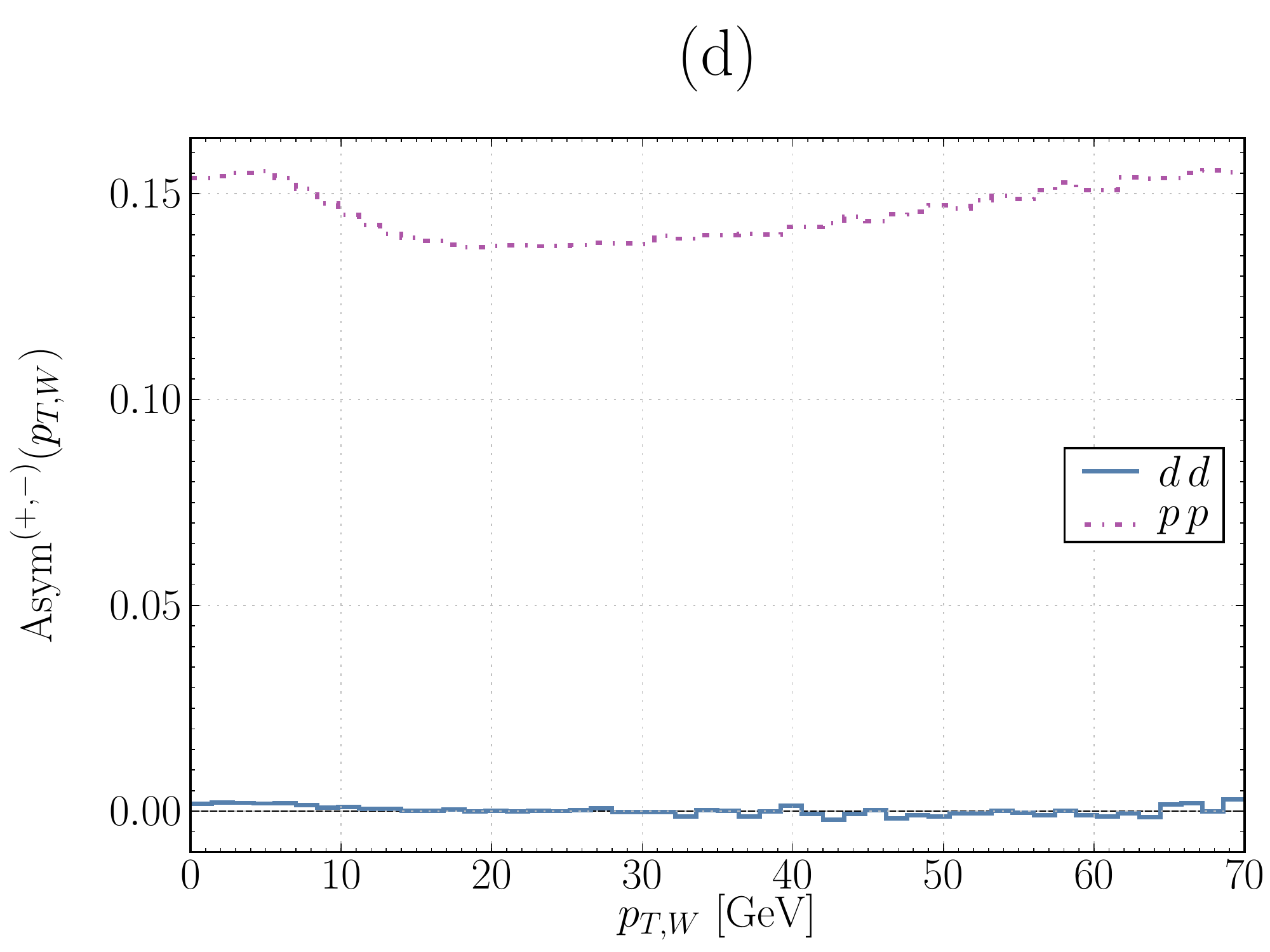}
    \vfill
    \includegraphics[width=0.495\tw]{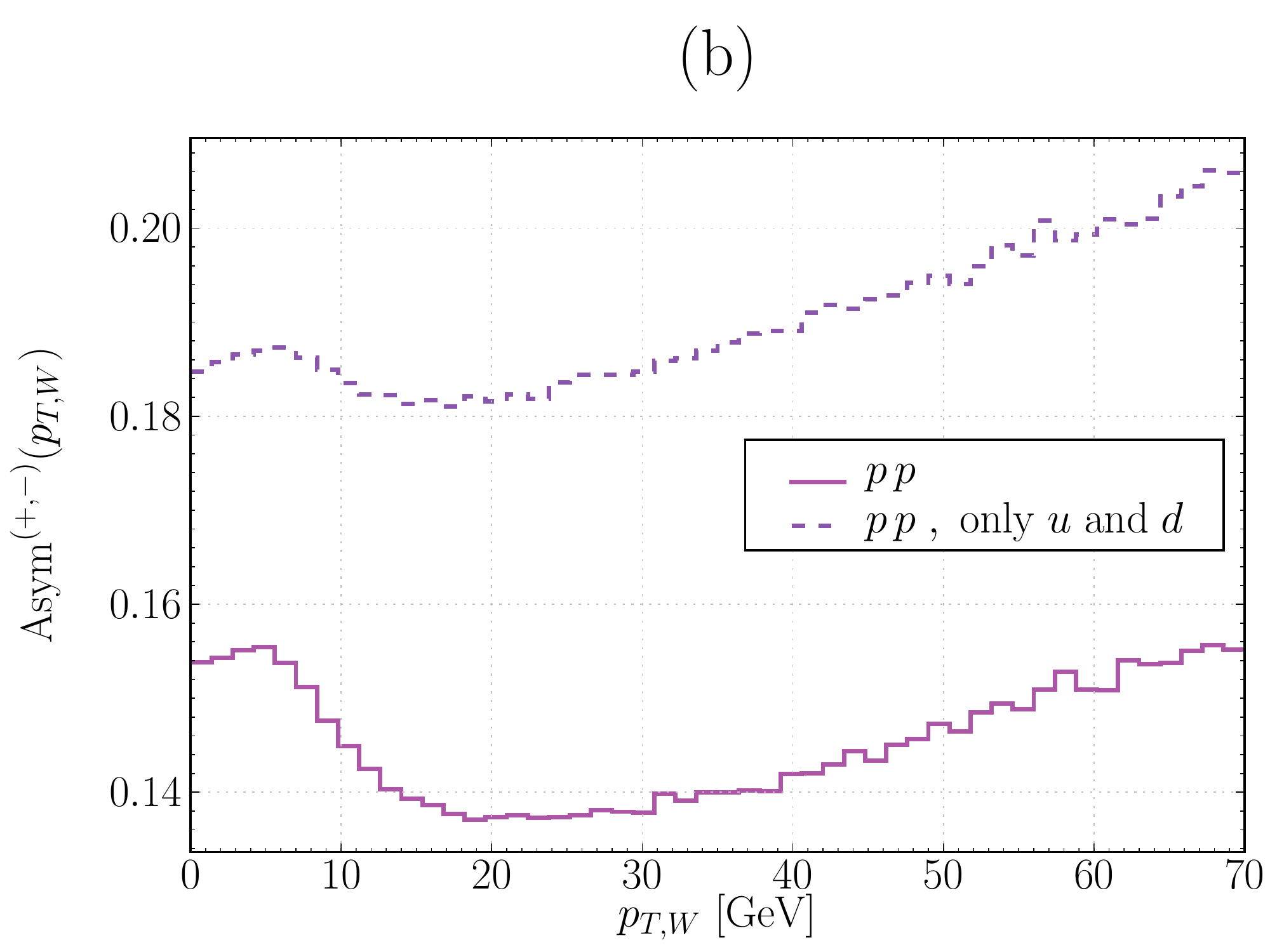}
    \hfill
    \includegraphics[width=0.495\tw]{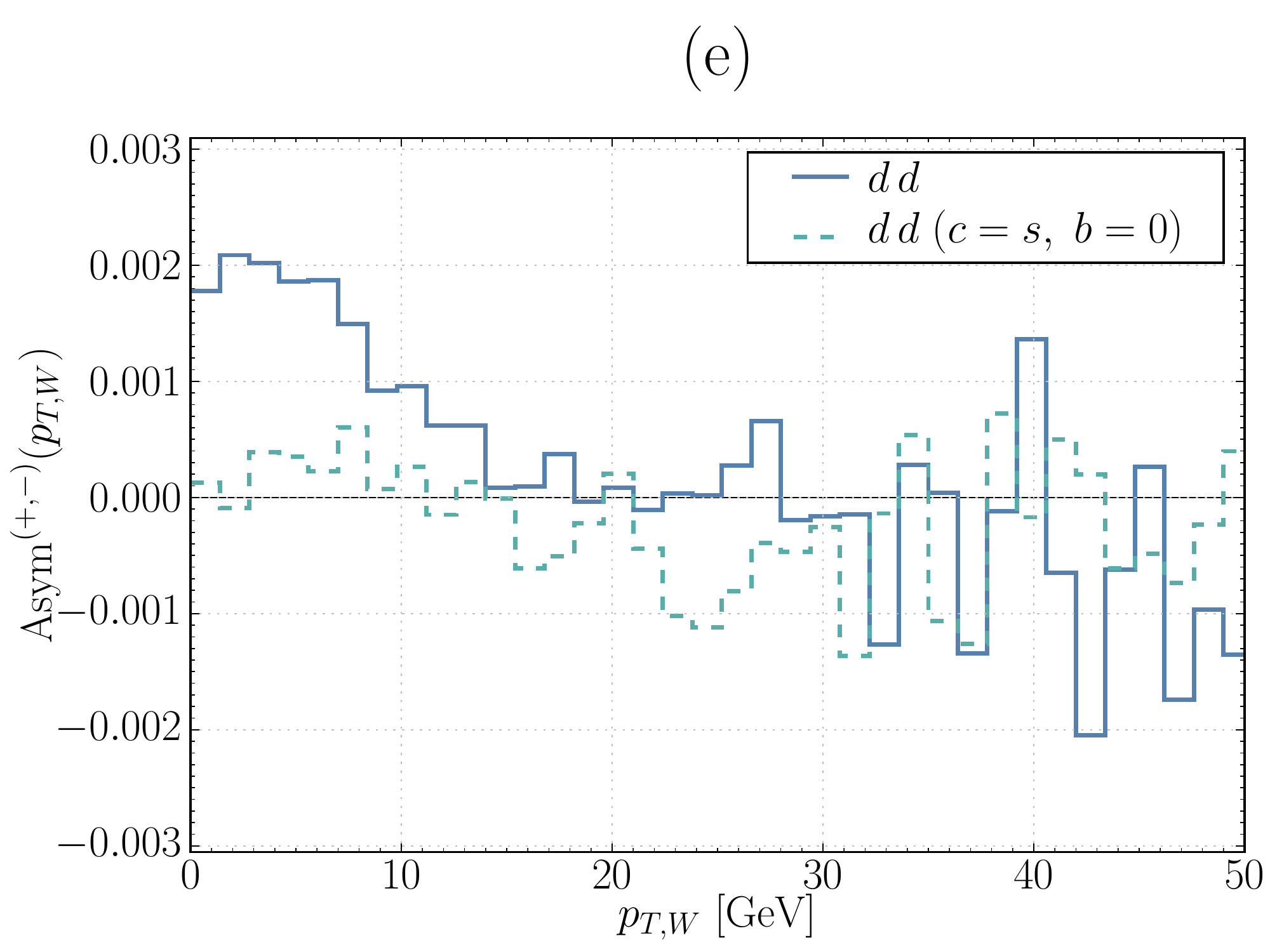}
    \vfill
    \includegraphics[width=0.495\tw]{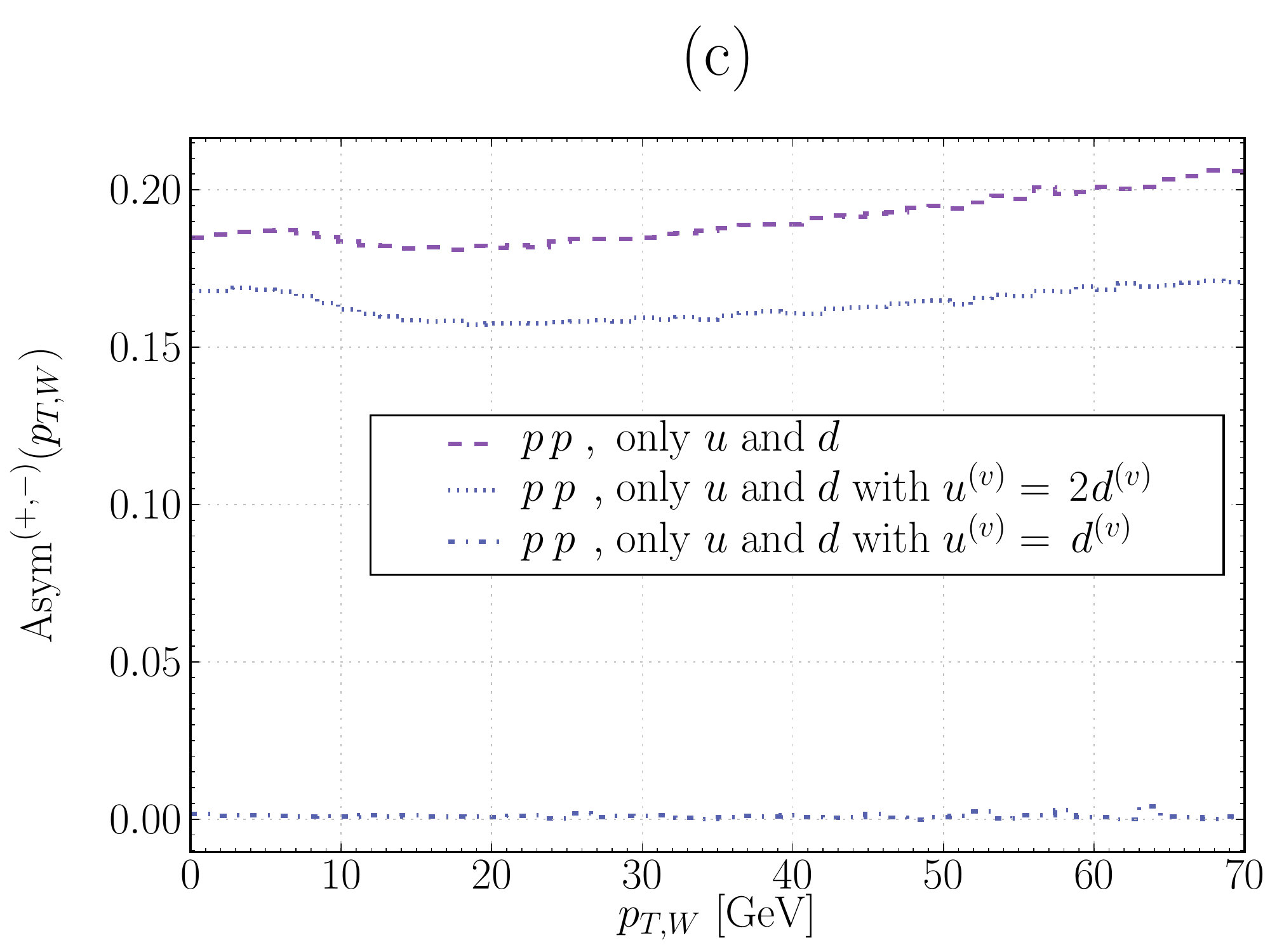}
    \hfill
    \includegraphics[width=0.495\tw]{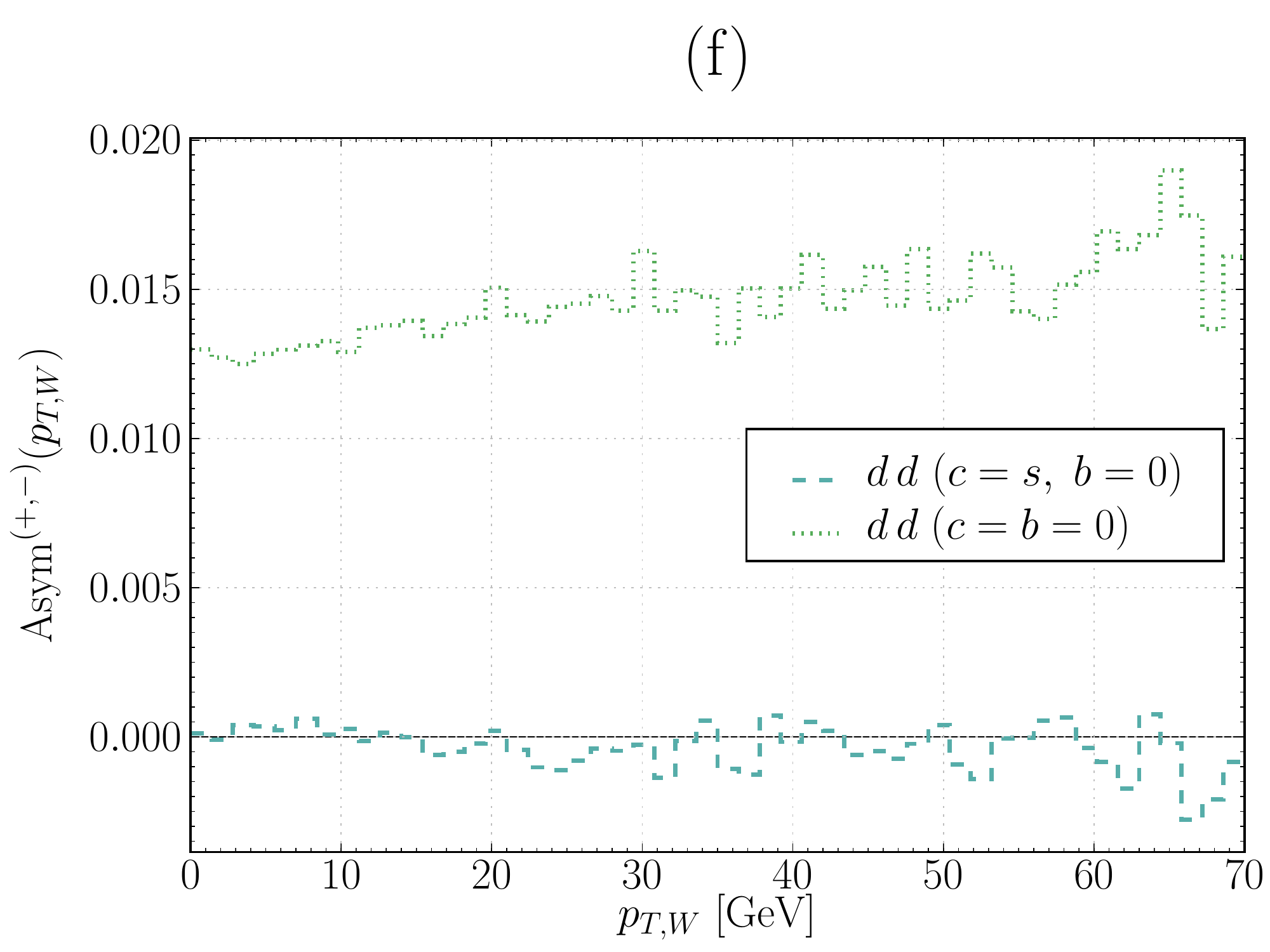}
    \caption[]
            {\figtxt{The charge asymmetries for $\pTW$\,:\;
              $\pp$ vs. $\ppbar$ (a),
              $\pp$ vs. $\pp$ ($u$ and $d$ quarks only) (b),
              $\pp$ ($u$ and $d$ quarks only) vs. $\pp$ ($u$ and $d$ quarks only and $u^{(v)}=2\,d^{(v)}$ )(c), 
              $\dd$ vs. $\pp$ (d),
              $\dd$ vs. $\dd$ ($c=s$ and $b=0$) (e),
              and finally $\dd$ ($c=s$ and $b=0$) vs. $\dd$ ($c=b=0$) (f).}
            }
            \label{fig_W_prod_asym}
  \end{center} 
\end{figure}

The results of our studies are presented  in 
Fig.~\ref{fig_W_prod_asym}.
In the first plot, (a), we compare the charge asymmetry
in the transverse momentum of the $W$-bosons   for the $\pp$ collision mode (solid line)
with the one for the $\ppbar$ collision mode  (dash-dotted line). 
The charge asymmetry for $\pp$ collision mode is large 
and varies as a function of $\pTW$. The next plot, (b), shows  the comparison of the $\pp$ asymmetry
from the previous plot (solid line) with the one in which the protons contain only $u$ and 
$d$ quarks (dashed line).
The observed increase of the asymmetry in the latter case 
indicates  that the charge asymmetry for the $\pp$ collisions
is driven mainly by the $u$ and $d$ quarks (the presence of the 
$s$ and $c$ quarks diminishes slighty its magnitude). 
In the plot (c) we demonstrate  that this asymmetry is slightly reduced when $u^{(v)}=2d^{(v)}$,
\ie{} when the valence $u$ and $d$ quark PDFs are assumed to have the same shape and differ only by
the normalisation factor corresponding to their numbers in the proton (dotted line).
In addition,  we note that the asymmetry becomes flatter as a function of  $\pTW$, indicating 
the role of the relative $x$-shapes of the $u$ and $d$ quarks PDFs. 
The asymmetry is reduced drastically  when we put $u^{(v)}=d^{(v)}$
(dash-dotted line), as it would be the case for the 
isoscalar beams. Therefore,  in the following plots
we analyze the simplest isoscalar beam-collision mode, the $\dd$ one.
The plot (d) shows that the charge asymmetry for $\dd$ (solid line) is much smaller than
the one for $\pp$ (dash-dotted line). This results from two facts: (1) all the terms
contributing to the charge asymmetry for $\dd$ include the off-diagonal CKM matrix elements
and (2) contributions of the $s$, $c$ and $b$ quark PDFs are smaller than the ones from the $u$ and $d$ 
sea-quark PDFs. The remaining asymmetry is at the level of $0.002$,  as can be seen in the plot
(e) (solid line),  and it can be reduced to a statistically negligible level when 
we set $c=s$ and $b=0$ (dashed line).
The last plot, (f), shows the 
effect of the deviation of the charge asymmetry due to the difference of the 
masses and of the momentum distributions of the strange and charm quarks in the extreme case 
corresponding to  $c=0$ (dotted line). The asymmetry is flat but significantly
higher, at a level
of $0.015$. It is a factor of $\sim 10$ bigger than the asymmetry for the $\dd$ collisions assuming 
the present understanding of the relative asymmetry in the distribution of 
the strange and charm quarks.

\begin{figure}[!h] 
  \begin{center}
    \includegraphics[width=0.495\tw]{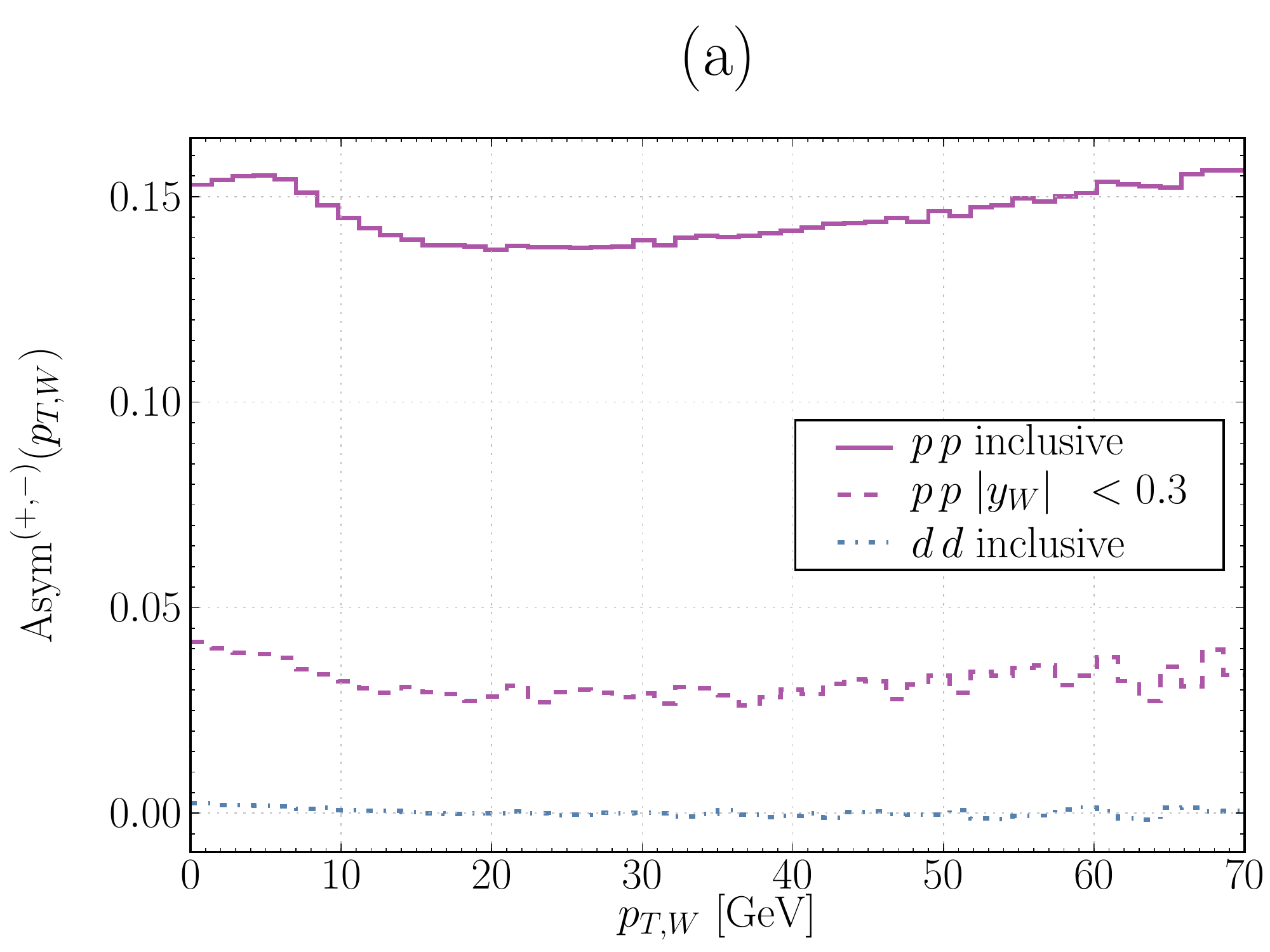}
    \hfill
    \includegraphics[width=0.495\tw]{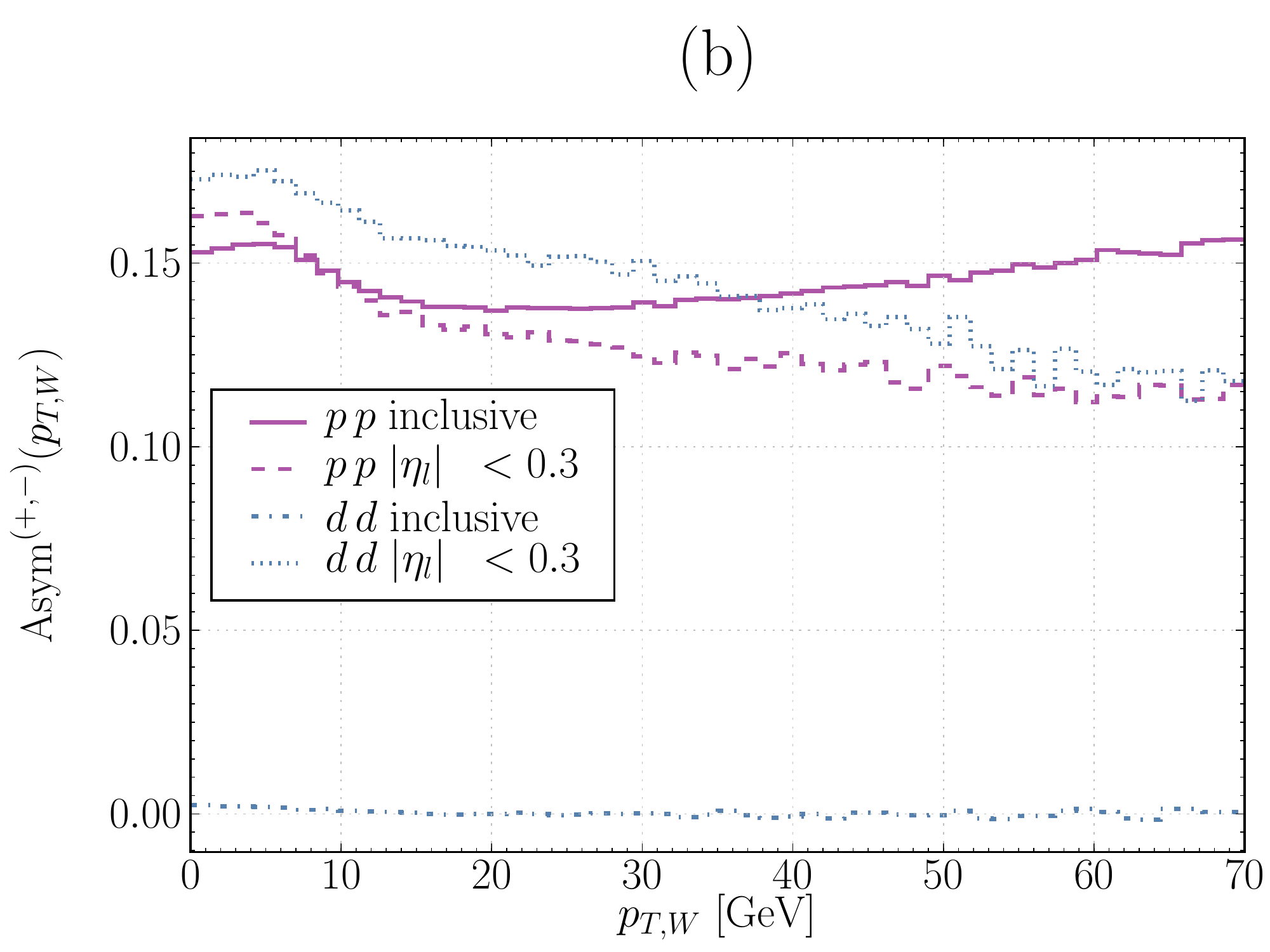}
    \caption[]
            {\figtxt{The charged asymmetries of $\pTW$ for inclusive $\pp,\,\dd$ collisions 
                     and with cuts\,:\; $|\yW|<0.3$ for $pp$ (a) and 
                     $|\etal|<0.3$ for  $\pp$ and $\dd$ (b).}}
            \label{fig_W_prod_asym_centr_yW_etal_bin}
  \end{center}
\end{figure}

In the analysis presented so far the $\pTW$ distribution have been integrated over the 
full range of kinematically allowed $x_q$ and $x_{\bar{q}}$. In order to optimise the measurement
strategy of the $W$-mass charge asymmetry, we now discuss the $\pTW$ distributions 
restricted to the selected kinematical regions.
The most obvious method to reduce the contributions of the valence quarks 
is to restrict the analysis to the  $\yW\sim 0$ region, 
where $x_q\sim x_{\bar{q}} \approx 6\times 10^{-3}$,  \ie{} where the valence quarks are
largely outnumbered by the sea quarks.
 
In Fig.~\ref{fig_W_prod_asym_centr_yW_etal_bin} we present the asymmetries
for the $\pp$ and $\dd$ collisions for the narrow central bins: (1) in the $W$ rapidity, 
(a), and (2) in the lepton pseudorapidity, (b). In the plot (a) we see that the $\pp$ charge asymmetry
is reduced and flattened by more than a factor of $3$ for the range of the $W$-boson rapidity $\yW<0.3$.
Unfortunately $\yW$ cannot be measured directly. It is thus natural to check if comparable  reduction 
can be obtained using the  
$\etal$ variable, which is correlated with $\yW$.
This turns out not be the case, 
as can be seen in plot (b).  The $\etal$ variable has thus significantly  lower discriminative power 
to reduce the valence quark contribution with respect to the $\yW$ variable. 
For the $\dd$ collision the asymmetry restricted to the narrow $\etal$ bin  increases considerably.
This observation draws out attention to the fact  
that the asymmetry in a decay mechanism  of the $\Wp$ and $\Wm$ 
bosons will have an important impact on the asymmetry of leptonic observables. 
This is  discussed in detail in the next section. 

There can also be a contribution to the charge asymmetry coming from the QED radiation from 
quarks, as upper and lower components of the $SU(2)_L$ quark-doublets have different electric charges. 
However, it has been found (with \Pythia{}) to be of the order of $2.5\times 10^{-4}$ 
in the $\Asym{\pTW}$ spectra, 
which is insignificant as compared to the above shown contributions to the
asymmetries for $\pp$ and $\dd$.
Another contribution to the charge asymmetry may come from the missing NLO QCD corrections.
These corrections are large for the  $\pTW$ distributions,  but of residual importance for the charge asymmetries,
in particular in  the kinematical region used in the determination of the masses of the  $\Wp$ and $\Wm$ bosons. 
This will be discussed further in Section \ref{sss:WBpBemittance}. 
\subsection{Decays of $\BFWp$ and $\BFWm$}\label{ss_W_prod_decay}
In this subsection we analyze the $W$-boson decay mechanism
and see how it influences the charge asymmetries 
in the lepton transverse momentum $\pTl$ and pseudorapidity $\etal$ distributions.

\begin{figure}[!h] 
  \begin{center}
    \includegraphics[width=0.495\tw]{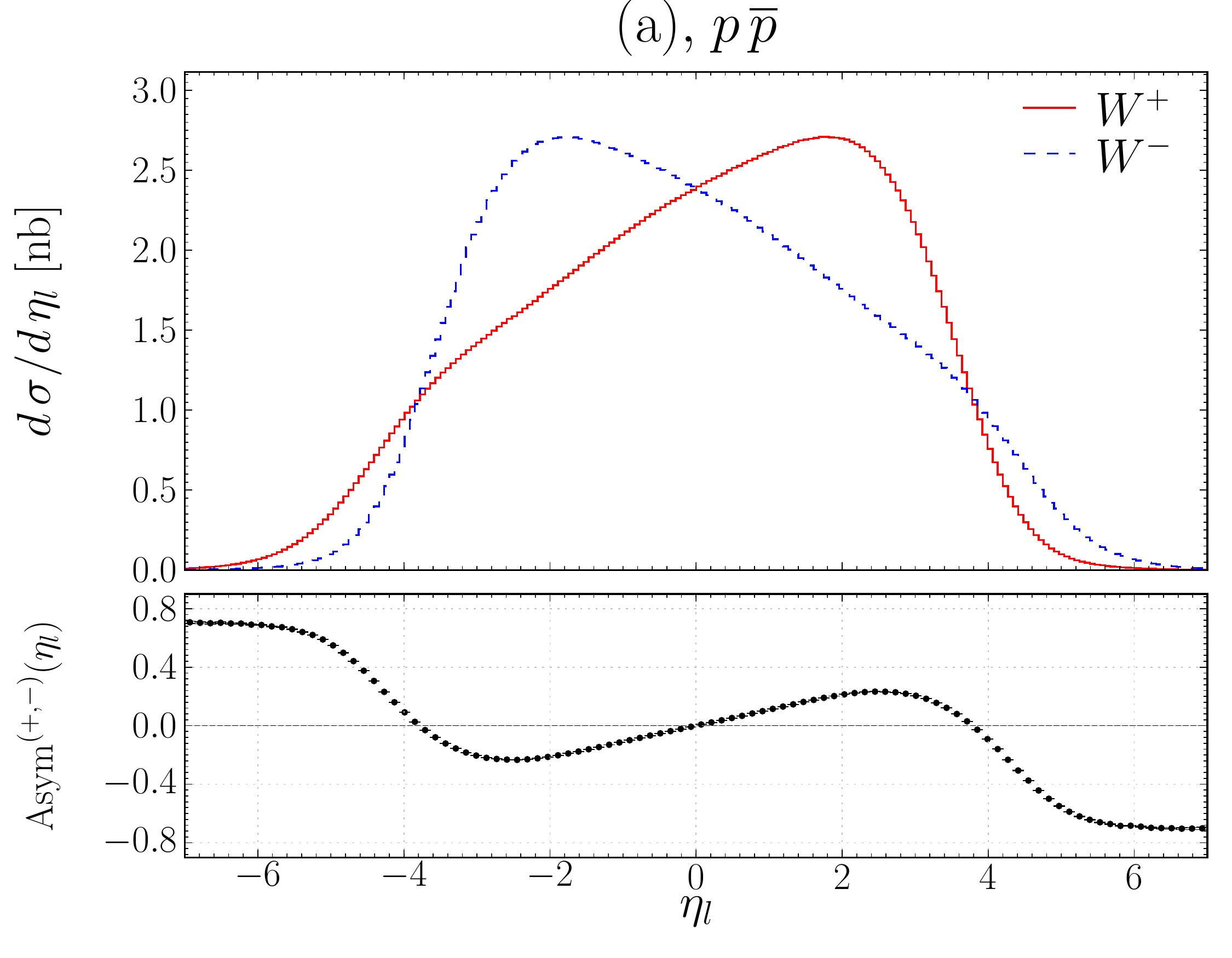}
    \hfill
    \includegraphics[width=0.495\tw]{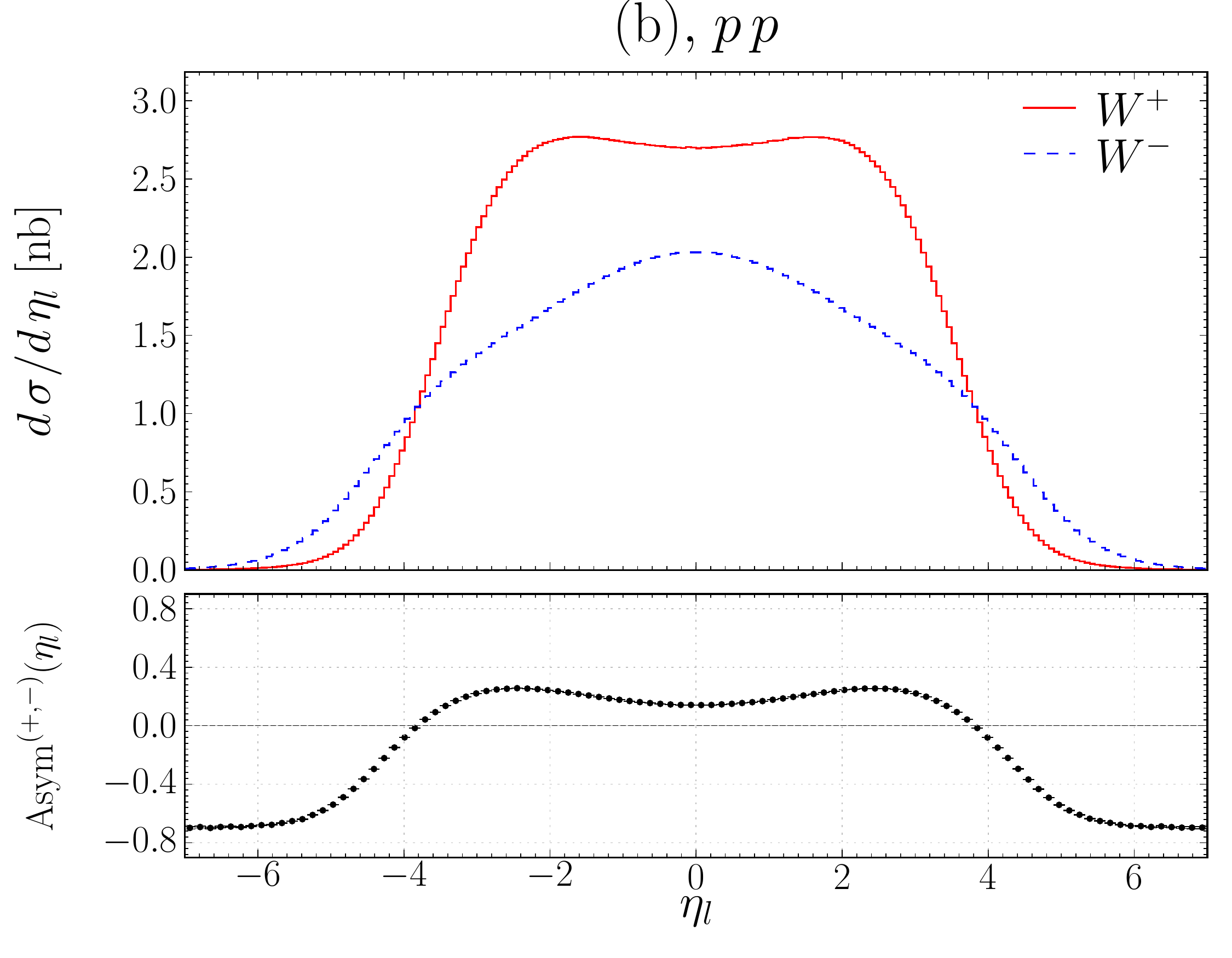}
    \vfill
    \includegraphics[width=0.495\tw]{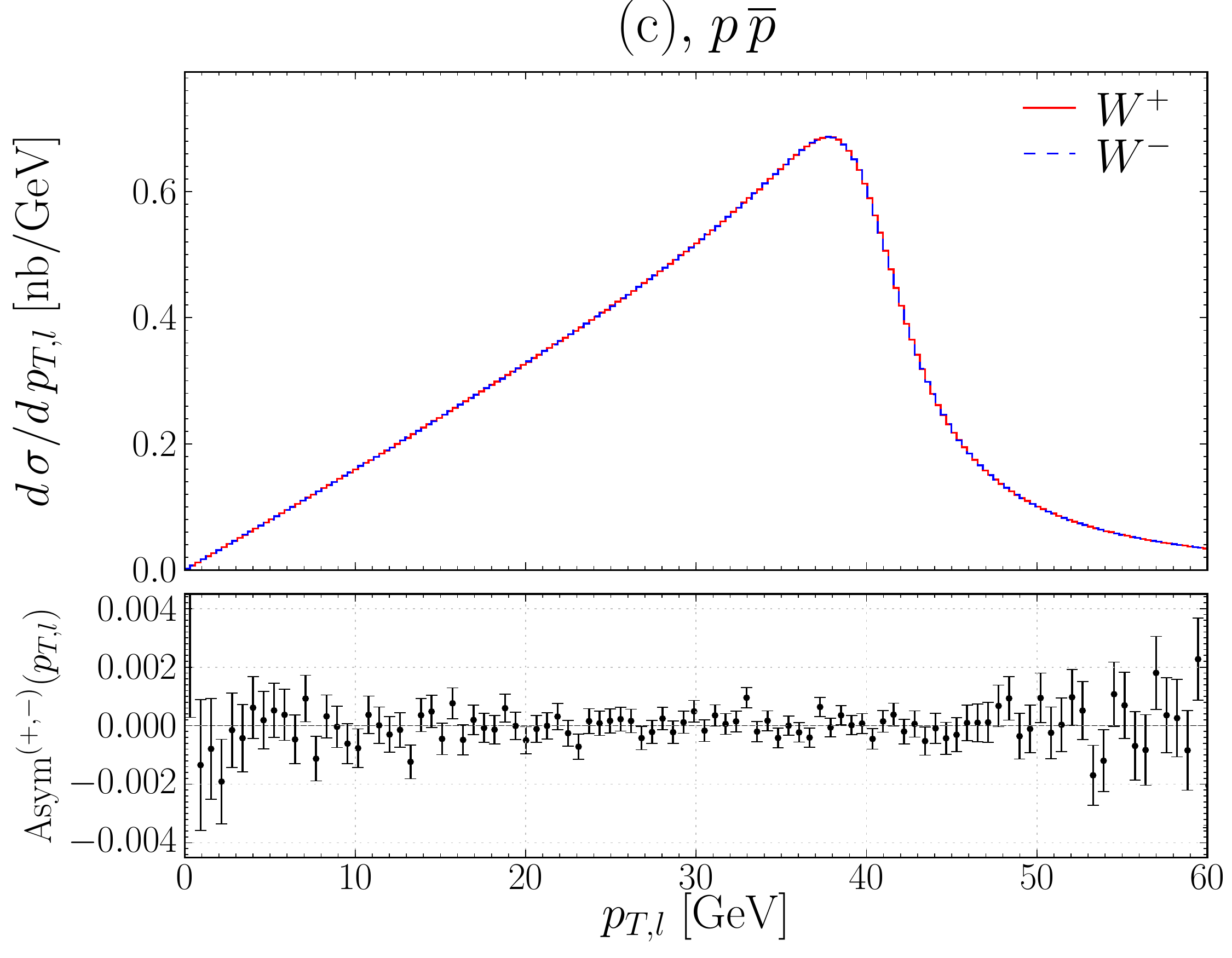}
    \hfill
    \includegraphics[width=0.495\tw]{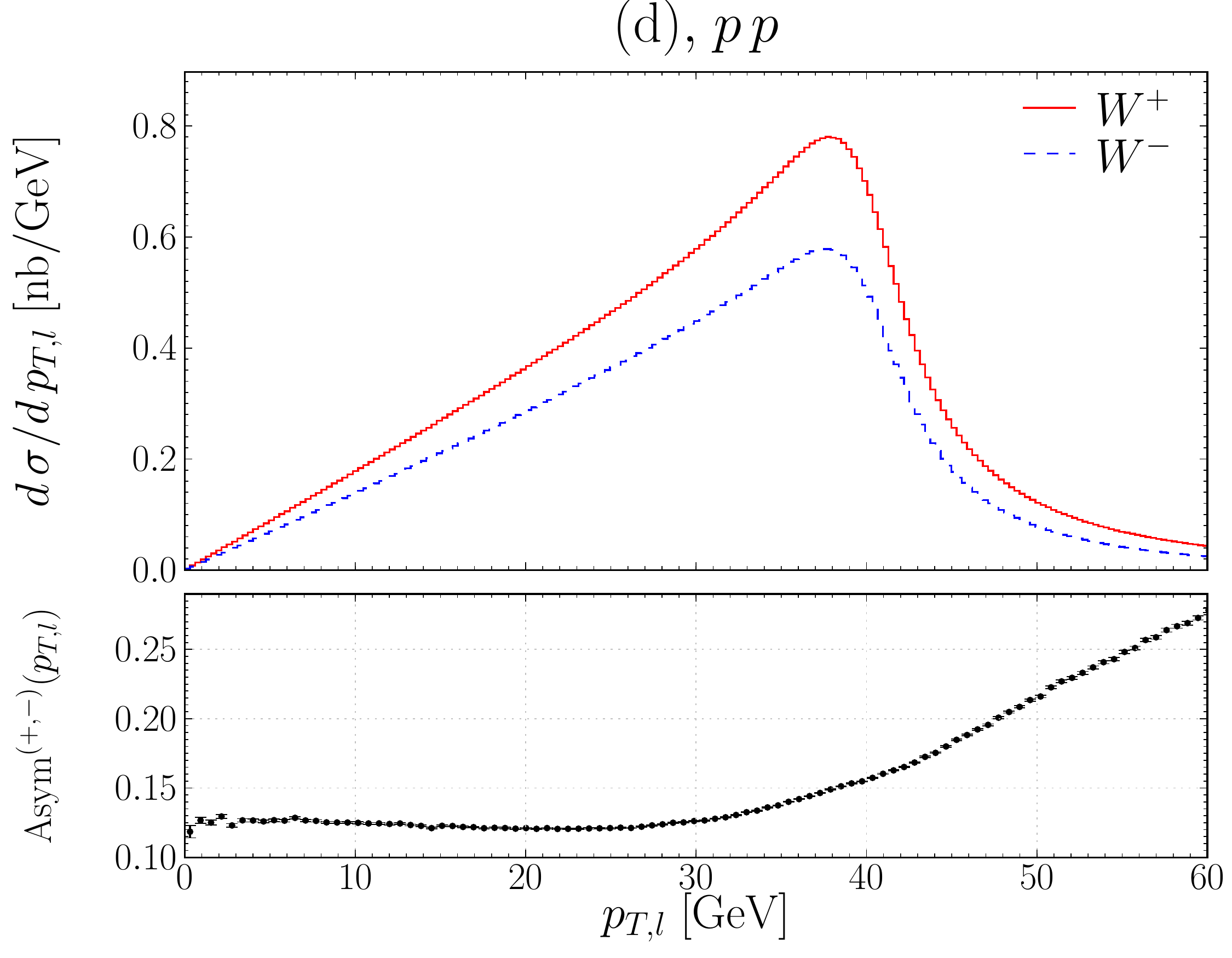}
    \caption[]
            {\figtxt{The pseudorapidity $\etal$ and transverse momentum $\pTl$ distributions for 
              charged leptons produced in $W$-boson decays for the $\ppbar$~(a,b) 
              and $\pp$~(c,d) collisions.}
            }
            \label{fig_etal_pTl_ppb_pp}
  \end{center} 
\end{figure}

The distributions of $\etal$ and $\pTl$ as well as their charge asymmetries for the
$\ppbar$ and $\pp$ collisions are shown in Fig.~\ref{fig_etal_pTl_ppb_pp}.
As can be seen in the LHS plots, in the $\ppbar$ case the $\pTl$ distributions for
positive and negative charges remain identical, while the $\etal$ ones are   
mirror reflected w.r.t. $\etal = 0$.
This behaviour is thus similar to the behaviour of the $W$-boson observables, 
discussed in the previous subsection.
For the $\pp$ collisions  the charge asymmetries (the RHS plots) 
are larger than those for the $W$-boson kinematical variables. 

The increase of asymmetry in the above plots can be explained by the
$V-A$ couplings of the $W$-bosons to leptons. The angular distributions of the final-state
leptons in the $W$-boson rest frame (WRF) can be expressed as follows:
\begin{eqnarray}
  d\,\sigma^{W_T^Q}/d\,\costhetaWlwrf 
  &\propto& \left(1 + \lambda\, Q\,\costhetaWlwrf\right)^2 \label{eq_WT_lep_decay},\\
  d\,\sigma^{W_L^Q}/d\,\costhetaWlwrf 
  &\propto& \sin^2 \theta_{W,l}^{\ast},
\end{eqnarray}
where $\thetaWlwrf$ is the charged lepton polar angle w.r.t. the $W$ momentum direction in the
laboratory frame, $Q$ is the $W$-boson electric charge in units of $|e|$ 
and $\lambda=0,\pm 1$ is the $W$-boson  helicity eigenvalue.
As can be seen, the angular distributions of $l^+$ and $l^-$ 
for longitudinally polarised $W$-bosons  ($W_L$ for $\lambda=0$)
are the same.
For the transversely polarised $W$-bosons ($W_T$ for $\lambda = \pm 1$),
they depend upon the $W$-boson helicity eigenvalue.  
Positively charged leptons,  coming from the decays of    
the left-handed $W$-bosons ($\lambda = -1$)), 
are emitted preferentially in the directions opposite to those  
of the parent $W$-bosons, 
while negatively charged leptons, emitted by the left-handed $W$-bosons, 
tend to retain the direction of their parents. 
The distributions for 
the right-handed $W$-bosons ($\lambda = +1$) decays show the opposite behaviour. 

\begin{figure}[!h] 
  \begin{center}
    \includegraphics[width=0.495\tw]{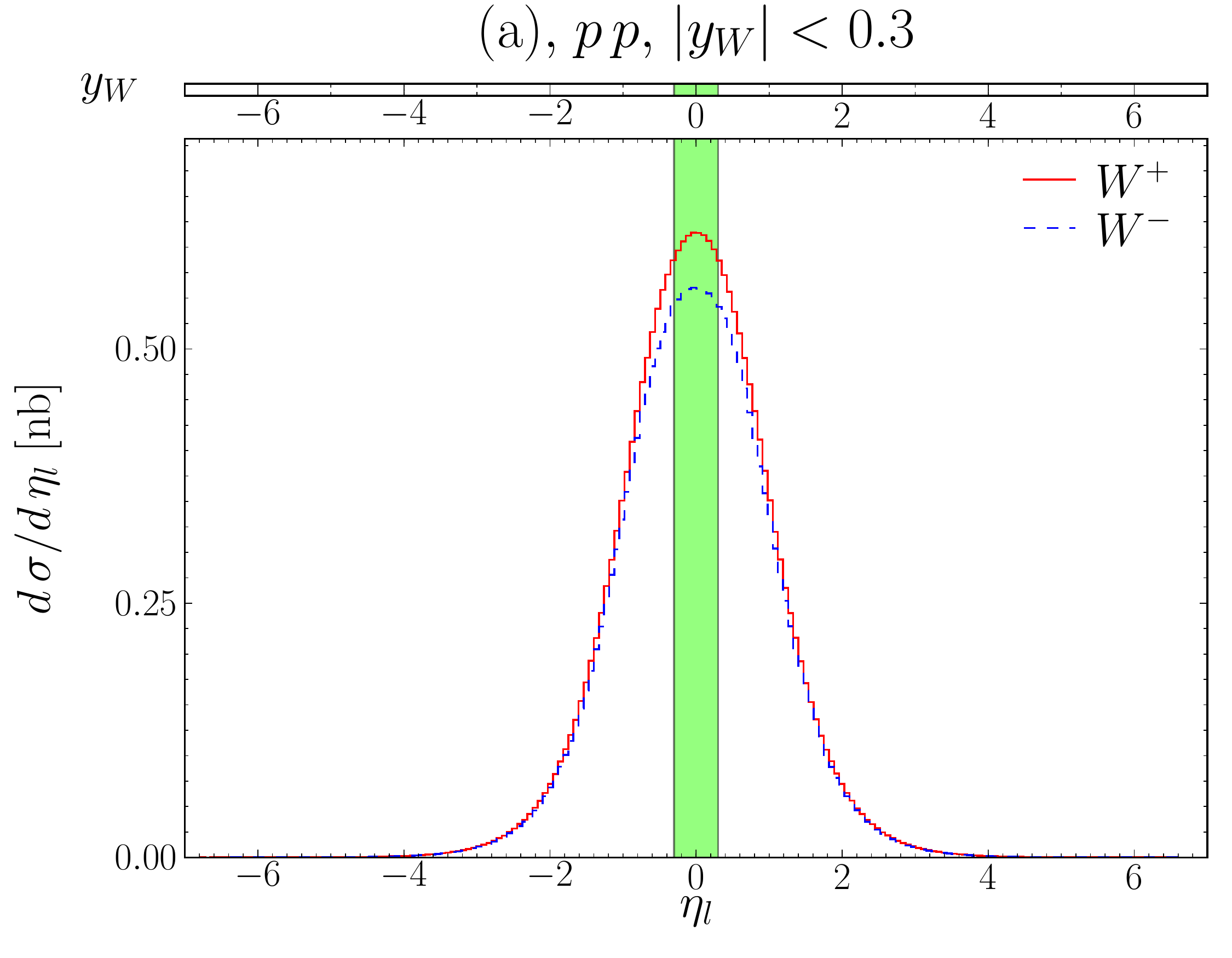}
    \hfill
    \includegraphics[width=0.495\tw]{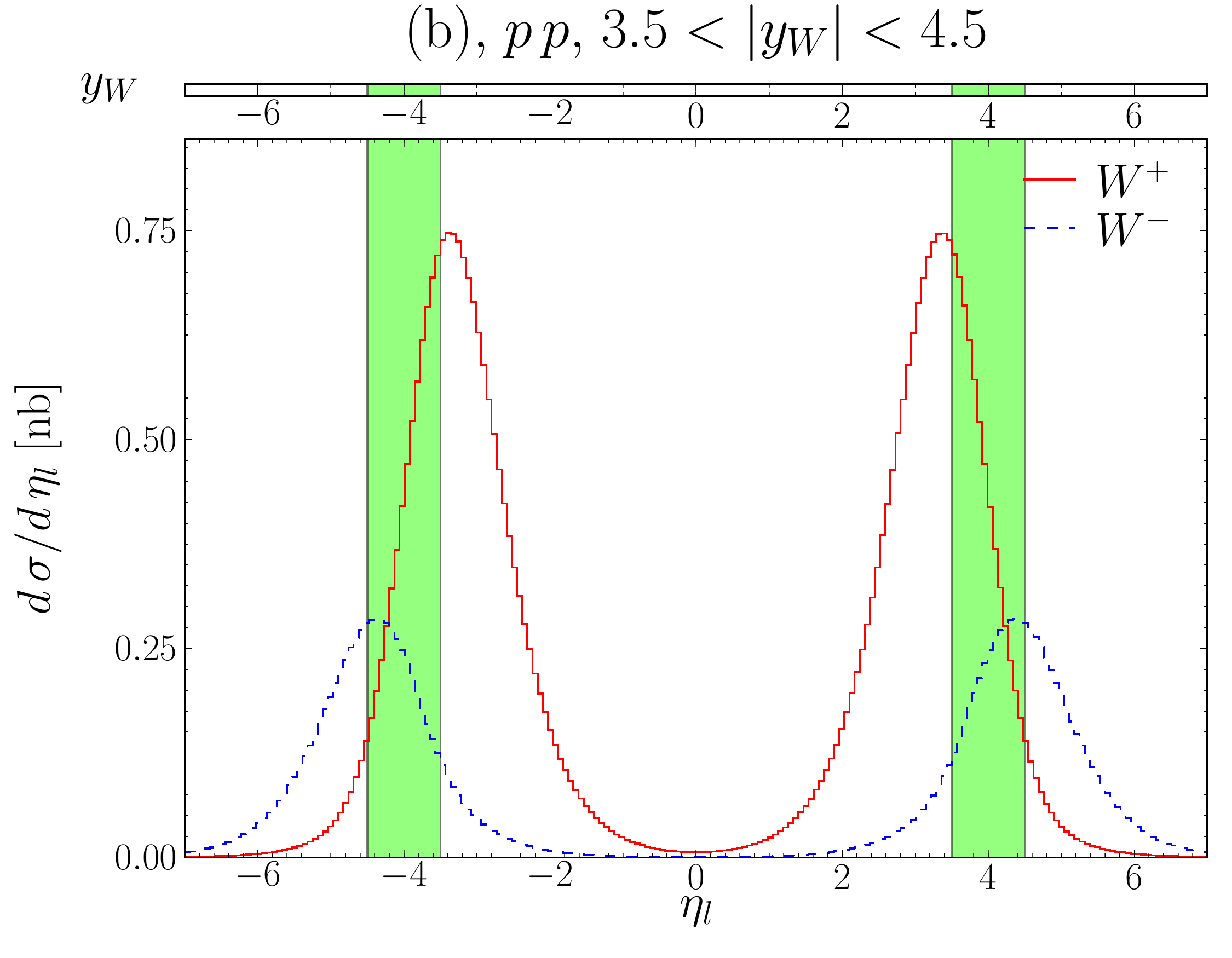}
    \vfill
    \includegraphics[width=0.495\tw]{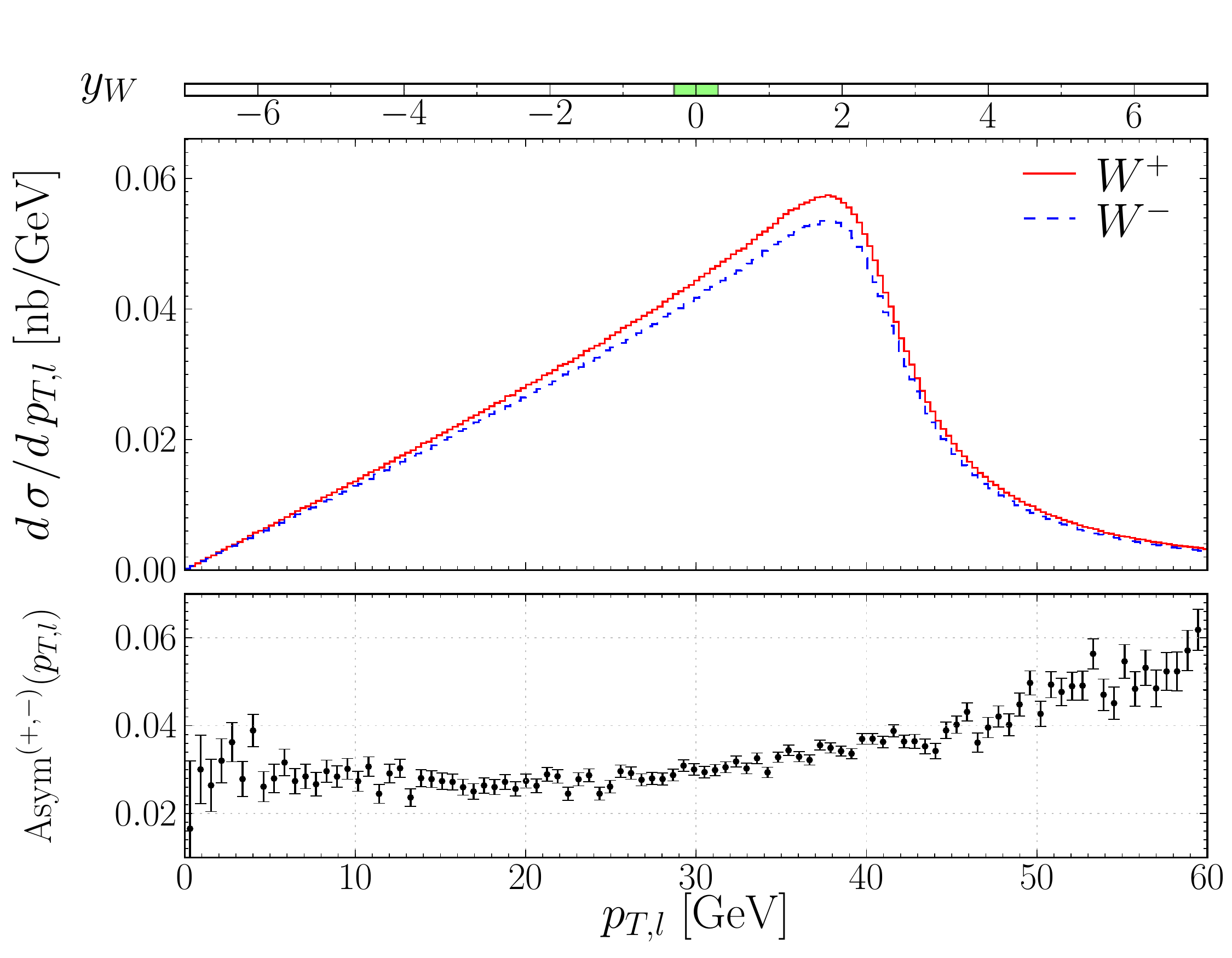}
    \hfill
    \includegraphics[width=0.495\tw]{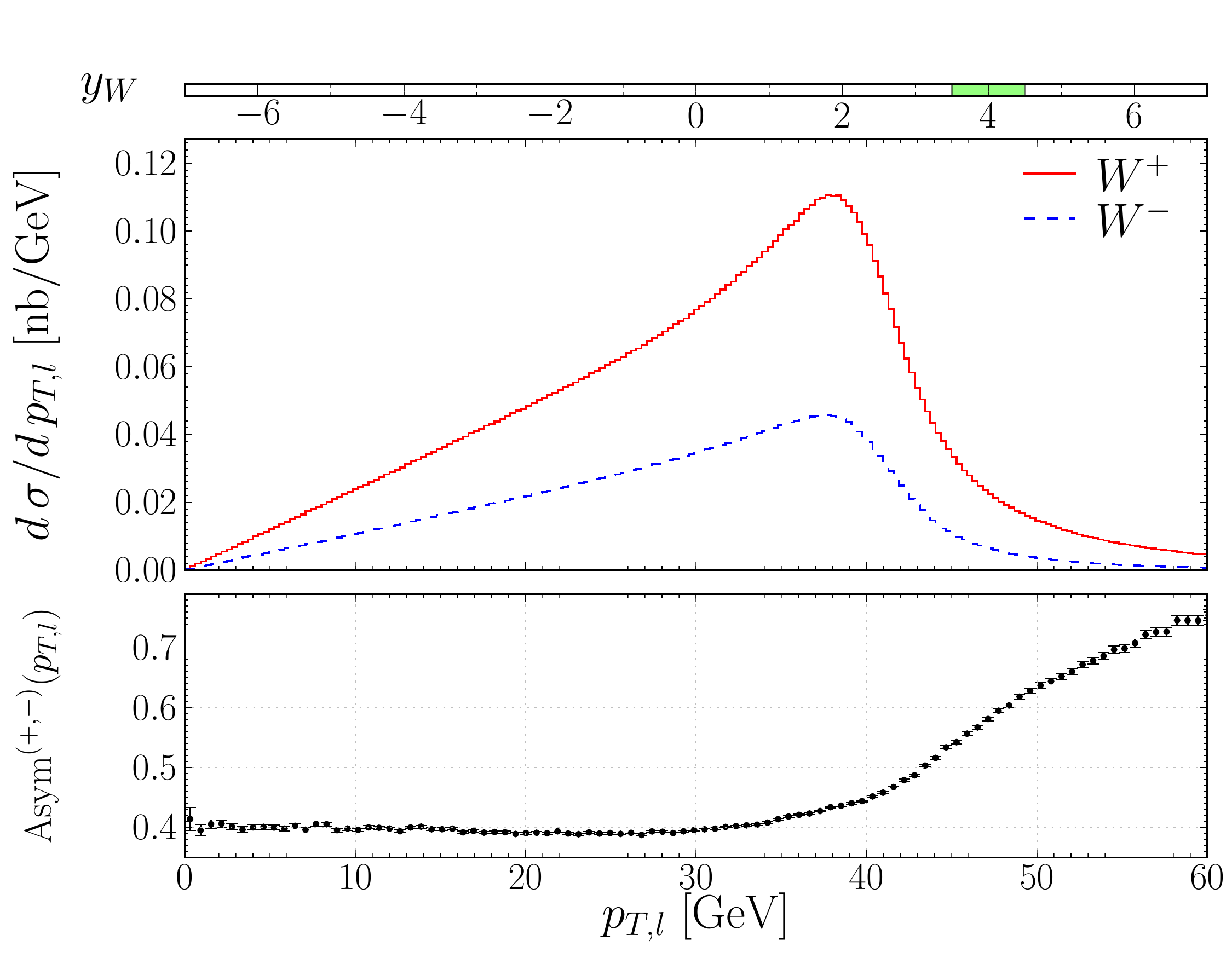}
    \caption[]
            {\figtxt{The distributions of $\etal$ and $\pTl$ for the two different ranges 
              of the $W$ rapidity: (a) $|\yW|<0.3$ and (b) $3.5<|\yW|<4.5$, 
              for the $\pp$ collisions.}
            }
            \label{fig_etal_pTl_in_yW_bin}
  \end{center} 
\end{figure}

Each colliding mode and each centre-of-mass energy determine the 
specific, $W$-boson rapidity-dependent polarisation asymmetry.
The LHC-specific effective polarisation asymmetry of the produced $W$-bosons will 
drive the observed charge asymmetries of the leptonic observables
in a more complicated way than in the case of the Tevatron.   
The magnitude of this contribution and its phase-space dependence 
is illustrated in Fig.~\ref{fig_etal_pTl_in_yW_bin}, 
where the $\etal$ and $\pTl$ distributions in the two distinct
$W$-boson  rapidity regions, corresponding to small and high $|\yW|$ values,  are shown.  
In the region of $|\yW|<0.3$ (LHS plots), the $W$-bosons are produced from 
quarks--antiquark pairs having   $x_q\sim x_{\bar{q}} \approx 6\times 10^{-3}$. 
In this $x$-region the contribution of the valence quarks is small and,  
as a consequence, the relative yields of 
the left-handed and the right-handed $W$-bosons are almost equal. 
In this kinematical region the asymmetries in the $W$-production observables
are hardly modified while switching to the lepton observables.  

In the region of  $3.5<|\yW|<4.5$ the $W$-bosons are produced predominantly 
by the annihilation of the valence quarks with the sea antiquarks. 
The majority of the $W$-bosons in this kinematical region is left-handed
because they are produced predominantly by the left-handed valence quarks and 
because, in most cases, they follow the 
initial direction of the valence quarks. As a result, the negative leptons  are
emitted predominantly in the direction of the $W$-bosons, while 
the positive ones in the opposite direction. 
The Lorentz-boost from the WRF to the laboratory frame increase 
the pseudorapidity of $l^-$ and decrease the  pseudorapidity of $l^+$, 
with respect to the $W$-boson rapidity.
In the upper RHS plot the shadowed  bands represent the $\yW$ regions for selected events.
The observed differences between these bands and the corresponding $\etal$ 
distributions for the positive and negative leptons 
explain the widening of the $\etal$ spectrum
of $l^-$ and narrowing that for $l^+$, observed in Fig.~\ref{fig_etal_pTl_ppb_pp}.
This plot  demonstrates that $\etal$ has significantly weaker resolution power for the 
momenta of annihilating quarks than $\yW$, and, as a consequence, weaker resolving 
power of the $W$-boson polarisation. 

If the $W$-bosons were produced with zero transverse momentum, the  $V-A$
decay effect, discussed above,  would not contribute to the charge asymmetries in the 
distributions of the transverse lepton momentum and would play no role in 
the measurement of the charge asymmetry of the $W$-boson masses.  
In reality, the increase of the transverse momenta of the $W$-bosons produced at large 
rapidities amplifies the impact of the $W$-boson polarisation asymmetry on the 
charge asymmetry of its decay products.  

All the above effects have important consequences for the measurement strategies 
presented in this paper. They will give rise to important  
measurement biases which are absent in the $\ppbar$ collision 
mode but will show up in the measurement of the $W$-boson properties in the 
LHC $\pp$ collision mode. 

In order to illustrate qualitatively the impact of these effects on  the measurement 
of $\MWp$ and $\MWm$ from the position of the Jacobian peaks in the $\pTl$
distributions,  we have made local-parabola fits in the range  $37\,$GeV $<\pTl<40\,$GeV,
for the generated samples of the $W$-bosons produced in the $\pp$ and $\ppbar$
collisions. These events have been generated under the assumption  $\MWp = \MWm$.
The fits have been performed in two $\etal$ regions: $|\etal|<0.3$ and $3.5<|\etal|<4.5$.
While the peak position is lepton-charge independent for the  $\ppbar$  collision
mode, in the case of the  $\pp$ collision mode, the differences in the peak positions 
are $-230\,$ MeV and $+1900\,$MeV for, respectively,  $|\etal|<0.3$ and $3.5<|\etal|<4.5$ bins. 
All the effects contributing to such shifts will have to be understood to a percent level 
if one wants to improve the measurement precision achieved at the Tevatron collider.  

To summarize, the difference between the positive and negative lepton $\pTl$ spectra at the LHC 
result from the interplay of the following three effects:
\begin{itemize}
\baselineskip 1pt
\item the presence  of the valence quarks leading to the dominance of the left-handed $W$-bosons
      with respect to the right-handed ones,
\item the non-zero transverse momentum of the $W$-bosons,
\item the $V-A$ couplings of the $W$-bosons to fermions in electroweak interactions.
\end{itemize}
These differences will depend strongly on the choice of the kinematical 
region used in the analysis. If expressed in terms of leptonic variables, the differences   
are amplified due to induced  biases in the effective $x$-regions of
partons  producing positively and negatively charged $W$-bosons. 
As these differences could mimic the asymmetry in the 
masses of positively  and negatively charged $W$-bosons, all these 
effects must  be controlled to a high precision and/or, as advocated in this paper,  eliminated 
by using the LHC-dedicated measurement strategy.  
These aspects are presented in with more details in Ref.~\cite{Florent_PHD}.

\section{Measurement strategies}\label{s_measurement_method} 

\subsection{Observables}\label{ss_observables}

The values of the  $\Wp$ and $\Wm$ boson masses can be unfolded 
from the measured lepton-charge-dependent distributions of $\pTl$ and/or 
from the transverse mass of the lepton-neutrino system, $\mTlnu$. 
In this paper we discuss only the methods based on the measurement of $\pTl$. 
These methods are almost insensitive to the detector and modelling 
biases in the reconstructed values of the neutrino transverse momentum, $\pTnu$.
We are aware that, for the measurement of the average mass of the $W$-boson, this merit is 
outbalanced by the drawback of their  large sensitivity to the precise understanding 
of the distribution of the $W$-boson transverse momentum, $\pTW$.
However, for the measurement of the charge asymmetry of the masses this 
is no longer the case because QCD radiation, which drives  
the shape of the $\pTW$ distribution, is independent of the charge the produced $W$-boson.
In our view the $\pTl$-based methods will be superior with respect to 
the  $\mTlnu$-based ones, in particular for the first measurements 
of the $W$-mass charge asymmetries at the LHC. 

The most natural method
is to analyze separately the $\lp$ and $\lm$ event-samples and determine 
independently the masses of the $\Wp$ and $\Wm$-bosons. 
This method 
is based on independent measurements of the $\FlatDsigmaDobs{\pTlp}$
and  $\FlatDsigmaDobs{\pTlm}$ distributions. It 
will be called hereafter \emph{the classic method}.

A new  method proposed and evaluated in this paper is based on the 
measurement of the $\FlatAsym\pTl$ distribution.  
This method will be called hereafter \emph{the charge asymmetry method}. 
The distribution of $\FlatAsym\pTl$ is, by definition,  robust with respect to 
those of systematic measurement effects and those of model-dependent effects that are independent 
of the lepton charge. The acceptance, and the lepton-selection 
efficiency corrections for this observable will, in the leading-order approximation, 
reflect only their lepton-charge-dependent asymmetries.
In addition,  the $\FlatAsym\pTl$ observable  is expected to be robust with respect 
to the modeling uncertainty of the QCD and QED radiation processes.

If extrapolated from the experience gained at the Tevatron, 
the precision of \emph{the charge asymmetry method}
will be limited by the understanding of relative biases in the reconstructed
transverse momenta for  positively and negatively charged particles.
These biases, contrary to the lepton-charge averaged  biases,  cannot be controlled 
using the $J/\psi,\Upsilon,Z$ `standard candles'. 

The third measurement method 
proposed and evaluated in this paper is based on the double charge asymmetry 
defined as
\begin{equation}
  \DAsym{\rhol} \;=\; \frac{1}{2} \left[ 
    \mathrm{Asym}^{(+,-)}_{\vec B =  B\,\vec e_z}\left(\rhol\right) + 
    \mathrm{Asym}^{(+,-)}_{\vec B = -B\,\vec e_z}\left(\rhol\right) 
    \right],
\label{eq_def_dble_charge_asym}
\end{equation}
where the variable $\rhol$, defined as $\rhol=1/\pTl$, represents the radius of the track 
curvature at the $W$-boson production vertex 
in the plane $xy$ perpendicular to the beam collision axis,
and $B$ is the strength of the magnetic field.

The $\DAsym{\rhol}$ distribution is expected
to be robust with respect to the charge dependent track measurement biases
if the following two conditions are fulfilled: (1) the inversion of the $z$-component 
of the magnetic field in the tracker volume can be controlled to a requisite precision,
(2) the $\vec E \times \vec B$ relative corrections to the reconstructed hit positions 
for the two magnetic field configurations, 
in the silicon tracker could be determined to a  requisite precision.
The measurement method using the $\DAsym{\rhol}$ distribution will be called hereafter 
\emph{the double charge asymmetry method}.

\subsection{The machine and the detector settings }\label{ss_configuration}

The primary goal of the LHC experiments is to search for new phenomena at the 
highest possible collision energy and  machine luminosity.
It is obvious that, initially,  the machine and the detector operation modes will be optimised  
for the above research program.
The main target  of the work presented in this paper is to evaluate the
precision of the measurement of the $W$-mass charge asymmetry  
which is achievable in such a phase of the detector and machine operation.  

A natural extension of this work is to go further and investigate 
if, and to which extent,  the precision 
of measurement of the Standard Model parameters could be improved in dedicated 
machine- and detector-setting runs. 
In our earlier work \cite{Krasny:2007cy}
we discussed the role of: (1) dedicated runs with reduced beam collision 
energies, (2) dedicated runs with isoscalar beams and (3)  
runs with dedicated detector-magnetic-field settings;   
in optimizing the use of the $Z$-boson production processes as 
a `standard candle' for the $W$-boson processes. 
In this paper we shall discuss the possible improvement
in the measurement precision of the charge asymmetry of the $W$-boson mass
which can be achieved (1) by replacing 
the proton beams with light isoscalar-ion beams,  and (2) by  running the 
detectors for a fraction of time with the inversed direction of the solenoidal 
magnetic field. These and other dedicated operation modes could 
be tried  in the advanced,  `dedicated-measurement phase' of the LHC experimental program. 
Such a phase, if ever happens, could start following the running period  when the collected 
luminosity will become a linear function of the running time and 
the gains/cost ratio  of its further increase will be  counterbalanced by the gains/cost  
ratio of running dedicated machine and detector operation modes.        
 
\subsection{Event selection}\label{ss_event_selection}

The $W$-boson production events used in our studies 
are  selected by requiring 
the presence of an  isolated lepton, an electron or a muon, 
satisfying the following requirements: 
\begin{equation}
\pTl>20\GeV  \quad {\rm and} \quad |\etal|<2.5.
\label{pTl-etal-cuts}
\end{equation} 
In the  CDF experiment \cite{Aaltonen:2007ps} the  
additional requirement of the reconstructed total missing transverse energy   
in the detector,  $\ETmiss \geq 30\,$GeV, allowed to reduce the contribution  
of the background processes to a sufficient level such that the resulting uncertainty 
of the measured mass of the $W$-boson was negligible with respect to other
sources of systematic errors. In our studies we assume that by using a suitable 
$\ETmiss$ cut the impact of the uncertainty in the background contribution 
on the measurement of the charge asymmetry of the $W$-boson mass at the LHC 
can be made negligible. This allows us to skip the generation and simulation 
of the background event-samples for our studies. 
It has to be stressed, that the above assumption  is weaker for the measurement of 
the charge asymmetry than for the measurement of the average $W$-boson mass
\cite{Atlas_W,CMS_W}
because,  to a good approximation, only the difference of the background
for the positive and negative lepton samples will bias the measurement.
Presentation of the results of our studies is largely simplified by noticing   
that they are insensitive to the presence 
of the $\ETmiss$ cut in the signal samples. 
Therefore, we present the results based on selection of 
events purely on the basis of the reconstructed charged 
lepton kinematical variables. Our studies have shown that these results 
will remain valid whatever a value of the $\ETmiss$ cut will be 
used at the LHC to diminish the impact of the background contamination 
at the required level of precision.

\begin{figure}[!ht] 
  \begin{center}
    \includegraphics[width=0.495\tw]{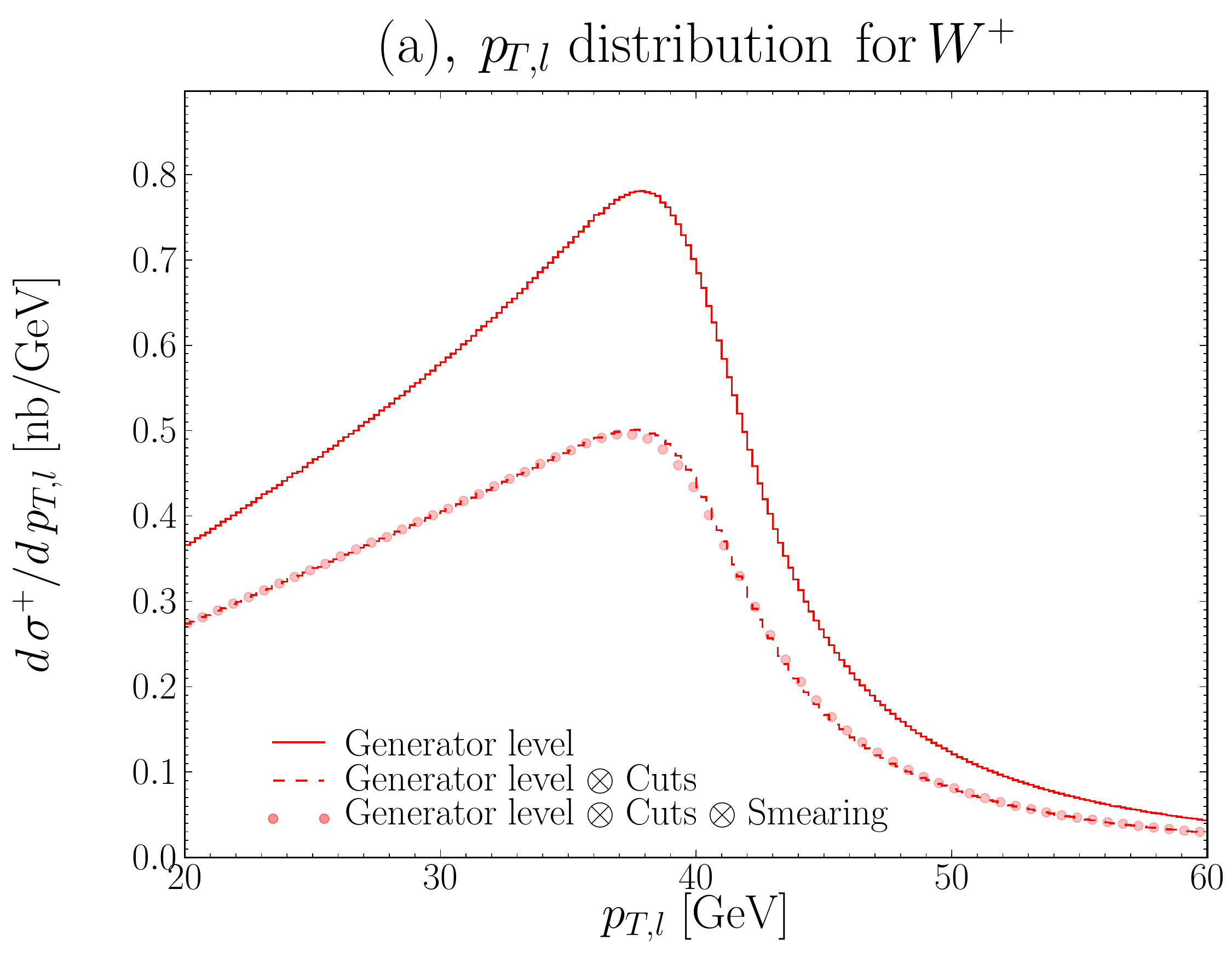}
    \hfill
    \includegraphics[width=0.495\tw]{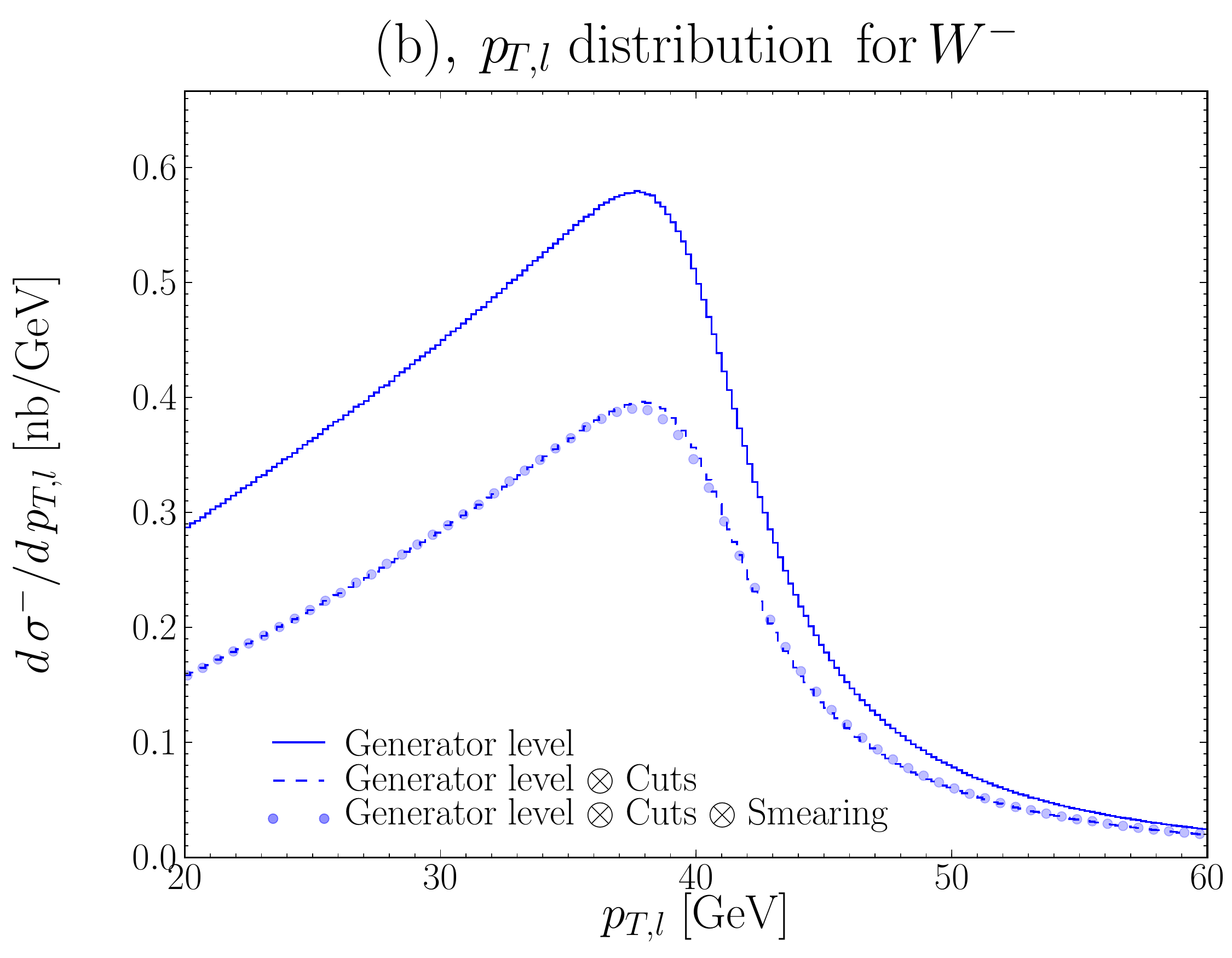}
    \vfill
    \includegraphics[width=1.\tw]{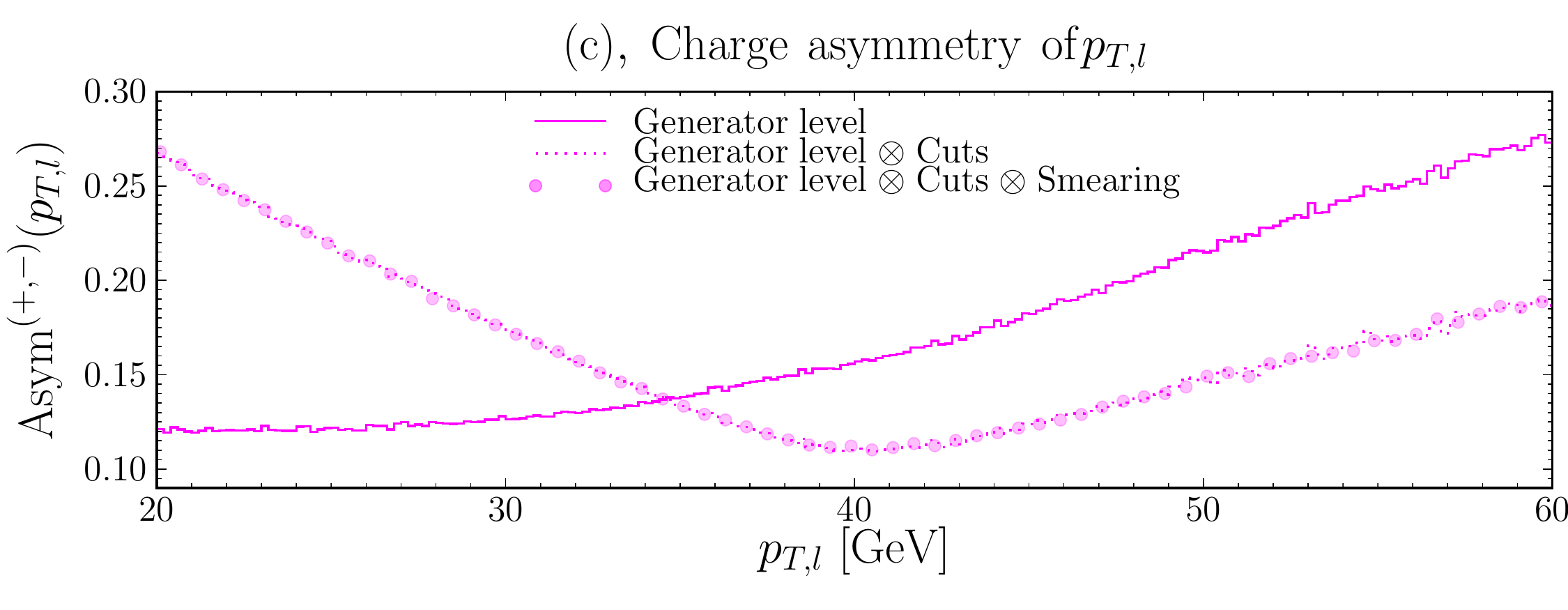}
    \caption[]
            {\figtxt{The $\pTl$ distributions for the positive (a) and negative (b) leptons
                and $\Asym{\pTl}$ (c) at
                the generator level, after the cuts ($\pTl>20\GeV$ and $|\etal|<2.5$), and
                finally, by adding the inner detector smearing.
               } }
            \label{fig_apparatus}
  \end{center} 
\end{figure}

In  Fig.~\ref{fig_apparatus} we show the  distributions of 
$\FlatDsigmaDobs{\pTl}$ and  $\Asym{\pTl}$ for: 
(1) the generated and unselected sample of events, 
(2) the generated and selected sample of events  and 
(3) the unbiased-detector-response-simulated and selected sample of events. 
Analysis of the systematic  biases affecting these distributions will allow us 
to evaluate the precision of the measurement methods discussed in this section.    
        
\section{The analysis method}\label{s_analysis_strategy}

In  this section we present the technical aspects of the analysis method
used in evaluation of the achievable precision of the measurement of the 
charge asymmetry of the $W$-boson masses,  $\MWp-\MWm$, denoted 
sometimes for convenience  as  $\DeltaPM$.

The shapes of the distribution shown in  Fig.~\ref{fig_apparatus} are sensitive 
to: (1) the assumed values of $\MWp$ and   $\MWp$, 
(2) the values of the other parameters of the Standard Model, 
(3) the modeling parameters of the $W$-boson production processes and 
(4) the systematic measurement biases. Our task is to evaluate the impact of the 
uncertainties of (2), (3) and (4) on the extracted values of $\MWp$ and   $\MWm$ for
each of the proposed measurement methods.  We do it by a {\it likelihood analysis}
of the distributions for the {\it pseudo-data} ($\cal{PD}$) event-samples 
and those  for the {\it mass-template} ($\cal{MT}$) event-samples.
Each of the $\cal{PD}$ samples represents 
a specific measurement or modeling bias, implemented respectively in the event-simulation 
or event-generation process.  Each of the $\cal{MT}$ samples 
was generated by assuming a specific $\MWp$ ($\MWm$) value or a value of their
charge asymmetry, $\DeltaPM$.
The $\cal{MT}$ samples are simulated using the unbiased 
detector response and  fixed values of all the parameters used in the modeling 
of the $W$-boson production and decays except for the mass parameters.   
The  likelihood analysis, explained below in more detail, 
allows us: (1) to find out which of the systematic measurement 
and modeling errors, 
could be falsely absorbed into the measured value of the $W$-boson masses and 
(2) to evaluate quantitatively the corresponding measurement biases.

\subsection{Likelihood analysis}\label{ssec_chiTwo}

Let us consider, as an example,  the impact 
of a systematic effect,  $\xi$,  on the bias  in the measured  value of the $\Wp$ mass
determined from the likelihood analysis of the $\FlatDsigmaDobs{\pTlp}$ distributions.  

The simulation of the pseudo-data event sample, $\cal{PD}$, representing a given systematic bias $\xi$,
is carried out  for a fixed value of the mass $M_{W^+}^{\mathrm{ref}}$. Subsequently, 
a set of the $2k+1$ unbiased (\ie{} $\xi=0$) template data samples, $\cal{MT}$, is simulated. 
Each sample $n$ of the $\cal{MT}$ set corresponds to a given value of 
$M_\Wp^{(n)} = M_{W^+}^{\mathrm{ref}} + \delta M_\Wp^{(n)}$, $n=-k,\ldots,k$.     
The likelihood between the binned $\FlatDsigmaDobs{\pTlp}$ distributions   
for the $n^\mathrm{th}$ $\cal{MT}$ sample  and the $\xi$-dependent $\cal{PD}$ sample is quantified
in terms of the  $\chiD$ value: 
\begin{equation}
  \chiD(\pTlp;\,\xi, n ) \, = \, \sum_i 
  \frac{\left(d\,\sigma_{i;\,\xi} - d\,\sigma_{i;\,\xi=0, n}\right)^2}
       {\Delta d\,\sigma_{i;\,\xi}^2 + \Delta d\,\sigma_{i;\,\xi=0,n}^2},
\label{eq:chi2}
\end{equation}
where $d\,\sigma_i$ is the content of the $i^\mathrm{th}$ bin 
of the  $\FlatDsigmaDobs{\pTlp}$ histogram and $\Delta d\,\sigma_i$ is the 
corresponding statistical error. 
The bulk of the results presented in this paper has been obtained
using a  bin size corresponding to $\sigma$ of the anticipated measurement resolution 
of the track curvature \cite{AtlasInnerDetector:1997fs} 
and the summation range satisfying the 
following condition: $30\GeV <\pTl<50\GeV $.

The $\chiD(\pTlp;\,\xi, n)$ dependence upon $\delta M_\Wp^{(n)}$ is  
fitted by a polynomial of second order. The position of the minimum, $\MWp(\xi)_{min}$, of the 
fitted function determines the systematic mass shift  
$\Delta\MWp(\xi) =\MWp(\xi)_{min} - M_{W^+}^{\mathrm{ref}}$ due to the systematic 
effect $\xi$. If the systematic effect under study 
can be fully absorbed by a shifted value of $\Wp$, then 
the expectation value of $\chiDmin/\dof$, where $\dof=\sum_i$,  is equal to $1$  
and the error of the estimated value of the mass shift, 
$\delta\,\left(\Delta\MWp(\xi)\right)$, can be determined from the 
condition 
$\chiD(\MWp(\xi)_{min}+\delta\,\left(\Delta\MWp(\xi)\right)) =\chiDmin+1$. 

Of course, not all the systematic and modeling effects can be absorbed into a variation 
of a single parameter, even if the likelihood is estimated in a narrow bin-range,
purposely chosen to have the highest sensitivity to the mass parameters. In such a case 
the value of $\chiDmin/\dof$ can be substantially different from $1$, and 
$\delta\,\left(\Delta\MWp(\xi)\right)$ looses its statistical meaning. 
This can partially be recovered by introducing supplementary degrees of freedom 
(the renormalisation of the $\cal{PD}$ samples, discussed in this section, 
is an example of such a procedure). However, even in such a case 
the estimated value of $\delta\,\left(\Delta\MWp(\xi)\right)$ will  
remain dependent 
upon the number of the $\cal{MT}$ samples, $2k+1$, their $\MWp$ spacing in the vicinity 
of the minimum and the functional form of the fit. 
Varying these parameters in our analysis procedure in a $\xi$-dependent 
manner would explode the PC farm CPU time and was abandoned. 
Instead, we have calibrated the propagation of statistical bin-by-bin errors
into the $\delta\,\left(\Delta\MWp(\xi)\right)$ error, and checked 
the biases of all the aspects of the above method using the statistically 
independent ``$\cal{PD}$-calibration samples'' in which, instead of varying 
the $\xi$ effects, we varied the values of $\MWp$.

\subsection{The $\cal{MT}$ and $\cal{PD}$ event samples}

\subsubsection{Classic method}
\label{sss:classic-method}

In the classic method the bias of $\DeltaPM(\xi)$ resulting from the systematic effects 
$\xi$ is determined in the following three steps:   
\begin{enumerate}
\baselineskip 1pt
\item We determine  $\Delta\MWp(\xi)\,\pm\,\delta\,\left(\Delta\MWp(\xi)\right)$
  using $\chiD(\flatDfDx{\sigma^\plus}{\pTl};\,\xi)$. 
\item We determine  $\Delta\MWm(\xi)\,\pm\,\delta\,\left(\Delta\MWm(\xi)\right)$ 
  using $\chiD(\flatDfDx{\sigma^\minus}{\pTl};\,\xi)$.
\item We combine these results and derive:

  \begin{equation}
    \DeltaPM(\xi)  = \Delta\MWp(\xi) - \Delta\MWm(\xi).
  \label{eq:Deltapm} 
  \end{equation}
\end{enumerate}

In the generation of the $\cal{PD}$ samples we assumed 
$M_{W^+}^{\mathrm{ref}}=M_{W^-}^{\mathrm{ref}}= 80.403\GeV$.
The $\cal{MT}$ samples have been generated for:
$\delta  M_\Wp^{(n)}  = \delta M_\Wm^{(n)}= n \times 5\,$MeV with $n=\pm 1,\ldots,\pm6$
and for   
$\delta M_\Wp^{(n)}=\delta M_\Wm^{(n)}= \pm 40, \pm 50, \pm 75, \pm 100, \pm 200\,$MeV
with $ n=\pm 7,\ldots,\pm 11$. In total $46$ $\cal{MT}$ samples, corresponding to $k=11$,
have been generated.

\subsubsection{Charge asymmetry method}
\label{sss:charge-asym}

In this method the biases $\DeltaPM(\xi)$ resulting from the systematic effects 
are  determined by a direct analysis of  the $\Asym{\pTl}$ distributions for the 
$\cal{MT}$ and $\cal{PD}$ event samples. The only difference with 
respect to the classic method is that the $\cal{MT}$
samples used in our studies represent the variation of the  $\DeltaPM$ values rather than 
uncorrelated variations of $\MWp$ and $\MWm$. 
The charge asymmetry method was first verified using two charge-symmetric procedures. 
In the first one the variation of $\DeltaPM$ was made  by fixing  
$\MWp=M_{W^+}^{\mathrm{ref}}$ and by changing $\MWm$.
In the second one we inverted the role of $\MWp$ and $\MWm$.
The results obtained with these two charge-symmetric methods were 
found to agree within the statistical errors.
 
The first measurement of the charge asymmetry of the $W$-boson masses at the LHC 
will have to use, as the first iteration step, the best existing constraints 
on the $W$-boson masses. The best available constraint is the average mass of the $\Wp$ and 
$\Wm$ bosons: $\MW=M_W^{\mathrm{ref}}$.
To mimic the way how the measurement will be done at the LHC, 
we thus fixed the  
$\MW=M_W^{\mathrm{ref}}$ value and varied, in a correlated way, 
both the $\MWp$ and $\MWm$ values when constructing the $\DeltaPM$-dependent $\cal{MT}$ samples.   

\subsubsection{Double charge asymmetry method}
\label{sss:double-asym}

The $\cal{MT}$ event samples for this methods are exactly the same as for the 
charge asymmetry method.
The $\cal{PD}$ event samples have been simulated  in two steps
corresponding to the two half-a-year periods of data taking with the two magnetic field configurations.

\begin{figure}[!ht] 
  \begin{center}
    \includegraphics[width=0.495\tw]{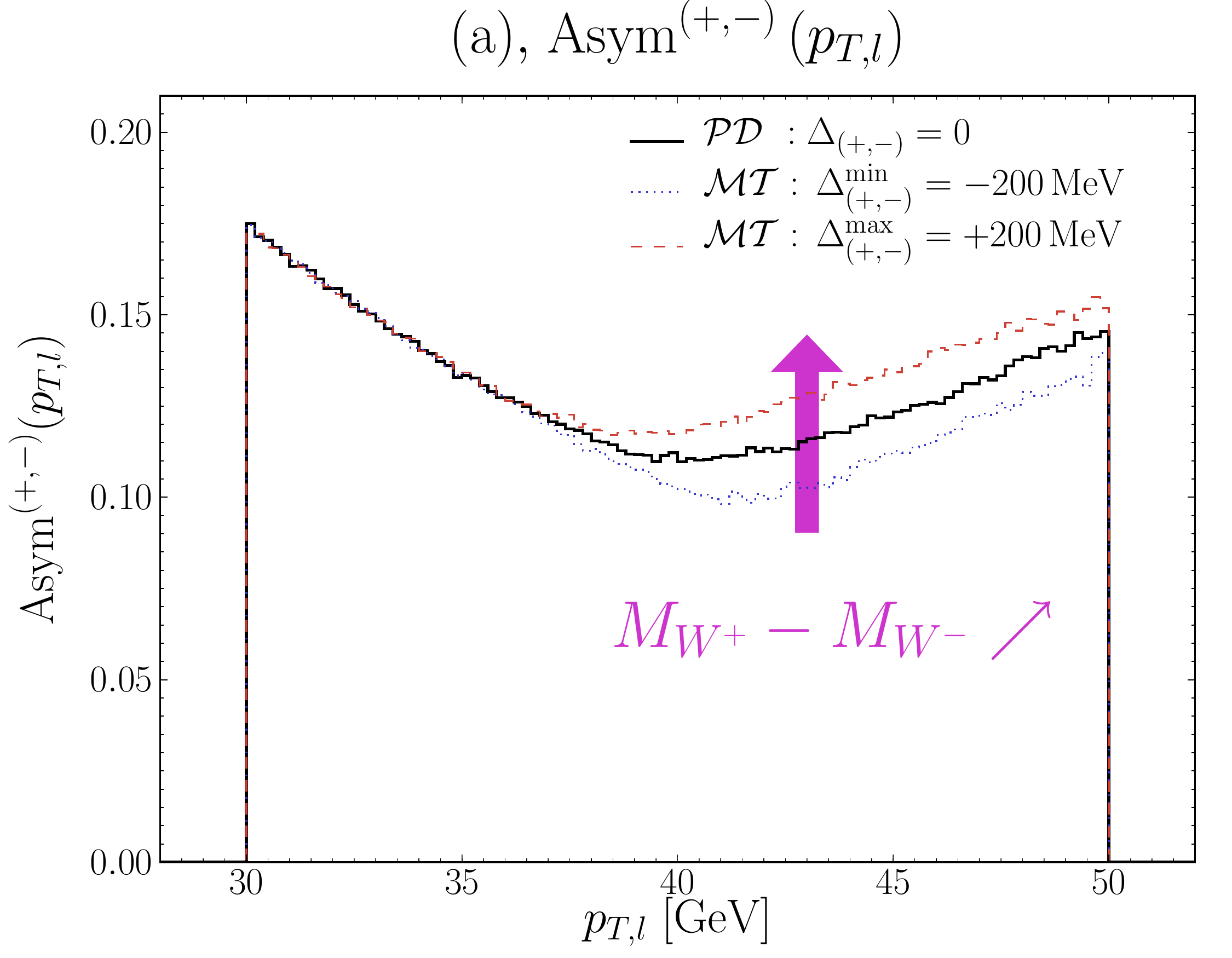}
    \hfill
    \includegraphics[width=0.495\tw]{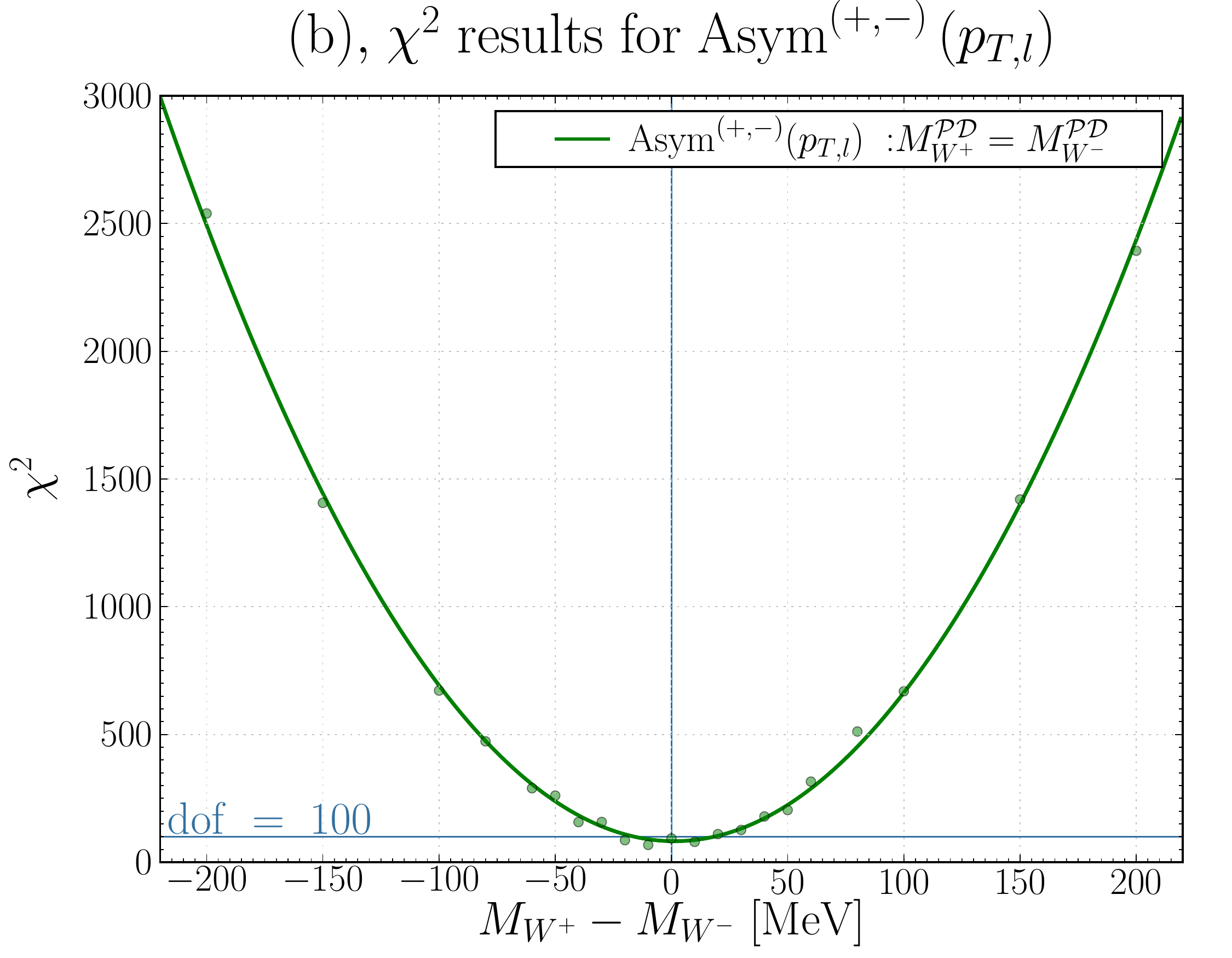}
    \caption[]
            {\figtxt{The charge asymmetry of $\pTl$ for the three values of $\DeltaPM$:
              $-200,~0,~+200\,$MeV (a) and 
              the $\chiD$ dependence  ($\cal{MT}$ points and their the polynomial fit)   of $\DeltaPM$ 
              (b).}
            }
            \label{fig_chi2_exp_cent_asym}
  \end{center} 
\end{figure}

The procedures discussed above are illustrated in Fig.~\ref{fig_chi2_exp_cent_asym}
for the case of the charge asymmetry method.
The $\Asym{\pTl}$ distribution is plotted in Fig.~\ref{fig_chi2_exp_cent_asym}a as a function of 
$\pTl$ for three values $\DeltaPM$. This plot illustrates the sensitivity of 
the $\Asym{\pTl}$ distribution to the $\DeltaPM$ value.
In Fig.~\ref{fig_chi2_exp_cent_asym}b the $\chiD$ variable is plotted 
for the $\cal{PD}$-calibration  sample 
corresponding to $\Delta_{(+,-)}^{\mathrm{ref}} =0$ and to an unbiased detector response, 
as a function of $\DeltaPM$. The 
position of the minimum is $\DeltaPM(\xi=0)= (1~\pm~4\,)$MeV  and the corresponding 
$\chiDmin/\dof =0.82$. This plot illustrates the calibration procedure.
It shows that no systematic biases are introduced by the proposed
analysis method. It calibrates the statistical precision 
of the measurement for the integrated luminosity of $10\,\mathrm{fb}^{-1}$. 
Our goal will be to estimate
the systematic biases of in the measurement of $\DeltaPM$ with the comparable precision.

\subsection{Scaling distributions for quarks flavors systematics}
\label{ss_scaling_trick}

Most of systematic measurement and modeling biases discussed in this paper lead to a distortion of 
the distributions and do not change their overall normalisation. The notable 
exception,  discussed in more detail in the next section, are the biases driven by the PDFs 
uncertainties. These biases cannot be `absorbed' by  the corresponding $\DeltaPM(\xi)$ shifts 
and require an adjustment of the event/nb normalisation of the corresponding $\cal{PD}$ samples
to obtain acceptable $\chiD$ values. 

The most natural method would be to extend the one-dimensional analysis presented in this 
section into two-dimensional analysis of both the mass and the normalisation parameters. Such 
an analysis would have, however, `square' the necessary computing time of the 
$\cal{MT}$ samples and, therefore, was not feasible in our time-scale. 
Instead,  we have tried  to `uncorrelate' the adjustment of the normalisation 
parameter and the mass parameters. As shown in Fig.~\ref{fig_chi2_exp_cent_asym}a,
the $\Asym{\pTl}$ distribution is, in the region of small $\pTl$, independent of 
the variations of $\DeltaPM$ over the range discussed in this paper. We use this observation 
and modify correspondingly the likelihood  analysis method. 
Before calculating $\chiD$, the $\cal{PD}$ and $\cal{MT}$ distributions
are integrated as follows:
\begin{equation}
\int_{20\GeV }^{35\GeV } d\,\pTl\; \left(\DfDx{\mathrm{Asym}^{(+,-)}}{\pTl}\right),
\label{eq:interal-pTl}
\end{equation}
giving, respectively, two scalars: $\alpha$ and $\beta(n)$.
Then we re-normalize the $\cal{MT}$ distributions  by factors $\alpha/\beta(n)$
and calculate the $\chiD$ values for the rescaled distributions. 
We have checked that the above procedure improves significantly the resulting $\chiD$
values for each of the three analysis methods. By changing the integration region 
we have verified that the above procedure does not introduce significant biases in the estimated 
$\DeltaPM(\xi)$ values.  

\section{Systematic error sources}\label{s_model_impact_sys} 

In this section we identify and model the systematic error sources that 
will limit  the precision of the $\MWp-\MWm$ measurement
at the LHC. 
Each of these errors sources will be modeled and reflected in the corresponding 
$\cal{PD}$ sample of events.
   
These error sources are of two kinds: (1) those reflecting  
uncertainties in modeling of the $W$-boson production and decay 
processes, (2) those reflecting the event selection and 
event reconstruction biases. A large fraction of the  error sources have been 
identified \cite{Aaltonen:2007ps}  and reevaluated in the context of the measurement 
of the average mass of the $W$-boson at the LHC \cite{Atlas_W,CMS_W}. 
In this paper we focus our discussion on {\it the dominant errors }
for the measurement of the $W$-mass charge asymmetry, in particular
on those that are specific to the LHC environment and have not been identified
in the earlier studies. We shall not discuss here: (1) the measurement errors 
reflecting the uncertainties in the background estimation and in the efficiency 
of the events selection, (2) other measurement uncertainties which can be studied to the 
required level of precision only once the real data are collected. As 
demonstrated  in the analysis of the Tevatron data \cite{Aaltonen:2007ps}, 
they are of secondary importance.     
A more complete presentation of the analysis of the 
error sources and their modeling is given in Ref.~\cite{Florent_PHD}.

\subsection{Modeling uncertainties}\label{s_model_impact_sys_pheno} 

The uncertainties in modeling of the production and decay of the $W$-bosons include: 
(1) the uncertainties in modeling of nonperturabative effects, 
(2) the approximations present in theoretical 
modeling of the perturbative EW and QCD effects,
(3) the uncertainties in the parameters of the Standard Model, and (4) a possible  
presence of the ``Beyond-the-Standard-Model'' (BSM) effects, 
affecting both the production and decay mechanisms of the $W$-bosons. 
The first two of them limit  our present 
understanding of the Wide-Band-partonic-Beam (WBpB) at the LHC.

\subsubsection{WBpB at LHC}
\label{sss:WBpB}

The measurement precision of the $W$-mass charge asymmetry  
will depend upon the level of understanding 
of the flavour structure, the momentum spectrum and the emittance 
of the WBpB at the LHC collision energy. 
The hard-scale dependent emittance of the WBpB
is defined here, in analogy to the emittance of the parent hadron beam,
in terms of its transverse momentum distribution
and in terms of its transverse and longitudinal beam-spot sizes.      
The above dynamic properties of the WBpB are highly correlated.
Only their scale dependence can be controlled by the 
Standard Model perturbative methods. In addition,  
several aspects of such a control, in particular the 
precise modeling  of the 
correlations between the flavour, the longitudinal and the transverse
momentum degrees of freedom of the WBpB  have not so far been 
implemented in the  Monte-Carlo generators available 
for the initial  phase of the LHC experimental program.

The present  `folklore' of understanding of the WBpB at the LHC
is driven by the presently available Monte Carlo (MC) generator tools. 
Within this `folklore',  the flavour-dependent longitudinal momentum distribution of the WBpB, 
specified by  `collinear' PDFs, is fed to one of the available  
parton-shower MC generators. The transverse momentum distribution of the WBpB  is 
then derived from the assumed longitudinal one. This procedure 
depends upon a particular evolution-scheme-dependent form of the parton shower
and upon the order of the perturbative expansion.
It depends as well upon the modeling method of the quark-flavour (quark-mass) 
effects in the parton-shower generation.   
The effects of the flavour dependence 
of the beam-size in the transverse plane are partially controlled 
using  auxiliary, impact-parameter
dependent resummation procedures. Finally, the scale-dependent evolution of 
the longitudinal beam-spot size is presently assumed to be driven by the 
DGLAP evolution.     

It is obvious that the precision of the present understanding of the WBpB at the LHC 
is difficult to asses within the 
above modeling environment. Since its impact on the precision measurements
of electroweak processes will be significantly higher for the LHC WBpB
with respect to the Tevatron one, some novel measurement and/or modeling  schemes
must be developed. They must assure either better theoretical control  of the WBpB 
parameters or,  as proposed in this paper, reduce  
their impact on the measured observables
to such an extent that their detailed modeling becomes irrelevant.   
For the latter strategy it is sufficient to rely on crude modeling methods of 
the WBpB at the LHC which are available within the \WINHAC\ generator.

\subsubsection{Uncertainty of PDFs}
\label{sss:PDFuncertainty}

The uncertainties in PDFs are, most often, 
propagated to the measurement errors of 
the physics observables by varying  
the PDF sets chosen  in
the event generation process. Alternatively, the 
uncertainties of the QCD fit parameters of a given PDF set are propagated 
by re-weighting the generated events
with ``min'' and ``max'' weights,  
$\mathrm{PDF}_{\Max/\Min}=\mathrm{PDF}_\Cen \pm \delta\mathrm{PDF}$,
where $\mathrm{PDF}_\Cen$ are the central-value-distributions of 
a given PDF set and
$\delta\mathrm{PDF}$ is computed according to the 
method described in Ref.~\cite{Pumplin:2002vw}.
We followed the latter method, mostly because  
of the computing-time constraints. 
We have used the CTEQ6.1 PDF set \cite{CTEQ6.1:2003}
in modeling of the standard PDFs uncertainties. 
The above methods, in our view, largely underestimate the influence of the 
PDFs uncertainty on the measurement precision of the $W$-boson mass.
 
As discussed in Section~\ref{s_charge_asym}, the charge 
asymmetry of the $W$-boson production and decay processes is 
sensitive: (1) to the presence of valence quarks in the 
WBpB, (2) to the flavour asymmetry of their 
longitudinal momentum distribution (called hereafter  
{\it the $u^\val-d^\val$ asymmetry}), and (3)
to the asymmetry in the relative momentum distribution of the strange and 
charm quarks ({\it the $s-c$ asymmetry}). The corresponding uncertainties 
must be modeled directly using the existing (future)  experimental constraints rather 
than be derived from the uncertainty of the PDF-set parameters. 
This is because they are driven almost entirely 
by the nonperturbative effects, and because the QCD-evolution
effects are, except for the quark-mass dependency, flavour independent.

\subsubsection{Uncertainty of $\mbf{u^\val-d^\val}$  asymmetry.}
\label{sss:ud-asym}

We assume the following two ways of modeling 
the uncertainty in the $u^\val-d^\val$  asymmetry:
\begin{equation}
u^\val_{\Max/\Min} = u^\val \pm 0.05\,u^\val,\qquad
d^\val_{\Min/\Max} = d^\val \mp 0.05\,u^\val,
\label{eq:ud1}
\end{equation}
and 
\begin{equation}
u^\val_{\Max/\Min} = u^\val \pm 0.02\,u^\val,\qquad
d^\val_{\Min/\Max} = d^\val \mp 0.08\,d^\val.
\label{eq:ud2}
\end{equation}
The first one preserves the sum of the distribution of the 
the $u$ and $d$ quarks and is constrained, to a good 
precision, by the measured singlet structure function  
in neutrino and anti-neutrino Deep-Inelastic-Scattering (DIS) of  isocalar 
nuclei. At the LHC the sum of the distributions will be constrained
by the rapidity distribution of the $Z$-bosons 
($d$ quarks and $u$ quarks contribute with similar weights).
The second one assures  the correct propagation 
of the measurement errors of the sum of the charge-square-weighted distributions of 
the $u$ and $d$ quarks, constrained by the  high-precision charged-lepton beam 
DIS data,  to the uncertainty of the individual distributions. 
The assumed uncertainties are compatible with those discussed in  
the recent review \cite{MRS2009}.   

While the sums of the  
distributions are well controlled by the existing and future data,
their mutually compensating shifts are not. The only experimental constraints on  
such shifts come from (1) the measurements of   
the ratio of the cross sections for deep inelastic scattering of charged leptons 
on proton and deuteron targets and (2) the measurements 
of the ratio of the neutrino-proton  to antineutrino-proton  
DIS cross sections. They determine the present uncertainty range of the  $u^\val-d^\val$
asymmetry. Improving this uncertainty range in the standard $\pp$ LHC colliding mode will 
be difficult and ambiguous. It will require simultaneous unfolding of the   
momentum distribution and the charge asymmetry of the sea quarks.

\subsubsection{Uncertainty of $\mbf{s-c}$ asymmetry}
\label{sss:cs-asym}

The $cs$ annihilation represent 
the $\approx 7\percent$ contribution to the total $W$-boson production cross section at the 
Tevatron collision energy. At the LHC collision energy it rises to  $\approx 25\percent$ and 
becomes charge asymmetric: $\approx 21\percent$ for the $\Wp$-boson and $\approx 28\percent$ for the $\Wm$-boson. 
The uncertainty in the relative distribution of the strange and charm quarks becomes an 
important source of systematic measurement errors of both the average $W$-boson mass and 
its charge asymmetry. 

We assume the two following ways of modeling 
the uncertainty in the $s-c$ asymmetry:
\begin{equation}
s_{\Max/\Min} = s \pm \gamma\,c,\qquad
c_{\Min/\Max} = c \mp \gamma\,c
\end{equation}
with $\gamma = \{0.2,\,0.1\}$ representing respectively  the present and future  uncertainty 
range for the relative shifts in the $s$ and $c$ quark distributions. 
The assumed uncertainties are compatible with those discussed in  
the recent review \cite{MRS2009}.   
As in the case of the $u^\val-d^\val$ asymmetry, we have assumed that the sum of the 
distribution of the $s$ and $c$ quarks will  be controlled to a very good precision 
by the $Z$-boson rapidity distribution. Therefore we have introduced only  unconstrained, mutually compensating 
modifications of the $s$ and $c$ quark distributions%
\footnote{In reality, the $s$ and $c$ quarks 
couple to the $Z$-boson with slightly different strength but the resulting effect will play no 
important role in the presented analysis.}.  
As seen previously in Eqs.~(\ref{eq_sWp_sWm_pp}) and  (\ref{eq_sWp_sWm_dd}), 
the valence quarks 
excess magnifies the contribution of the $s-c$ uncertainty to the 
measurement precision of $W$-boson mass.  

\subsubsection{WBpB emittance}
\label{sss:WBpBemittance}
The $u^\val-d^\val$ and  $s-c$ longitudinal momentum asymmetries 
would have no effect on the measured
$W$-boson mass asymmetry in the case of the collinear partonic beams.
The angular divergence (transverse-momentum smearing) of the WBpB at the LHC
is driven by the gluon radiation. Its parton-shower Monte Carlo model  
determines the relationship between  the longitudinal and the transverse 
degrees of freedom of the WBpB. It may  give rise to a the parton-shower-model-dependent 
asymmetries of the $\Wp$ and $\Wm$ boson transverse momenta. 

Instead of trying to estimate 
the uncertainties related to the precision of the parton-shower
modeling of the quark-flavour dependent effects, we allow 
for exceedingly large uncertainty in the size of the flavour-independent 
primordial transverse momentum Gaussian smearing of the WBpB: 
$\sigma_\kT= 4^{+3}_{-2}\GeV $ (the $\cal{PD}$ samples have been simulated for the following
values of the sigma parameter of the Gaussian smearing: 
$\sigma_\kT=2,3,4,5,6,7\GeV $).
Such a large uncertainty range, easily controllable using 
the $Z$-boson transverse momentum distribution, represents the effect of   
amplifying (small values of $\sigma_\kT$) or smearing out 
(large values of $\sigma_\kT$) the flavour dependent asymmetries
of the WBpB transverse momentum. 
It has to be stressed that  the resulting uncertainty of the distributions of the transverse momenta 
of the  leptons coming from decays $\Wp$ and $\Wm$ bosons is --  in our  mass measurement region -- 
sizably larger that the systematic effect due to the missing NLO QCD corrections.
In addition, the range has been chosen to be large enough to cover 
the uncertainties due to: (1) nonperturbative effects, e.g. those discussed in \cite{Gieseke:2007ad},
(2) the quark-mass effects and (3) resummation effects.  

\subsubsection{EW radiative corrections}
\label{sss:EWcor}

Out of the full set of the EW radiative corrections implemented in the \WINHAC\  
generator, those representing the emission of real photons could contribute
to the measured $W$-mass charge asymmetry. Two effects need to be 
evaluated: the charge-asymmetric interference terms between the photon emission in the 
initial and final states, and the radiation of the photons in the $W$-boson decays
in the presence of the $V-A$ couplings.
The above corrections are described to a high precision by \WINHAC, as has been shown
in Refs.~\cite{CarloniCalame:2004qw,Bardin:2008fn}. Therefore, their influence on the
$W$-mass charge asymmetry measurement can be modeled very accurately.  
In this paper we do not consider these effects, leaving a detailed study for our future
works.

\subsection{Experimental uncertainties}

\subsubsection{Energy scale (ES) of the charged lepton}
\label{sss:ES}
The uncertainty in the lepton energy scale is the most important source of 
the $\MW$ measurement error for the Tevatron experiments. 
At the LHC, producing unequal numbers of the $\Wp$ and $\Wm$ bosons,
its impact on the overall measurement precision will be amplified. 
For the measurement methods discussed in this paper the lepton 
energy scale error will be determined: (1) by the curvature radius (sagitta)
measurement errors, (2) by the uncertainties in the magnetic field maps 
within the tracker volume, and (3) by the modeling precision 
of the physics processes which drive the link between the measurements
of the particle hits in the tracker and the reconstructed particle 
momentum. While the first two sources of the 
measurement error are independent of the lepton flavour, 
the third one affects the electron and muon samples differently. 
In the following we shall assume, on the basis of the Tevatron 
experience, that modeling of physics processes of particle 
tracking will be understood at the LHC to the required level of precision, 
on the basis of dedicated auxiliary measurements\footnote{For example,  
the energy loss of the electrons in the dead material within the tracker 
volume will be understood using a conjugate process of the photon conversion.}.
This simplification allows us to discuss  the muon and electron track measurement 
simultaneously. We assume as well that the solenoid magnetic field strength in the 
volume of the tracker will be understood to better than $0.1\percent$ of its 
nominal value. We base this assumption on the precision of $0.01\percent$ achieved 
\eg{} by the H1 experiment at HERA \cite{H1:1990} and by the ALEPH experiment at LEP
\cite{ALEPH:1995}. If this condition is fulfilled, 
the energy scale error $\es_{l}$ is driven by the curvature radius measurement error:  
\begin{equation}
\rhol^\rec = \rhol^\smear \,(1+\es_{l}),
\end{equation}
where $\rhol^\rec$ and $\rhol^\smear$ are, respectively, the reconstructed   
and the true curvature smeared by the unbiased detector response function.

Based on the initial geometrical surveys, the initial scale 
of $\rho_l$ will be known to the precision of $0.5\percent$.
This precision will have to be improved at least by a factor of 10 
to match the precision of the Tevatron experiments,
if the same measurement strategy is applied. 
To achieve such a precision,
the local alignment of the tracker elements and/or average biases of 
the reconstruction of the trackers space-points must be known to the $\approx 3\, \mu\mm{m}$
precision. In addition, the global deformation of the tracker elements 
assembly must be controlled to a precision  which is beyond the reach  
of the survey methods. 

Several modes of the global deformations have to be considered.
They  have been discussed in details in Ref.~\cite{Brown:2006zz}.
The main difference between the measurements of the $W$-boson properties at the Tevatron 
and the LHC boils down to their sensitivity to the different types of the global deformation 
modes.
Both for the Tevatron and LHC measurements  
the $\Delta z$ translations are of no consequences since they  do not  affect the shape of the 
transverse projection of the particle helix. The $\Delta r$ deformations
(the radial expansion $r\,\Delta r$, the elliptical flattening $\phi\,\Delta r$ and the bowing $z\,\Delta r$)
give rise to common biases for positive and negative particle tracks. 
On the other hand, 
the $\Delta\phi$ curl and twist deformations 
give rise to biases which are opposite for negative and positive particles.
In the case of the Tevatron $\ppbar$ collisions, producing equal numbers 
of  the $\Wp$ and $\Wm$ bosons,  the 
dominant effect of $\pm z$-coherent curling of the outer-tracker layers with 
respect to the inner-tracker layers has  residual influence on the uncertainty of the
average $W$-boson mass, leaving the residual effect of relative twist of the $+z$ and $-z$  
sides of the tracker volume as the principal source of the measurement error. 
For the LHC $\pp$ collisions, producing unequal numbers of the $\Wp$ and $\Wm$ bosons, 
both deformation modes influence the measurement biases of the average $W$-boson mass.
In the case of the LHC there is no escape from the necessity of precise understanding 
of the lepton-charge-dependent biases on top of the lepton-charge-independent biases.  

In the presence of the above 
two sources of biases  the energy scale bias  $\es_{l}$
can be expressed in the limit of small deformations as follows:
\begin{eqnarray}
\es_\lp &=& \:\:\; \es_{\mathrm{curl}} + \es_{\Delta r},\\
\label{eq:epsilonp}
\es_\lm &=& - \es_{\mathrm{curl}} + \es_{\Delta r},
\label{eq:epsilonm}
\end{eqnarray}
where $\es_\mathrm{curl}$ represents the particle charge-dependent  $\Delta\phi$-type bias
and $\es_{\Delta r}$  represents the charge-independent $\Delta r$-type bias.

While the $\es_{\Delta r}$-type biases can be controlled with the help of the $Z$-boson, 
$\Upsilon$ and
$J/\Psi$ `standard candles', \eg{} using the CDF procedures, the global charge-dependent 
and symmetric $\es_\mathrm{curl}$ biases cannot. At the Tevatron these biases were investigated 
using the electron samples by  studying the charge dependent $E/p$ distribution, where $E$ is 
the energy of the electron (positron) measured in the calorimeter and $p$ is its  reconstructed
momentum. The relative scale error of positive and negative electrons was recalibrated using 
the mean values of the  $E/p$ distributions. The achieved precision was the principal 
limiting factor of the measurement of $\DeltaPM$.  Even if the statistical precision 
of such a procedure can be improved significantly at the LHC, this method is no longer unbiased.
This is related to the initial asymmetry of the transverse momentum distribution for 
positive and negative leptons in the selected $W$-boson decay samples. 
As a consequence, both the positive and negative lepton events, 
chosen for the calibration on the basis of the energy deposited in the calorimeter, will 
no longer represent charge-unbiased samples of tracks. The biases will be driven both by the 
influence of distribution shape and by the migration in and out of the chosen energy range. 
A partial remedy consists of using a statistically less-precise sample of positive and negative 
lepton tracks in a selected sample of $Z$-boson decays. However, due to the different 
weights of the $V-A$ and $V+A$ couplings of the $Z$-boson to leptons, even these track samples
are biased. In both cases these biases can be corrected for, but the correction factor will 
be sensitive to the uncertainty in the momentum spectra of the valence quarks.

Given the above sources of the uncertainties, we assume the following two values of the 
size of the biases, both for the charge-independent and charge-dependent scale shifts:  
\begin{eqnarray}
\es_\lp=+\es_\lm&=&\pm 0.5\percent,\,\pm 0.05\percent,\\
\label{eq:epspval}
\es_\lp=-\es_\lm &=&\pm 0.5\percent,\,\pm 0.05\percent. 
\label{eq:epsmval}
\end{eqnarray}
The first value correspond to the precision which can be achieved on the basis of the initial 
geometrical survey and 
the initial measurement of the field maps. The second one corresponds to what, in our view,
can be achieved using the above data based on the calibration methods -- given all 
the LHC-specific effects, which make this procedure more difficult at the LHC 
than at the Tevatron.  

\subsubsection{Resolution (RF) of the charged lepton track parameters }
\label{sss:resolution}
The finite resolution of measuring the lepton track 
parameters  may lead to biases in the measured value of 
$\MWp-\MWm$. We model the possible biases introduced by the 
ambiguity in the assumed size of the $\sigma_{1/\pT}$ and $\sigma_{\cotan\theta}$
smearing by decreasing and increasing the widths of their gaussian distributions by 
the factor $\mathrm{RF}=0.7,\,1.3$.

\section{In search for optimal measurement strategy}\label{sec_results}

\subsection{Reducing impact of systematic measurement errors}

In Section~\ref{s_measurement_method} three measurement methods 
of $\DeltaPM$ have been presented: \emph{the classic, charge asymmetry and 
double charge asymmetry} methods. The basic merits of the two latter 
methods is that they use the dedicated observables which are meant to be  largely insensitive 
to the precise understanding of the event selection and reconstruction efficiency, the background 
contamination level, understanding to the absolute calibration and the biases
of the reconstruction of the neutrino transverse momentum, the internal and 
external (dead-material) radiation. It will remain to be proved, using the data 
collected at the LHC, that all these error sources have negligible impact on the 
precision of the $\DeltaPM$ measurement. At present, such a statement must rely on  
the extrapolation of the Tevatron experience. 

The impact of the remaining measurement errors specified in previous section 
and quantified using the analysis methods discussed  in Section~\ref{s_analysis_strategy}  
is presented in Table~\ref{table_exp_sys_classic_vs_casym_vs_dcasym}.
The precision of estimating the systematic shifts of $\DeltaPM(\xi)$ for each of 
the systematic effect $\xi$ and each  
measurement method  is assessed using the validation procedures 
described in Section~\ref{s_analysis_strategy}. The resulting $\delta\left[\DeltaPM(\xi)\right]$ 
of $\approx 5\,$MeV corresponds the collected luminosity of $10\,\mathrm{fb}^{-1}$. 
The first observation is that the precise understanding of the measurement 
smearing of the track parameters does not introduce any bias in the 
measured values of $\DeltaPM$. The impact of the energy scale errors 
on the $\DeltaPM$ biases differs  for each of the discussed methods.

\begin{table}[]
\begin{center}
\begin{TableSize}
\renewcommand\arraystretch{1.45}
\begin{tabular}{|c|c||r@{\kern\tabcolsep}>{\kern-\tabcolsep}l|r@{\kern\tabcolsep}>{\kern-\tabcolsep}l|r@{}l|}
  \cline{3-8}
  \multicolumn{2}{c|}{} & \multicolumn{6}{c|}   {$\MWp-\MWm\quad[\mm{MeV}]$} \\ 
  \cline{2-8}
  \multicolumn{1}{c}{}  & \multicolumn{1}{|c||} {Systematic $\xi$}
                        & \multicolumn{2}{c|}   {``Classic'' Method}
                        & \multicolumn{2}{c|}   {$\Asym{\pTl}$} 
                        & \multicolumn{2}{c|}   {$\DAsym{\rhol}$}  \\

  \hline
  \multicolumn{1}{|c}{MC truth}   & \multicolumn{1}{|c||}{$\xi=0$} 
  & \multicolumn{2}{c|}{$\!\!\!\!-2 \,\pm\, 5$ }
  & \multicolumn{2}{c|}{$\!\!\!\!\!-1 \,\pm\, 3$ }
  & \multicolumn{2}{c|}{$ 0 \,\pm\, 3$ } \\
  \multicolumn{1}{|c}{Cent. Exp.} & \multicolumn{1}{|c||}{$\xi=0$} 
  & \multicolumn{2}{c|}{$1 \,\pm\, 6$ } 
  & \multicolumn{2}{c|}{$1 \,\pm\, 4$ }
  & \multicolumn{2}{c|}{$0 \,\pm\, 4$ } \\
  \hline\hline
  \multirow{8}{*}{ES [\%]}
  & $\es_\lp=+\es_\lm=+0.05\,\%$ 
  &   \qquad\quad\;\;$3$& 
  &   \qquad\quad\;$2$&
  &   \multicolumn{2}{c|}{} \\
  & $\es_\lp=+\es_\lm=-0.05\,\%$ 
  &   $-2$& 
  &    $0$&
  &   \multicolumn{2}{c|}{ \multirow{2}{*}{$\huge\times$}} \\
  \arrayrulecolor{Greymin}
  \cline{2-6}
  \arrayrulecolor{Black}
  & $\es_\lp=+\es_\lm=+0.50\,\%$ 
  &   $16$&
  &    $8$& 
  &   \multicolumn{2}{c|}{}  \\
  & $\es_\lp=+\es_\lm=-0.50\,\%$ 
  &  $-36$&
  &   $-6$& 
  &   \multicolumn{2}{c|}{}  \\
  \arrayrulecolor{Greymax}
  \cline{2-8}
  \arrayrulecolor{Black}
  & $\es_\lp=-\es_\lm=+0.05\,\%$ 
  &  $-56$& 
  &  $-57$& 
  &  \multicolumn{2}{c|}{\multirow{2}{*}{$1$}} \\
  & $\es_\lp=-\es_\lm=-0.05\,\%$ 
  &   $57$& 
  &   $57$& 
  &  \multicolumn{2}{c|}{} \\
  \arrayrulecolor{Greymin}
  \cline{2-8}
  \arrayrulecolor{Black}
  & $\es_\lp=-\es_\lm=+0.50\,\%$
  & $-567$& 
  & $-611$& 
  & \multicolumn{2}{c|}{\multirow{2}{*}{$\!\!\!\!-1$}} \\ 
  & $\es_\lp=-\es_\lm=-0.50\,\%$
  &  $547$& 
  &  $515$&
  &  \multicolumn{2}{c|}{} \\ 
  \hline\hline
  \multirow{2}{*}{RF}
  & $0.7$ 
  &  $1$& 
  & $-2$& 
  & \multicolumn{2}{c|}{ \multirow{2}{*}{$\huge\times$}} \\
  & $1.3$ 
  & $-3$& 
  &  $3$& 
  & \multicolumn{2}{c|}{} \\
  \hline
\end{tabular}
\renewcommand\arraystretch{1.45}

\end{TableSize}
  \caption[Systematic shifts of $\MWp - \MWm$ induced by  the measurement biases discussed in the text, 
                for the classic, charge asymmetry and  
    double charge asymmetry methods.]
          {\figtxt{Systematic shifts of $\MWp - \MWm$ induced by  the measurement biases discussed in the text, 
                for the classic, charge asymmetry and  
    double charge asymmetry methods.
    The  "MC truth" and the "Cent. Exp." represent the unbiased pseudo data samples (1) at the generator,  and (2) at the   
                  detector level. They are used  here  both as a cross-check of the estimation method and as a calibration of the 
                  statistical errors of  the presented systematic shifts.
            }}
            \label{table_exp_sys_classic_vs_casym_vs_dcasym}
\end{center}
\end{table}

For the lepton-charge-independent shift even the `initial' ($0.5\percent$)
scale error has no statistically significant impact on the measurement precision 
for the charge asymmetry method. For the classic method the scale error 
has to be reduced to the `ultimate' value of ($0.05\percent$) to achieve 
a comparable  measurement precision of $\DeltaPM$.

\begin{figure}[!h] 
  \begin{center}
    \includegraphics[width=0.495\tw]{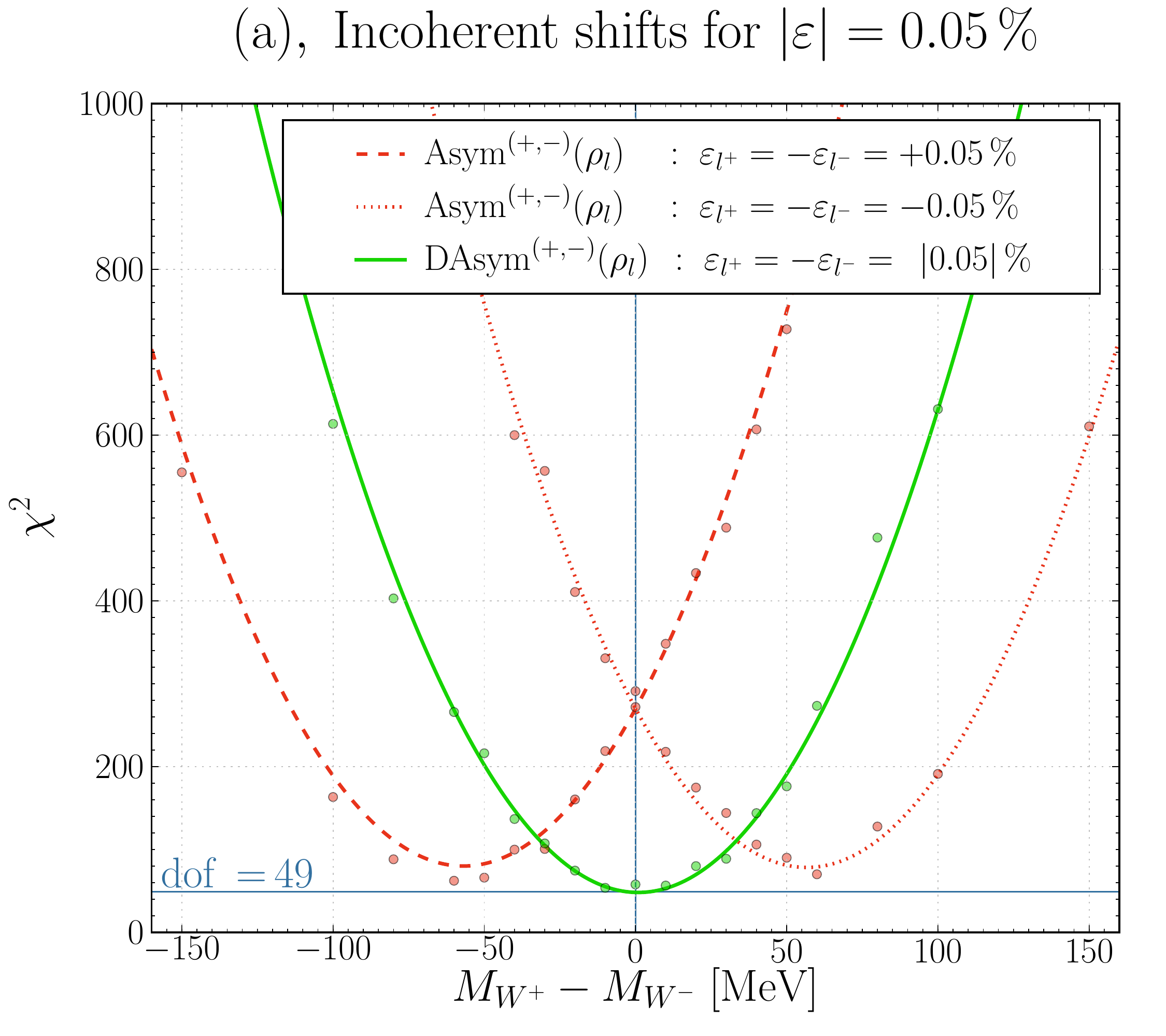}
    \hfill
    \includegraphics[width=0.495\tw]{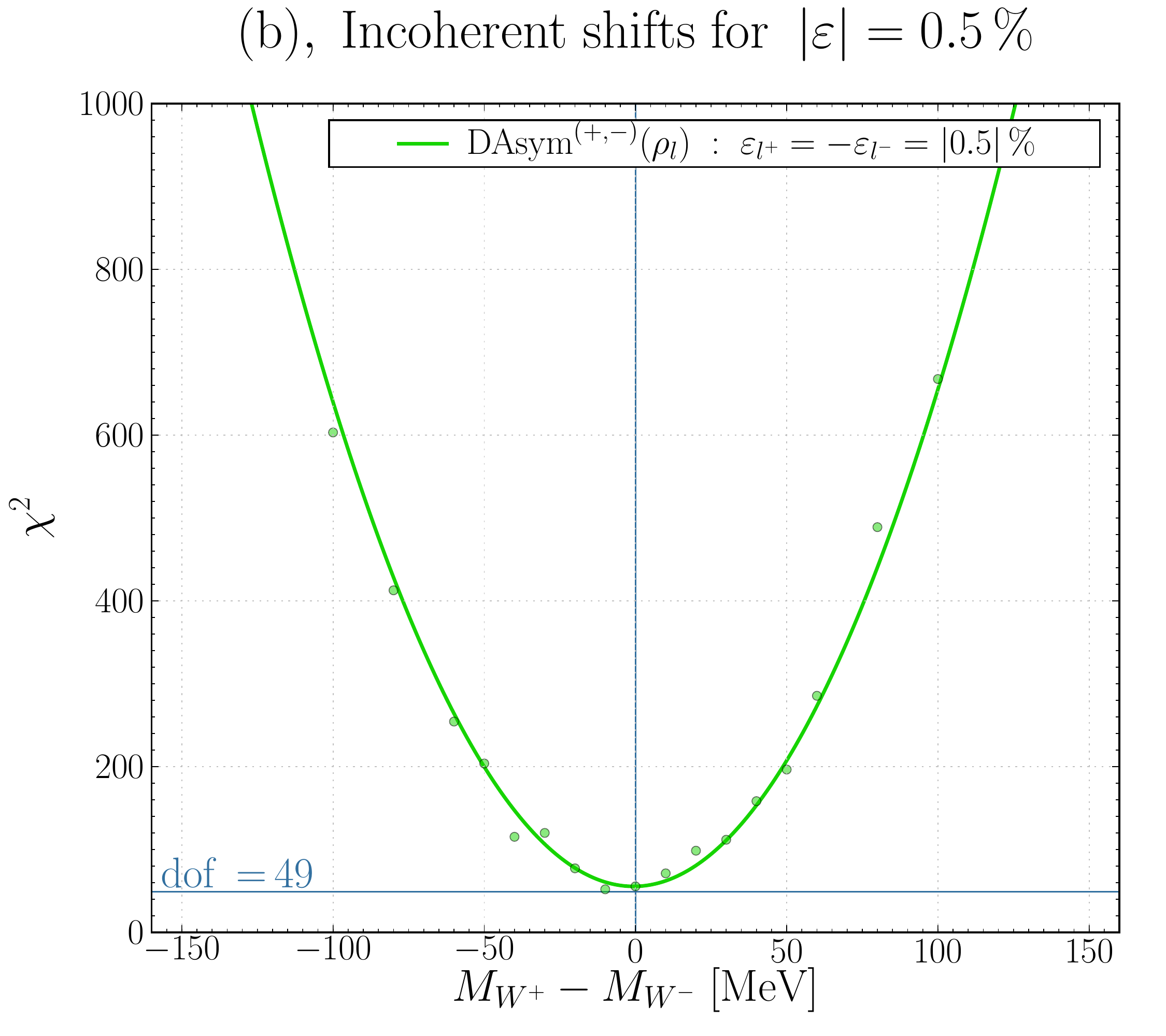}
    \caption[]
            {\figtxt{The $\chiD$ values and their polynomial fits for the incoherent shifts of the energy scale
                for the charge asymmetry and double charge asymmetry methods 
                using $|\varepsilon|=0.05\percent$  (a)
                and (only for the double charge asymmetry) using $|\varepsilon|=0.5\percent$ (b).
              }
            }
            \label{fig_chi2_ES_asym_dasym}
  \end{center} 
\end{figure}

For the lepton charge-dependent shifts the classic and asymmetry methods
provide similar measurement precision. The measurement error remains to be 
large ($\approx 60\,$MeV), even if the ultimate precision of controlling the 
energy scale biases is reached.  The double charge asymmetry method reduces 
the measurement error to the extend that the resulting  bias is statistically 
insignificant, even for the initial scale uncertainty. This is illustrated 
in Figures~\ref{fig_chi2_ES_asym_dasym}a and \ref{fig_chi2_ES_asym_dasym}b. 
These plots show the comparison of 
the $\chiD$ fits for the charge asymmetry method and the double charge 
asymmetry method for the lepton charge-dependent scale error of 
$|\es|=0.05\percent$ (a), and the $\chiD$ fit 
corresponding to $|\es|=0.5\percent$ for the double charge asymmetry method (b). 
The results for the double charge asymmetry correspond to $\es_\lp=-\es_\lm>0$ for the first 
running period with the standard magnetic field configuration  and $\es_\lp=-\es_\lm<0$ for the 
running period with the inverted direction of the $z$-component of the magnetic field.
 
The above reduction of the measurement 
sensitivity to the energy scale error can be achieved for the initial survey precision
of the tracker alignment. Such a survey will have 
to be made at the beginning of each of the two running periods. A special 
care will have to be taken to understand the relative curl and twist deformations
induced by reversing the current in the solenoid. 
It has to be stressed that the precision of the double charge asymmetry method
is insensitive to the relative $\vec E\times \vec B$ biases of the reconstructed 
hit-positions for the two data-taking periods,  provided that they are 
not larger than 10 times the average hit reconstruction precision achieved
in the standard-field-configuration running period. 
Worsening of the hit-position resolution for the inverted-field configuration, 
driven by the geometrical layout of the silicon tracker modules,  
have no significant effect on the measurement precision. 
Similarly,  the required level of precision of understanding the hysteresis effects, leading to 
inequality of the absolute field strength in the two running periods, corresponding 
to reverse solenoid current directions can be achieved with the standard field 
mapping methods. Note that the precision required for the asymmetry measurement may be up to 
$10$ times worse with respect to the one needed for the measurement of the average $W$-boson mass.   
The reduced sensitivity to all the above effects 
is due to the fact that the impact of each of these effects is, to a large 
extent, canceled in each of the running periods. This is done in the same way as canceling the  
time-dependent effects of the detector response, calibration and alignment procedures.   
Note, that the residual impact of all the above effects can be reduced further
(if necessary)  using the 
$B$-field configuration-dependent analyses of straight track residua and/or 
the position of the reconstructed $Z$-boson mass peak.

\subsection{Reducing impact of systematic modeling errors}

As discussed in the previous section, by using the double charge asymmetry method the systematic measurement precision of 
$\DeltaPM$ could  be reduced to the level of ${\cal O}(10)\,\MeV$. 
In this section we discuss the impact of the 
modeling uncertainties described in Subsection~\ref{s_model_impact_sys_pheno} 
on the measurement 
precision of $\DeltaPM$ for the charge asymmetry method%
\footnote{From the point of view of the modeling uncertainties,
the charge asymmetry and the double charge asymmetry methods are equivalent
and the discussed results are the same for both methods.}.

In Table~\ref{table_charge_asym_results} we show, in the first column, the expected measurement 
biases of $\DeltaPM$ due to the dominant modeling uncertainties, discussed in the 
previous section, for $\pp$ collisions at the LHC energy. 
We do not see a significant impact of the coherent shifts 
of the partonic distributions, defined in the previous section and denoted as 
the PDFs uncertainty. It would be, however,  misleading to conclude 
prematurely that the  $\DeltaPM$ biases are  insensitive to the uncertainties in the partonic 
distributions.  

\begin{table}[]
\begin{center}
\begin{TableSizeTwo}
\renewcommand\arraystretch{1.5}
\begin{tabular}
{|c|l||r@{\kern\tabcolsep}>{\kern-\tabcolsep}l|r@{\kern\tabcolsep}>{\kern-\tabcolsep}l|r@{\kern\tabcolsep}>{\kern-\tabcolsep}l|r@{\kern\tabcolsep}>{\kern-\tabcolsep}l|}
  \cline{3-10}
  \multicolumn{2}{c|}{} & \multicolumn{8}{c|}  {$\MWp-\MWm\;[\mm{MeV}]$ results using the Charge Asymmetry of $\pTl$} \\ \cline{2-10}
  \multicolumn{1}{c|}{} & \multicolumn{1}{c||} {Systematic $\xi$}
                        & \multicolumn{2}{c|}   {$\pp$ - $|\etal|<2.5$}
                        & \multicolumn{2}{c|}   {$\pp$ - $|\etal|<0.3$  } 
                        & \multicolumn{2}{c|}   {$\pp$ - $|\yW|<0.3$}
                        & \multicolumn{2}{c|}   {$\dd$ - $|\etal|<2.5$}  \\

  \cline{2-10}\hline
  \multicolumn{1}{|c|}{MC truth}   
  & \multicolumn{1}{c||}{$\xi=0$} 
  & \multicolumn{2}{c|}{$\!\!\!\!-1 \,\pm\, 3$ }
  & \multicolumn{2}{c|}{$0  \,\pm\, 1$ }
  & \multicolumn{2}{c|}{$0  \,\pm\, 1$ }
  & \multicolumn{2}{c|}{$\!\!\!\!\!-2 \,\pm\, 4$ } \\
  \multicolumn{1}{|c|}{Cent. Exp.} 
  & \multicolumn{1}{c||}{$\xi=0$} 
  & \multicolumn{2}{c|}{$1 \,\pm\, 4$ }
  & \multicolumn{2}{c|}{$0 \,\pm\, 4$ }
  & \multicolumn{2}{c|}{$1 \,\pm\, 4$ }
  & \multicolumn{2}{c|}{$0 \,\pm\, 4$ } \\
  \hline\hline
  \multirow{5}{*}{$\sigma_\kT$ [GeV]}
  & \multicolumn{1}{c||}{$2$} 
  & \hspace*{.75cm} $8$&
  & \hspace*{.75cm} \cellcolor{hl}{$0$}&\cellcolor{hl}{} 
  & \hspace*{.85cm} \cellcolor{hl}{$2$}&\cellcolor{hl}{} 
  & \hspace*{.6cm} $24$& \\
  & \multicolumn{1}{c||}{$3$}
  &  $7$&
  &  \cellcolor{hl}{$3$}&\cellcolor{hl}{}
  & \cellcolor{hl}{$-2$}&\cellcolor{hl}{} 
  & $17$& \\
  & \multicolumn{1}{c||}{$5$}
  & $-4$&
  & \cellcolor{hl}{$-3$}&\cellcolor{hl}{}
  & \cellcolor{hl}{$-6$}&\cellcolor{hl}{} 
  &$-15$& \\
  & \multicolumn{1}{c||}{$6$}
  & $-8$&
  &  \cellcolor{hl}{$2$}&\cellcolor{hl}{}
  & \cellcolor{hl}{$-5$}&\cellcolor{hl}{} 
  &$-34$&  \\
  & \multicolumn{1}{c||}{$7$}
  & $-16$& 
  &   \cellcolor{hl}{$2$}&\cellcolor{hl}{}
  &  \cellcolor{hl}{$-8$}&\cellcolor{hl}{} 
  & $-52$& \\
  \hline\hline
  \multirow{2}{*}{PDF}
  & \multicolumn{1}{c||}{Min.} 
  & $-4$&
  &  $6$&
  &  $0$& 
  & $-3$& \\
  & \multicolumn{1}{c||}{Max.}
  &  $4$&
  & $-8$&
  &  $5$&
  &  $2$& \\
  \hline\hline
  \multirow{8}{*}{$u^\val,\,d^\val$}
  & \twolinebox{${u^\val_\Max=1.05\,u^\val}$}{${d^\val_\Min\,=d^\val-.05\,u^\val}$} 
  & $115$&
  &  $69$&
  & $-38$&
  &  \cellcolor{hl}{$2$}&\cellcolor{hl}{} \\
  \arrayrulecolor{Greymin}
  \cline{2-10}
  \arrayrulecolor{Black}
  & \twolinebox{${u^\val_\Min\,=0.95\,u^\val}$}{${d^\val_\Max=d^\val+.05\,u^\val}$} 
  & $-139$&
  &  $-87$&
  &   $60$&
  &  \cellcolor{hl}{$3$}&\cellcolor{hl}{} \\
  \arrayrulecolor{Greymax}
  \cline{2-10}
  \arrayrulecolor{Black}
  & \twolinebox{${u^\val_\Max=1.02\,u^\val}$}{${d^\val_\Min\,=0.92\,d^\val}$} 
  &  $84$& 
  &  $53$& 
  & $-31$& 
  &   \cellcolor{hl}{$4$}&\cellcolor{hl}{} \\
  \arrayrulecolor{Greymin}
  \cline{2-10}
  \arrayrulecolor{Black}
  & \twolinebox{${u^\val_\Min\,=0.98\,u^\val}$}{${d^\val_\Max=1.08\,d^\val}$}
  & $-89$&
  & $-57$& 
  &  $44$& 
  &  \cellcolor{hl}{$4$}&\cellcolor{hl}{} \\
  \hline\hline
  \multirow{6}{*}{$s,\,c$}
  & \twolinebox{${c_\Min\,=0.9\,c},$}{${s_\Max=s+0.1\,c}$}
  & $17$&
  & $10$&
  & \cellcolor{hl}{$7$}&\cellcolor{hl}{}
  & $19$& \\
  \arrayrulecolor{Greymin}
  \cline{2-10}
  \arrayrulecolor{Black}
  & \twolinebox{${c_\Max=1.1\,c},$}{${s_\Min\,=s-0.1\,c}$}
  & $-11$& 
  & $-10$& 
  &  \cellcolor{hl}{$0$}&\cellcolor{hl}{} 
  & $-19$& \\
  \arrayrulecolor{Greymax}
  \cline{2-10}
  \arrayrulecolor{Black}
  & \twolinebox{${c_\Min\,=0.8\,c},$}{${s_\Max=s+0.2\,c}$}
  & $39$& 
  & $25$& 
  &  \cellcolor{hl}{$6$}&\cellcolor{hl}{} 
  & $38$& \\
  \arrayrulecolor{Greymin}
  \cline{2-10}
  \arrayrulecolor{Black}
  & \twolinebox{${c_\Max=1.2\,c},$}{${s_\Min\,=s-0.2\,c}$}
  & $-29$& 
  & $-24$& 
  & \cellcolor{hl}{$1$}&\cellcolor{hl}{}
  & $-39$& \\
\hline
\end{tabular}
\renewcommand\arraystretch{1.45}

\end{TableSizeTwo}
  \caption[]
          {\figtxt{Systematic shifts of $\MWp - \MWm$ induced by  the modeling  biases discussed in the text, 
                for the charge asymmetry method.
                 The  "MC truth" and the "Cent. Exp." represent the unbiased pseudo data samples (1) at the generator,  and (2) at the   
                  detector level. They are used  here  both as a cross-check of the estimation method and as a calibration of the 
                  statistical errors of  the presented systematic shifts. }}
            \label{table_charge_asym_results}
\end{center}
\end{table}

Indeed, the  present uncertainty of the relative distribution of the 
the $u$ and $d$ valence quarks (the  $u^\val-d^\val$  asymmetry) 
leads to large shifts in the $\DeltaPM$ values. These shifts 
are specific to the LHC $\pp$ collider and are largely 
irrelevant for the Tevatron $\ppbar$ collisions. This is perhaps 
the reason why they were neglected in the previous studies \cite{Atlas_W,CMS_W},
in spite that they concern the average $W$-boson mass measurement. 
There are three origins for these shifts. The effects due to each of them add up 
and result in the amplification of the biases.
Firstly, increasing the $u^\val$ content 
of the proton shifts downwards the average momentum of the $\bar d$ antiquarks.
This leads to increase of the average transverse momentum of  the $\bar d$ antiquarks 
producing $\Wp$,  mimicking the increase of the $\Wp$-boson mass. Simultaneous 
decreasing of the $d^\val$ acts in the opposite direction for $W^-$,  
leading to large and positive values of $\DeltaPM$. Secondly, 
at the LHC, contrary to the Tevatron,  the presence of the $d^\val$ quarks 
leads to an asymmetry in the production rate of the $W$-boson from the $c$
quarks and $\bar c$ antiquarks. Since the average transverse 
momentum of the charm quarks is higher with respect to the 
light quarks,  this asymmetry shows up  in the relative shifts 
in the $\pTl$  distributions for positive and negative leptons.
Increasing the density of the $d^\val$ quarks mimics 
thus the effect of increasing the mass of the $\Wm$ with respect to the 
$\Wp$ boson. 
The above two effects are amplified by the bias in 
the degree of the transverse polarisation 
of $\Wm$ with respect to $\Wp$,   
induced by the event selection procedure based on the 
lepton kinematics. The relative movements of the $d^\val$ 
and $u^\val$ amplify (attenuate) the initial event selection 
procedure bias.  
What must be stressed is that if the $d^\val$ shifts are  
compensated by the corresponding shifts of the $u^\val$ distributions,
they cannot be constrained to a better precision by the present data,
and they will not affect the rapidity distributions of the $Z$-boson. 
Thus, it will be difficult to pin them down using the standard 
measurement procedures. 

The uncertainties of the relative density of the strange and charm quarks, 
the $s-c$ asymmetry, gives rise to smaller but significant biases  in the $\DeltaPM$ values,
as shown in the first column of  Table~\ref{table_charge_asym_results}. 
Since the transverse momentum of the $c$ quarks is significantly higher than the corresponding 
momentum for the  $s$ quarks, this effect, even if  Cabbibo-suppressed,  cannot be neglected. 
What must be stressed again is that if the $c$ shifts are  
compensated by the corresponding shifts of the $s$ distributions,
they will not affect the rapidity distributions of the $Z$-bosons. 
Thus, it will be difficult to pin them down using the standard 
measurement procedures. This asymmetry can be constrained unambiguously only  
by dedicated measurements, \eg{} by measuring the associated production of the $W$-bosons 
and charmed hadrons.

Compared to the above, the biases corresponding to the 
uncertainties in the flavour independent smearing of the intrinsic transverse 
momentum distribution of partons are smaller in magnitude and can be neglected, 
if the intrinsic transverse momentum of partons is controlled to the precision of 
$2\GeV$.

It is obvious from the above discussion that using the standard measurement 
procedures, the modeling uncertainties will be the dominant  source of the 
measurement errors of the $W$-boson mass asymmetry, already for the collected luminosity 
$100$ times smaller than the one considered in this paper.
In order to diminish the impact of the modeling errors on the measurement of $\DeltaPM$
to a level comparable to statistical and experimental measurement errors, some dedicated 
measurement methods must be applied. Two such procedures are proposed and evaluated below:
(1) {\it the narrow bin method} and (2) {\it the isoscalar beams method}.

\subsubsection{Narrow bin method}
\label{sss:narrowbin}
As discussed previously, the dominant source of  
large uncertainties in  $\DeltaPM$ comes from the presence of the 
valence quarks in the WBpB and from the uncertainties in their 
flavour-dependent momentum distributions. In order to reduce this effect 
we propose to profit from the large centre-of-mass energy of the LHC and 
measure  $\DeltaPM$ using a selected fraction of the $W$-bosons which are 
produced predominantly by the sea rather than by the valence quarks. These 
$W$-bosons are produced with small longitudinal momentum in the laboratory frame.
 
Two methods of selecting such a sample are discussed below. The first
is based on restricting the measurement region to  $|\etal|<0.3$. 
The merit of the $\etal$-cut based selection is that it uses a directly 
measurable kinematical variable. Its drawback is that  rather broad 
spectrum of the longitudinal momenta of annihilating partons is accepted due to 
the large mass of the $W$-boson. The second is based on
restricting the measurement region to $|\yW|<0.3$. Here, only a narrow 
bin of the longitudinal momenta of annihilating partons is accepted in the region 
where the sea quarks outnumber the valence quarks. However, $\yW$ cannot 
be measured directly. It has to be unfolded from the measured transverse momentum 
of the charged lepton and the reconstructed transverse momentum of the neutrino.
The unfolding procedure \cite{Bodek:2007cz} neglects the width of the 
$W$-boson and depends upon the initial assumption of the relative momentum spectra
of the valence and sea quarks. However, in  the selected kinematical region 
the above approximation are expected to lead to a negligible measurement bias.
It has to be stressed that the narrow bin measurements will require a $10$ times
higher luminosity to keep the statistical error of $\DeltaPM$ at the level of 
$5\,$MeV. Therefore, the results presented below  for the narrow bin method
correspond to an integrated luminosity of $100\, \mathrm{fb}^{-1}$ and  
are based on the dedicated set of the simulated mass-template and pseudo-data 
samples. Each sample contains $N_\Wp=1.74\times 10^9$ and $N_\Wm=1.14\times 10^9$
simulated events, respectively. 

The systematic biases of $\DeltaPM$ due to modeling uncertainties discussed 
in the previous sections are presented in columns 2 and 3 of  
Table~\ref{table_charge_asym_results} respectively for the $|\etal|<0.3$
and $|\yW|<0.3$ selections. The $\etal$-cut based method reduces slightly  
the biases related to the uncertainties in the $u^\val-d^\val$ and $s-c$ asymmetries.
The gain in the measurement precision is clearly seen for the $\yW$-cut based method
which reduces to a negligible level the $s-c$ biases. It is interesting 
to note that the  $u^\val-d^\val$ shifts in $\DeltaPM$ change their signs for the above two 
methods, reflecting the importance of the $W$-boson polarisation effects discussed earlier.

The narrow bin method allows, thus to reduce the impact of the $W$-boson modeling 
uncertainties on the $\DeltaPM$ biases to the level comparable to the statistical 
precision for all the effects, except for the $u^\val-d^\val$ asymmetry effect.
Here another remedy has to be found.

\subsubsection{Isocalar beams}
\label{sss:isobeams}

Isoscalar targets have been successfully used in most of  the previous fix-target
deep inelastic scattering experiments at SLAC, FNAL and CERN,
but this aspect has been rarely discussed in the context of the electroweak physics at the LHC.
The merits of the ion beams for the generic searches 
of the electroweak symmetry breaking mechanism at the LHC have been 
discussed in \cite{Krasny:2005cb}. Their use as carriers of the parasitic electron 
beam, to measure the emittance of the WBpB at the LHC,  
has been proposed  in  \cite{krasny_ebeam}.
In the presented series of papers we shall strongly advocate 
the merits of the isoscalar beams in improving the measurement 
precision of the parameters of the Standard Model. In this 
section we discuss their role in increasing the precision of
the measurement of $\DeltaPM$. We shall 
consider light ions: deuterium or helium.
As far as the studies of the $W$-boson asymmetries are concerned,
they are equivalent because  
shadowing corrections are quark-flavour independent.
The energies of the LHC ion beams satisfy the equal magnetic 
rigidity condition. For the isoscalar beams the nucleon energy 
is thus two times lower that the energy of the proton beam.
In the presented studies we assume that the ion--ion luminosity 
is reduced by the factor $A^2$ with respect to the $\pp$ luminosity. 

In column 4 of Table~\ref{table_charge_asym_results}
we present the impact of the modeling uncertainties on the 
$\DeltaPM$ biases. The isosacalar beams allow to 
reduce the measurement biases due to the $u^\val-d^\val$ asymmetry effect
to a negligible level. This colliding beam configuration          
allows to profit from the isospin symmetry of the strong interactions 
which cancels the relative biases in the momentum distribution 
of the $u$ and $d$ quarks. It is interesting to note that, 
as expected, the $s-c$ biases are similar for the proton and 
for the light isoscalar beams. On the contrary, the biases 
related to the flavour-independent intrinsic momentum of the 
quarks and antiquarks are amplified due to the reduced centre-of-mass
collision energy%
\footnote{In order to amplify this effect, we have kept 
the same central value of the intrinsic transverse momentum 
smearing in the reduced collision energy as for the nominal collision
energy.}.   

\subsection{Two complementary strategies}

Two complementary strategies to achieve the ultimate measurement precision
of $\DeltaPM$ will certainly be tried. The first one will be based on 
an attempt to reduce the size of the systematic measurement and 
modeling uncertainties, discussed in the previous section. 
In our view, such a strategy will quickly reach the precision brick wall
-- mostly due to the a lack of data-driven constraints on modeling the 
flavour dependent $W$-boson production at the LHC energy. 
The second one, proposed in this paper,  instead of reducing the size of the uncertainties, 
attempts to reduce their impact on the systematic error of the measured quantity by 
applying the dedicated methods. Such a strategy requires
running the dedicated machine and detector configurations. It is thus time 
and luminosity consuming. However, in our view, only such a strategy 
allows to measure the $W$-mass charge asymmetries at the precision comparable 
to the one achieved in the muon decay experiments.  
  
Let us recollect the main elements of the proposed dedicated measurement strategy
that allow to reduce the systematic errors to the level shown in 
the shaded areas of Table~\ref{table_charge_asym_results}:
\begin{itemize}
\item
The charge asymmetry method allows 
to reduce the impact of most of the systematic measurement errors, 
except for the relative momentum-scale errors for the positive and negative leptons.
If they cannot be experimentally controlled to the level of ${\cal O} (10^{-4})$,
their impact can be drastically reduced  in the dedicated LHC running periods  
using the double charge asymmetry method. 
\item
The impact of the uncertainty in modeling of the 
intrinsic transverse momentum of the WBpB can  be reduced to a negligible level  using the 
narrow-bin measurement method. 
\item
The impact of the $s-c$ uncertainty can be attenuated  
using the $\yW$-selection-based narrow bin method. 
\item
Finally, the impact of the $u^\val-d^\val$
uncertainty can be annihilated in the dedicated LHC runs with light isoscalar beams. 
\end{itemize}

\section{Summary and outlook}\label{s_summary}

This paper is the  second in our  series of papers devoted 
to optimisation of measurement strategies of the Standard
Model parameters at the LHC.  
It presents a dedicated strategy for the precision 
measurement of the charge asymmetry of the $W$-boson mass at the LHC.

This measurement must, in our view, 
precede the measurement of 
the charge-averaged mass of the $W$-boson and the measurement of 
$\sin\theta _W$, in order to diminish the risk of false absorption of 
variety of unknown Beyond-the-Standard-Model effects within  the 
Standard Model parameter space. 
This measurement is of particular importance for the 
two following reasons. Firstly, at the LHC  -- contrary to the Tevatron 
$\ppbar$ collider -- the measurement of the averaged mass of the $W$-boson cannot 
be dissociated from the measurement of the masses of its charge 
states. Secondly, the precision of verification of the  
charge-universality of the 
Fermi coupling constant $\GF$, measured via the charge asymmetry of the 
muon life-time, must be matched by the precision of  verification  
of the charge universality of the $SU(2)$ coupling strength $g_W$.
This can be achieved only if the mass difference $\MWp - \MWm$
can be determined with the precision of a few MeV, 
\ie{} a factor of $\sim 20$ better than the best present measurement.   
  
The Tevatron $\ppbar$ collision mode, as far as the 
systematic and modeling errors are concerned,  is better suited 
for the precision measurement of the $W$-boson charge asymmetry.
However, the measurement will be limited by the statistical 
precision, affecting both the $W$-boson samples  and, 
more importantly,  the $Z$-boson calibration sample. 

At the LHC the requisite statistical precision can be achieved already 
for the integrated luminosity of $10\,{\rm pb}^{-1}$, \ie{} in the first 
year of the LHC operation at the `low', 
${\cal O}(10^{33}\,{\rm cm}^{-2}{\rm s}^{-1})$, 
luminosity.
However, in order to achieve a comparable systematic precision
in an analysis based on the calibration and measurement strategies 
developed at the Tevatron,  
the charge-dependent biases in the energy (momentum) scale of 
positive (negative) leptons must be controlled to the precision 
of $0.005\percent$. As we have argued in this paper, it will be extremely    
hard, if not impossible, to achieve such a precision using 
the calibration methods developed at the Tevatron. 
Moreover, we have identified the LHC-specific sources of errors,
related to the uncertainties in the present knowledge 
of the flavour composition of the WBpB
which limit,  at present,   the measurement precision to 
${\cal O}(100\,{\rm MeV})$.
Certainly, this uncertainty will be reduced once the high-statistic 
$W$-boson and $Z$-boson samples are collected. Nevertheless, 
it will be hard, if not impossible,  to  
improve by a factor of 10 or more the precision
of the $u^\val$--$d^\val$ and $s$--$c$ quark-momentum distribution asymmetries, 
as they are hardly  detectable in the $Z$-boson production processes.  
Whether or not the requisite precision target will be reached using
the standard measurement strategies remains to be seen. In our view, 
it will be indispensable to use the dedicated 
LHC-specific measurement strategies.

The strategy proposed and discussed here makes a full use of  
the flexibility of the machine and detector configurations
which, we hope, will be exploited  
in the mature phase of the LHC experimental program. 
It requires: (1) running for a fraction of time the inverted-polarity 
current in the detector solenoid, (2) the dedicated trigger 
and data-acquisition  configuration 
in the `high', ${\cal O}(10^{34}\,{\rm cm}^{-2}{\rm s}^{-1})$, 
luminosity LHC operation mode
and (3) replacing the LHC proton beams by the light isoscalar-ion beams.  

The underlying principle of the proposed dedicated strategy is that,  
instead of diminishing the systematic measurement 
and modeling uncertainties, it  minimizes their influence 
on the measured value of $\MWp -\MWm$. We have demonstrated
that already for the modest 
(easy to fulfill) measurement and modeling precision
requirements,  the resulting uncertainty of $\MWp -\MWm$
can be kept at the level comparable to the statistical 
measurement uncertainty, \ie{} at the level of ${\cal O}(5\,{\rm MeV})$.

It remains to be demonstrated that the remaining  systematic measurement 
errors, of the secondary importance at the Tevatron and not discussed here,
could be neglected at this level of the measurement precision. 
This can, however, be proved only when the real data are collected
and analysed, both for the standard and for the dedicated 
measurement strategies.

\vspace{10mm}
\noindent
{\large\bf Acknowledgements} \\
We are indebted to F.~Lediberder for his  encouragement and support 
for our {\it ``Generic Searches and Dedicated Measurements at LHC''}  
research program.


\end{document}